\documentclass{aa}  

\usepackage{graphicx}
\usepackage{txfonts}
\usepackage{lipsum}
\usepackage{verbatim}
\usepackage{subcaption}         
\usepackage{lscape}             
\usepackage{placeins}           
                                
\begin{document}

   \title{JWST/NIRCam observations of HD~92945 debris disk: An asymmetric disk with a gap}

   \author{C. Lazzoni\inst{1}
        \and R. Bendahan-West\inst{2}
        \and S. Marino\inst{2}
        \and K. D. Lawson\inst{3}
        \and A. Carter\inst{4}
        \and V. Squicciarini\inst{2}
        \and G. Strampelli \inst{4}
        \and S. Hinkley\inst{2}
        \and G. Kennedy \inst{5,6,7}
        \and A. D. James\inst{2}
        \and J. Milli \inst{8}
        \and S. Ray \inst{9}}

   \institute{INAF, Astronomical Observatory of Padua, Vicolo dell'Osservatorio 5, I - 35122, Padua, Italy\\
             \email{cecilia.lazzoni@inaf.it}
            \and Department of Physics and Astronomy, University of Exeter, Stocker Road, Exeter EX4 4QL, UK
            \and NASA-Goddard Space Flight Center, Greenbelt, MD 20771, USA
            \and Space Telescope Science Institute, 3700 San Martin Dr, Baltimore, MD 21218, USA
            \and Centre for Exoplanets and Habitability, University of Warwick, Coventry CV4 7AL, UK
            \and Department of Physics, University of Warwick, Coventry CV4 7AL, UK
            \and Malaghan Institute of Medical Research, Gate 7, Victoria University, Kelburn Parade, Wellington, New Zealand
            \and Univ. Grenoble Alpes, CNRS, IPAG, F-38000 Grenoble, France
            \and School of Mathematics and Physics, University of Queensland, St Lucia, QLD 4072, Australia}

   \date{Received September 30, 20XX}

 
  \abstract
   {}
   {We present the first observations of the HD\, 92945 debris disk obtained with the James Webb Space Telescope (JWST), targeting this nearby K0V star located at 21.54 pc from the Sun. The main objectives are to characterize the disk's morphology in the near infrared, compare it with previous data from ALMA and HST, and place new constraints on the presence and properties of potential planetary companions shaping the disk.} 
   {High-contrast coronagraphic imaging was performed using JWST/NIRCam in the F200W and F444W filters. Advanced postprocessing techniques were employed, including reference differential imaging (RDI) with custom-built point spread function (PSF) libraries, and forward modeling of the disk using synthetic PSFs and MCMC optimization. After subtracting the disk contribution, the residuals were analyzed to identify candidate point sources. From these, we derived contrast curves and constructed detection probability maps for substellar companions.}
   {The disk is clearly detected in both NIRCam filters and reveals a broad, inclined structure with a gap, consistent with previous scattered-light and ALMA observations. The modeling confirms the presence of a gap at $\sim$80 au and shows a scale height and scattering properties compatible with a dynamically active disk.
   A significant brightness asymmetry is observed in the southwestern inner ring at both 2 and 4.4 $\mu$m, consistent with previous ALMA results. Observing this feature across different wavelengths and epochs strongly supports a scenario where one or more unseen planetary companions are perturbing the disk. No comoving sources are detected, and all candidate objects in the field are consistent with background stars or galaxies. The derived detection limits exclude planets more massive than $\sim$0.4-0.5 M\textsubscript{Jup} beyond 100 au and more massive than $\sim$1 M\textsubscript{Jup} beyond 20-40 au. This, in turn, rules out the possibility of a single planet placed beyond $\sim20$ au as responsible for the astrometric signal observed by Gaia. 
   These results, combined with the observed disk features, support a scenario in which a single or multiple sub-Jupiter planets dynamically shape the system through mechanisms such as secular apsidal resonances, providing a coherent explanation for the gap, the asymmetric brightness distribution and the astrometric signal.}
   {}


   \maketitle

\nolinenumbers
\section{Introduction}

Direct imaging of exoplanets is emerging as a pivotal technique for advancing our understanding of planetary systems, particularly the outer extents of their architectures. This method, which isolates the faint light of a planet from its much brighter host star, is uniquely positioned not only to provide critical insights into planetary systems but also to outline the structures of circumstellar debris remnants of planetary formation.

Despite the potential of direct imaging, especially in the context of young, self-luminous planets, ground-based surveys have historically encountered significant challenges. Detection rates for young, widely separated planets have been lower than initially expected \citep{Nielsen1,Vigan2021}. These results are now attributed to the rarity of such planets coupled with the limitations inherent in ground-based observations, such as bright speckles, generated by the atmospheric interference which dominate the noise close to the star where planets are more likely to be found. 

Targeted searches using indicators such as structured circumstellar debris disks have been attempted in the past years \citep[e.g.,][]{Meshkat,Meshkat1,Dahlqvist,Tamura}, based on the idea that dynamical sculpting of these disks may be caused by unseen planetary companions. Features such as gaps, warps, asymmetries, or sharp edges can trace the presence of planets, making debris disk systems especially promising targets for direct imaging efforts. However, these surveys struggled to yield significant results, highlighting the complexities of predicting and identifying these distant worlds. In any case, features in disks,  when detected and characterized, can help to put strong constraints on the mass and orbital parameters of hidden planets \citep[see e.g., ][]{ Tabeshian, Marino4, Yelverton, Friebe, Sefilian,Pearce1}, especially in multiwavelength surveys.

The James Webb Space Telescope (JWST) offers a new eye on the sky, giving for the first time the chance of performing high-contrast imaging from space at both near and mid-infrared wavelengths, accessing the yet unexplored range of fainter and lower-mass directly imaged planets. The capabilities of this observatory were assessed during the ERS program \citep{Hinkley1, Carter} and were later confirmed through the detection of further faint and cold companions \citep[see for example][]{Lagrange6, Matthews2}.

Moreover, JWST proved high sensitivity to faint debris disks, revealing details which are beyond the capabilities of previous telescopes. Striking examples of JWST's capabilities include the observation of the debris disk around $\beta$ Pictoris \citep{Rebollido}, revealing new structures in the disk—such as the "cat's tail"—believed to result from radiation acting on small dust particles in the aftermath of collisions. Another example is the detection of small dust particles in the Fomalhaut system \citep{Gaspar1}, resolved for the first time in the mid-infrared with the Mid-Infrared Imager (MIRI).

In this paper, we present the first images of the HD\, 92945 system observed with JWST with the Near-Infrared Camera (NIRCam). HD\, 92945 is a K0V spectral type star located at a distance of $21.54 \pm 0.02$ pc from the Sun \citep{GAIA2}. Its age is estimated to be between 100 and 300 Myr \citep{Song, Plavchan}, with the possibility of being a member of the AB Dor stellar group, which has an estimated age of $149^{+31}_{-49}$ Myr \citep{Nielsen2}. Spectroscopic analysis provided an estimation of the stellar mass of $0.86 \pm 0.01$ $M_{\odot}$ and of the stellar radius of $0.75 \pm 0.02$ $R_{\odot}$ \citep{Mesa5}.
The debris disk around HD\, 92945 was initially identified due to its flux at 60 $\mu$m measured by IRAS \citep{Silverstone}, significantly higher than the expected photospheric flux at that wavelength. Infrared spectroscopic observations with the Spitzer Space Telescope showed excess emission compared to the photospheric flux in the mid-infrared \citep{Chen1, Plavchan}. 

The Hubble Space Telescope (HST), equipped with the Advanced Camera for Surveys (ACS), provided the first resolved images of the disk in scattered light \citep[observations acquired on 2004-12-01 and 2005-07-12,][]{Golimowski}. In the same work, \cite{Golimowski} also presented a detailed characterization of the disk, coupling the HST images with Spitzer's photometric data at 24 and 70 $\mu$m. As a result, HD\, 92945 showed an axisymmetric and inclined disk, with an inner ring at 43-65 au from the star and an outer disk extending from 65 au, with a surface brightness that slowly declines with increasing radius and suddenly drops at 110 au. 

Observations with the Atacama Large Millimeter/submillimeter Array (ALMA) revealed further details about the structure of the disk at 0.86 mm. Analysis of the data showed a gap at $73.7^{+1.8}_{-1.7}$ au, $27.6^{+8.7}_{-9.9}$ au wide and with a fractional depth of $0.73^{+0.08}_{-0.10}$ \citep[observations acquired on 2016-12-13/18,][]{Marino2,Marino3}.

The Spectro-Polarimetric High-contrast Exoplanet REsearch \citep[SPHERE,][]{Beuzit} instrument at the Very Large Telescope (VLT) was used to obtain high-contrast images of the system in the near infrared. Observations did not reveal any disk features or planetary companion candidates \citep{Mesa5}.

In Section 2, we describe the observations and data reduction procedures. Section 3 presents the results of the disk detection and forward modeling. In Sect. 4, we discuss the detection limits and contrast curves derived from the postprocessed images, along with the characterization of the point sources detected in the field. Finally, we summarize our findings and their implications in Section 5.

\section{Observations and data reduction}

\renewcommand{\arraystretch}{1.5}

\begin{table*}
\caption{Observation parameters.}
   \centering
    \begin{tabular}{ccccccccccc}
        \hline
        Filter & $\lambda_{mean}$ ($\mu$m) & $W_{eff}$ ($\mu$m) & Mask & Readout & $N_{groups}$ & $N_{int}$ & $t_{exp} (s)$ & $N_{dithers}$ & $N_{rolls}$ & $t_{total}$ (s) \\ 
        \hline
       \textbf{HD\, 92945} &  & &&&&&&&& \\
       F444W & 4.397 & 0.979 & MASK335R & DEEP8 & 18 & 5 & 1865.577 & 1 & 1 & 1865.577  \\ 
       F200W & 1.988 & 0.461 & MASK335R & DEEP8 & 18 & 5 & 1865.577 & 1 & 1 & 1865.577 \\ 
       \textbf{HD\, 92921} &  & &&&&&&&& \\
       F444W & 4.397 & 0.979 & MASK335R & DEEP8 & 8 & 4 & 637.23 & 5 & 1 & 3186.15 \\ 
       F200W & 1.988 & 0.461 & MASK335R & DEEP8 & 8 & 4 & 637.23 & 5 & 1 & 3186.15 \\ 
        \hline
    \end{tabular}
    \tablefoot{Observation parameters for the scientific target (HD\, 92945) and the reference star (HD\, 92921), observed in date 2024-05-20.}    
    \label{tabF444}
\end{table*}

HD\, 92945 was observed as part of program 3989 (PI: Hinkley, co-PI: Lazzoni, Marino). NIRCam observations were performed in date 2024-05-20 with the wideband filters F200W ($\lambda =1.988$ $\mu$m, $\Delta \lambda = 0.461$  $\mu$m) and F444W ($\lambda =4.402$ $\mu$m, $\Delta \lambda = 1.024$  $\mu$m) in coronagraphic mode (MASK335R). Based on early science results for the HIP65426 system \citep{Carter}, it was determined that, for this instrument, the angular differential imaging \citep[ADI, ][]{Marois5} technique is not as impactful as the Reference Differential Imaging \citep[RDI, ][]{Lafreniere}. Moreover RDI only and the coupling of RDI+ADI yield very similar performances in terms of contrasts, indicating that ADI does not add relevant improvements to RDI for NIRCam observations.  Consequently, only one scientific exposure was taken, and the RDI only technique was employed using a dedicated reference star (HD\, 92921) observed on the same date as HD\, 92945 (2024-05-20), with a five point small-grid dither pattern. 

The data reduction was performed using the \texttt{spaceKLIP} pipeline \citep{Kammerer, Carter}, which provides a robust and tailored approach for processing NIRCam data, addressing the challenges of the instrument to facilitate accurate and reliable high-contrast imaging analysis.

\texttt{spaceKLIP} accepts as inputs the \textit{stage-0} (.uncal files) products of the JWST pipeline and performs dedicated preprocessing steps, including ramp fitting, background subtraction, bad pixel correction, alignment, and more, to address instrument-specific effects. Following these steps, the pipeline generates PSF-subtracted images using, as a postprocessing method, the \texttt{KLIP} algorithm \citep[e.g., ][]{Soummer2}. For HD\, 92945 reduction we explored Karhunen–Loeve (KL) modes 1, 2, 5, 10, 20.
\texttt{spaceKLIP} also allows the calculation of contrast curves, which include the KLIP and mask throughputs \citep{Adams} and the small sample statistics correction \citep{Mawet}, and precise measurements of photometry and astrometry for any source in the field of view.

\begin{figure*}
    \centering
    \includegraphics[trim=0.2cm 0.2cm 0.2cm 0.2cm, clip=true, width=\textwidth]{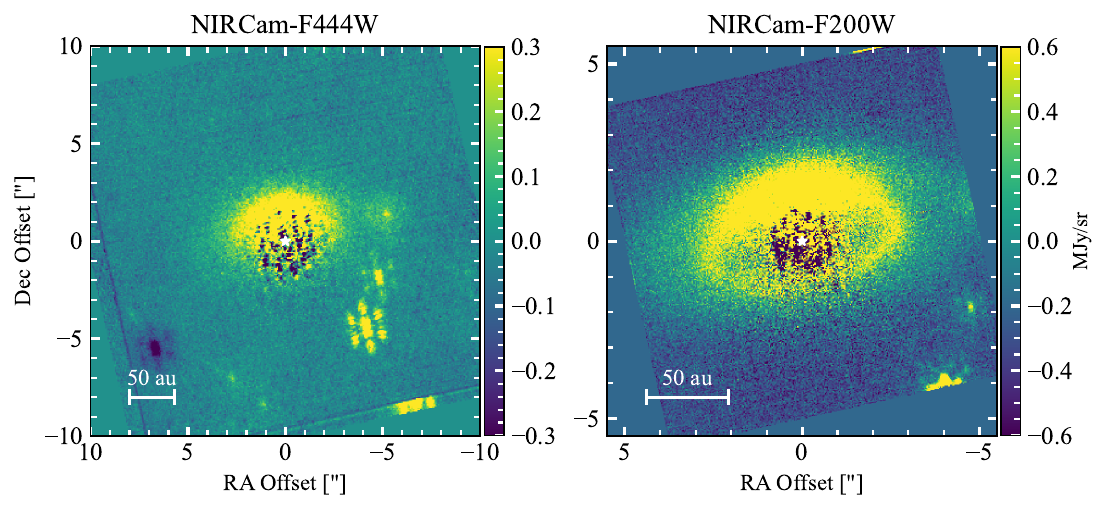}
    \caption{NIRCam reductions for HD\, 92945 for the F444W (left panel) and the F200W (right panel) filters. Both images where obtained performing RDI with HD\, 92921 as reference. The images are rotated to northeast, with the scale bar in the bottom left corner representing a projected distance of 50~au.}
    \label{HD92945red}
\end{figure*}

Figure \ref{HD92945red} displays the final products of the pipeline for NIRCam F444W (left panel) and F200W (right panel) filters, obtained applying RDI using HD\, 92921 as reference and KL=10 for which the S/N was the highest. In the F444W filter, we note the presence of a background object with negative excess at $\Delta RA\sim6.92''$ and $\Delta Dec \sim 5.68''$. This source is present in the reference star field of view and does not influence the scientific reduction. The characterization of the sources in the field of view is discussed in Section 4, while the disk modeling is addressed in the subsequent section.

When applying RDI, we either used the dedicated reference star HD\, 92921, or a more extensive library comprising reference stars acquired within program 3989 (HD\, 25945, HD\, 92921, HD\, 114642, HD\, 125161, HD\, 161915, HD\, 190580, and HD\, 197051). Some stars present few background objects which, generally, were faint enough not to affect the reduction. However, HD\, 190580 was excluded from the PSF library due to a bright source present in the field of view, which significantly impacted the performance of the library.
In the single-reference RDI approach, only the five dithers of HD\, 92921 are used to construct a PSF model, yielding a maximum KL mode of 20. This allows for some diversity in the reference PSF while maintaining high spectral and photometric fidelity with respect to the target, reducing the risk of chromatic mismatches and preserving extended astrophysical signals. In contrast, the PSF library approach expands the reference set by including dithers from multiple reference stars enabling a broader exploration of PSF variability. This approach enables a higher range of KL modes (max KL modes = 457) which in turn allows for a more effective suppression of the stellar PSF while preserving the underlying astrophysical signal.
The PSF library is initially constructed by collecting all available \textit{stage-0} files corresponding to suitable reference stars observed in program 3989. These files are then processed using \texttt{spaceKLIP}, as detailed in the previous section. After the preprocessing steps are completed, the final postprocessed products are obtained by subtracting from the science image an optimized combination of the PSF library images, selected to minimize the residuals.
In the following, we demonstrate how leveraging, multiple reference stars improves the subtraction of central regions enhancing the signal of the disk around HD\, 92945.


\section{Modeling and characterization of the disk}

The disk around HD\, 92945 is detected at both 4.4 and 2 $\mu$m with the NIRCam instrument. In this Section, we aim to model the disk in the F200W and F444W filters, then subtract it and reveal potential faint sources obscured by the disk's glare. 

For disk modeling and subtraction, we utilized the Python package \texttt{Winnie} \citep{Lawson, Lawson1}, which provides tools for disk model convolution using synthetic PSF grids, customized RDI implementations, and forward modeling for RDI. The disk model used by \texttt{Winnie} is based on the \textit{scattered\_light\_disk()} function provided by the \texttt{vip\_hci} package \citep{Gonzalez}. This function offers a simplified version of the \texttt{GRaTer} code \citep{Augereau}, modeling the radial distribution of dust as a smoothed two-power-law profile for a single-ring disk, combined with a linear combination of two Henyey-Greenstein (HG) phase functions to represent the forward and backward scattering of light by the dust.

However, HD\, 92945's disk has several features, already identified in previous analyses, both in scattered light with HST and at submillimeter wavelengths with ALMA, that may require a more complicated model.
Thus, to achieve a more detailed characterization, we adopted the same functional form used to model the surface density when modeling the ALMA data, incorporating a broad disk component with an embedded gap. The density distribution is described by the following equation 
\begin{equation}
\rho (r,z) = \Sigma(r) e^{-0.5\big(\frac{z}{hr}\bigr)^2}/(\sqrt{2\pi}hr),
\end{equation}
with $h$ being the disk's scale height and $\Sigma(r)$ the radial distribution, defined as in \cite{Marino3}:
\begin{equation}
\Sigma_r=f_g \frac{\Sigma_c}{4}\biggl(\frac{r}{r_{\rm in}}\biggr)^{\gamma}\biggl(1+\tanh\biggl(\frac{r-r_{\rm in}}{l_{\rm in}}\biggr)\biggr)\biggl(1+\tanh\biggl(\frac{r_{\rm out}-r}{l_{\rm out}}\biggr)\biggr)
\label{rden}
\end{equation}
Here, $r_{\text{\rm in}}$ and $r_{\text{\rm out}}$ represent the inner and outer radii, $\gamma$ the inner surface density slope, $l_{\text{\rm in}}$ and $l_{\text{\rm out}}$ control the steepness of the inner and outer edges, and $\Sigma_c$ is normalized to 1 at the reference radius. Finally, $f_g$ models the gap as a Gaussian function
\begin{equation}
f_g=1-\delta_g e^{-\frac{(r-r_g)^2}{2\sigma_g^2}}\\
\end{equation}
where $\delta_g$ represents the gap depth, $r_g$ the gap radius, and the full width at half maximum is defined as $w_g = 2\sqrt{2 \ln(2)} \sigma_g$.

The model described above was then implemented within the \textit{scattered\_light\_disk()} function to reproduce the more complex density distribution of the dust expected for HD\, 92945. To estimate the forward and backward scattering of dust particles, instead, we employed the same two-HG phase function provided by \textit{scattered\_light\_disk()} .

For both filters, the raw model of the disk obtained with our customized function was convolved within \texttt{Winnie} with a JWST PSF calculated using the \texttt{STPSF} package and rotated to match the position angle of the science acquisition. Next, an RDI reduction was performed, either with a single PSF or with the PSF library as described in Section 2.1, on the science image and on the convolved model, and the residuals were obtained by subtracting the two. 

\subsection{F200W filter}

\begin{table*}
    \centering
        \caption{Disk parameters for HD\, 92945.}

    \begin{tabular}{ccccc}
    \hline
        &\multicolumn{3}{c}{\textbf{NIRCam-F200W}} & \textbf{ALMA} \\ 
        &\textbf{Single Ref} & \textbf{PSF Lib} & \textbf{PSF Lib + Mask} \\ 
        \hline

        $r_{in}$ (au)  & $55.1^{+1.6}_{-1.4}$ & $57.4^{+3.0}_{-2.8}$ & $58.8^{+3.3}_{-2.6}$ &$50.3^{+3.0}_{-2.4}$\\ [0.5em]
        $r_{out}$ (au)  & $101.1^{+1.0}_{-1.0}$ & $98.4^{+1.9}_{-1.9}$ & $97.7^{+2.9}_{-3.6}$ &$128.3^{+7.3}_{-14.4}$\\ [0.5em]
        $\gamma$  &$-0.6^{+0.1}_{-0.1}$ & $-0.2^{+0.1}_{-0.2}$ & $-0.6^{+0.3}_{-0.2}$ &$-0.6^{+0.6}_{-0.7}$ \\ [0.5em]
        $l_{in}$   & $17.7^{+1.0}_{-0.8}$  & $22.2^{+1.6}_{-1.9}$ & $18.8^{+1.9}_{-1.6}$ & $6.1^{+2.3}_{-2.8}$ \\ [0.5em]
        $l_{out}$   &$6.2^{+1.0}_{-0.8}$ & $7.5^{+1.6}_{-1.4}$ & $10.7^{+2.6}_{-2.4}$ & $24.9^{+8.0}_{-7.6}$ \\ [0.5em]
        $\delta_{g}$  &$0.72^{+0.04}_{-0.04}$ & $0.77^{+0.03}_{-0.04}$ & $0.79^{+0.04}_{-0.05}$ & $0.73^{+0.08}_{-0.10}$ \\ [0.5em]
        $r_{g}$ (au)  &$80.9^{+0.6}_{-0.6}$ & $81.4^{+0.9}_{-1.0}$ & $79.8^{+1.1}_{-0.97}$ & $73.7^{+1.8}_{-1.7}$  \\ [0.5em]
        $w_{g}$ (au)  &$31.0^{+3.3}_{-2.9}$ &  $28.6^{+2.9}_{-2.4}$ & $26.5^{+3.8}_{-3.0}$ &  $27.6^{+8.7}_{-9.9}$\\ [0.5em]
        $h$  &$0.085^{+0.011}_{-0.011}$  & $0.087^{+0.012}_{-0.01}$ & $0.077^{+0.010}_{-0.011}$ 
        & $0.043^{+0.014}_{-0.014}$ \\[0.5em] 
        $pa$ $(^{\circ})$  &$-79.2^{+0.1}_{-0.1}$ & $-79.3^{0.2}_{-0.2}$ & $-79.3^{+0.2}_{-0.2}$ &$-80.0^{+0.6}_{-0.6}$ \\ [0.5em]
        $incl$ $(^{\circ})$ & $63.7^{+0.2}_{-0.2}$  &  $63.2^{+0.2}_{-0.2}$ &$64.9^{+0.4}_{-0.4}$ &$65.4^{+0.6}_{-0.6}$ \\ [0.5em]
        $g_1$   &$0.74^{+0.01}_{-0.01}$ & $0.72^{+0.01}_{-0.01}$& $0.87^{+0.05}_{-0.06}$ &\\ [0.5em]
        $g_2$  & $-0.10^{+0.01}_{-0.01}$ & $-0.16^{+0.01}_{-0.02}$& $-0.17^{+0.03}_{-0.03}$&\\ [0.5em]
        $w_{g_1}$ & $0.699^{+0.004}_{-0.004}$ & $0.714^{+0.004}_{-0.006}$ & $0.83^{+0.06}_{-0.06}$ &\\ [0.5em]
        $F$ & $5.2^{+0.1}_{-0.1}$  & $5.0^{+0.1}_{-0.1}$& $8.1^{+0.9}_{-0.8}$& \\ 
        \hline
    \end{tabular}
    \tablefoot{Disk parameters for HD\, 92945 at 2 $\mu$m and comparison with ALMA results \citep{Marino3} . For F200W filter we show the results obtained from MCMC simulations for a single reference (HD\, 92921), the entire PSF library and the PSF library plus a mask to exclude the bright region close to the minor axis, as shown in the last row of Figure \ref{F200model}. Except for $l_{\text{\rm out}}$, the values for the F444W filter were assumed to be the same as those derived for the F200W filter, as detailed in Section 3.2.}
    \label{tabdisk}
\end{table*}

To constrain the disk parameters and associated uncertainties at 2 $\mu$m, where we are able to pierce closer to the star thanks to the smaller pixel scale (0.031 arcsec/pixel), we used a Markov Chain Monte Carlo (MCMC) sampling of the posterior distribution with the \texttt{emcee} Python package \citep{Foreman-Mackey}. The log-likelihood function was calculated based on the $\chi^2$ of the residuals after subtracting an RDI-processed image of the convolved model from an RDI-processed image of the science data. The 15 parameters varied during the optimization were $r_{\rm in}$, $r_{\rm out}$, $\gamma$, $l_{\rm in}$, $l_{\rm out}$, $\delta_g$, $r_g$, $w_g$, $h$, position angle, inclination, the forward ($g_1>0$) and backward ($g_2<0$) scattering coefficients, the weight of the two HG coefficients ($w_{g1}$ and $w_{g2}=1-w_{g1}$), and the peak surface brightness of the disk ($F$). We considered projected radii in the range $[1, 4.8]$ arcsec to exclude the central oversubtracted area and sources at the field edge. The MCMC simulation ran for 4,000 iterations with 50 walkers. To analyze the results, we discarded the first 1500 burn-in steps and applied a thinning factor of 100 to reduce correlations between steps. 
\renewcommand{\arraystretch}{1.5}

\begin{figure*}
    \centering
\includegraphics[width=\textwidth]{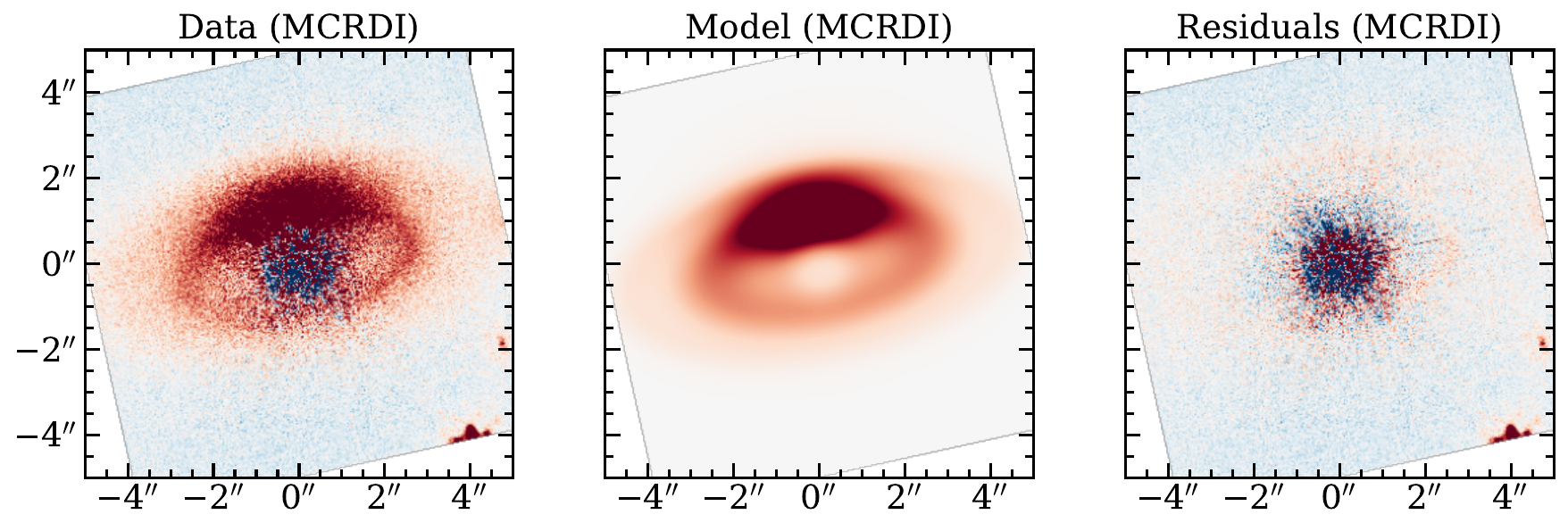}
\includegraphics[width=\textwidth]{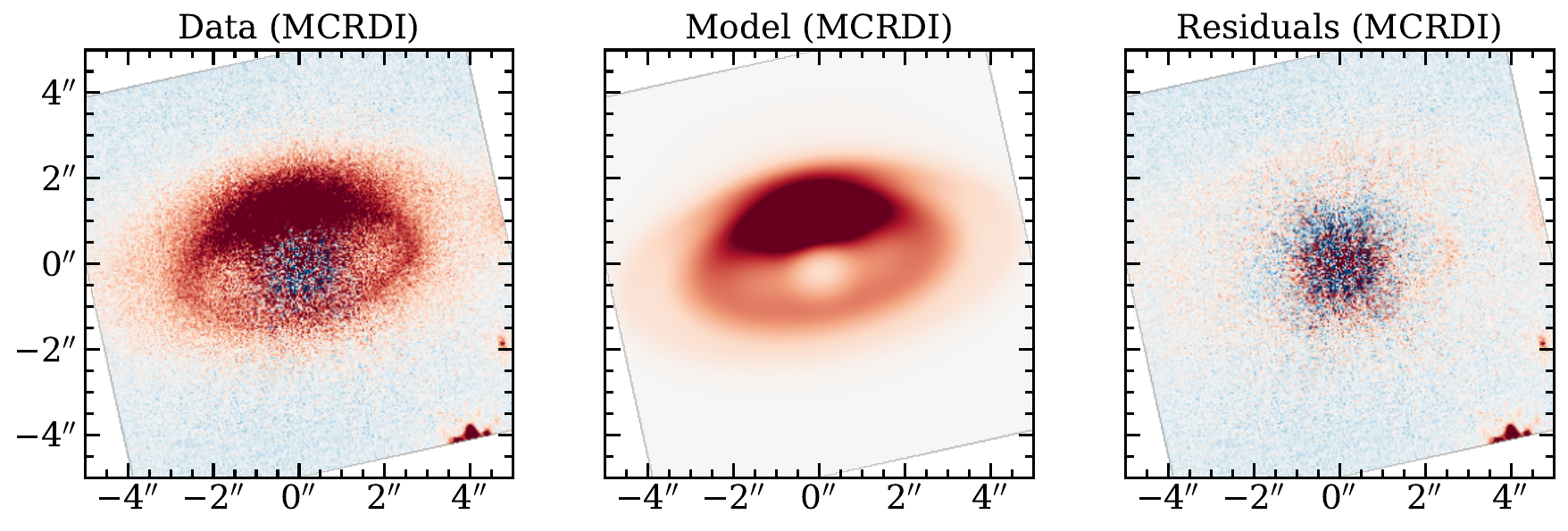}
\includegraphics[width=\textwidth]{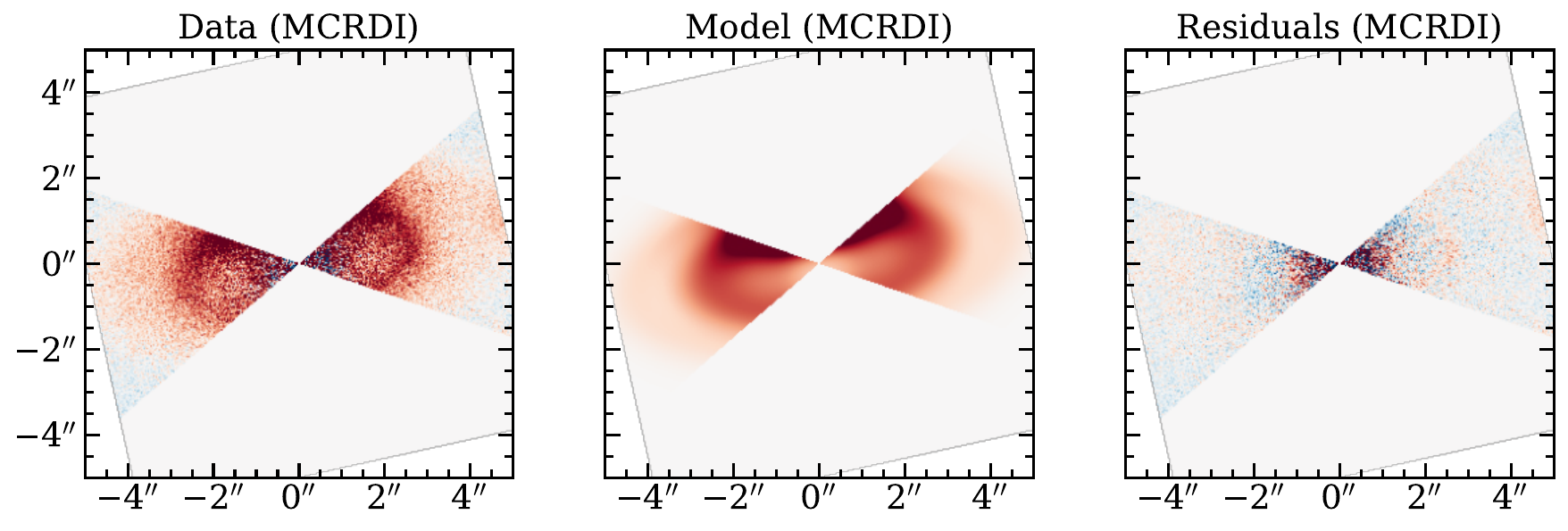}
\caption{F200W data, model, and residuals of HD\, 92945's disk using one reference, HD\, 92921, (top panels) and the PSF library considering the entire image (second row) or excluding the bright regions close to the minor axis (last row).}
\label{F200model}
\end{figure*}

We retrieve a model for the disk in both cases where we used either a single reference PSF or the entire library for the RDI. The second and third columns of Table \ref{tabdisk} list the median values and the 16th and 84th percentiles for each of the 15 parameters as obtained for one and multiple reference stars, respectively. An image of the disk, the model, and the residual as given by both simulations is shown in Figure \ref{F200model}, where the models were obtained using the mean parameters listed in Table \ref{tabdisk}.

The PSF library proved to be more effective in producing a model closer to the data, as proven by the contrast curves derived in Section 4.2. However, some of the parameters differ significantly from what was obtained with the same model in the submillimeter with ALMA (see last column of Table \ref{tabdisk}). This is especially true for the inner and outer slopes of the disk, $l_{\rm in}$ and $l_{\rm out}$, which differ by 5 and 2 $\sigma$, respectively, from the values obtained from ALMA data.
A potential source of bias could be the parameterization of the phase function that we assumed. Thus, to mitigate this potential bias, we applied a butterfly-shaped mask, oriented along the position angle ($-79^{\circ}$) with an opening angle of $\pm 30^{\circ}$, to exclude from consideration the disk's bright region near the minor axis (see last row of Figure \ref{F200model}).
We then reran the MCMC simulation on the masked images with the same settings described previously. Only the PSF library was taken into account in this case given the better subtraction achieved. Results are shown in the fourth column of Table \ref{tabdisk} and in the last row of Figure \ref{F200model}. The derived parameters do not differ significantly from those obtained in previous simulations—whether using a single reference star or the PSF library—but they show improved agreement with the ALMA results. Notably, we recovered a higher disk inclination ($64.9^{\circ} \pm 0.4^{\circ}$), which aligns remarkably well with its submillimeter counterpart ($65.4^{\circ} \pm 0.6^{\circ}$). This parameter serves as a valuable benchmark for assessing the quality of our modeling, as it is largely wavelength-independent and can be directly compared with the high-precision measurements provided by ALMA.

\begin{figure*}
    \begin{subfigure}{0.45\textwidth}
        \includegraphics[width=\textwidth]{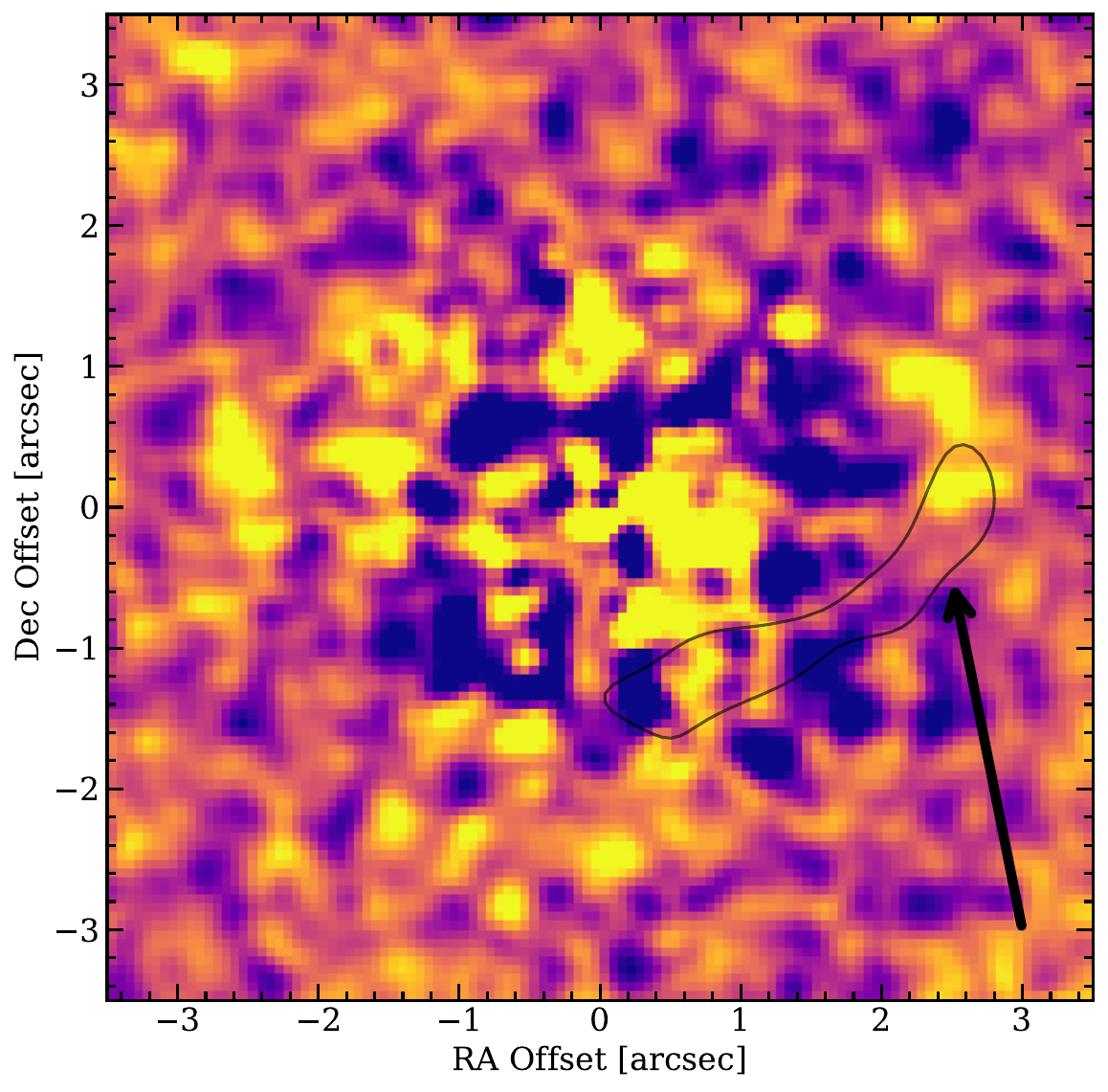} 
        \label{fig:sub4}
    \end{subfigure}
    \hfill
    \begin{subfigure}{0.45\textwidth}
        \includegraphics[width=\textwidth]{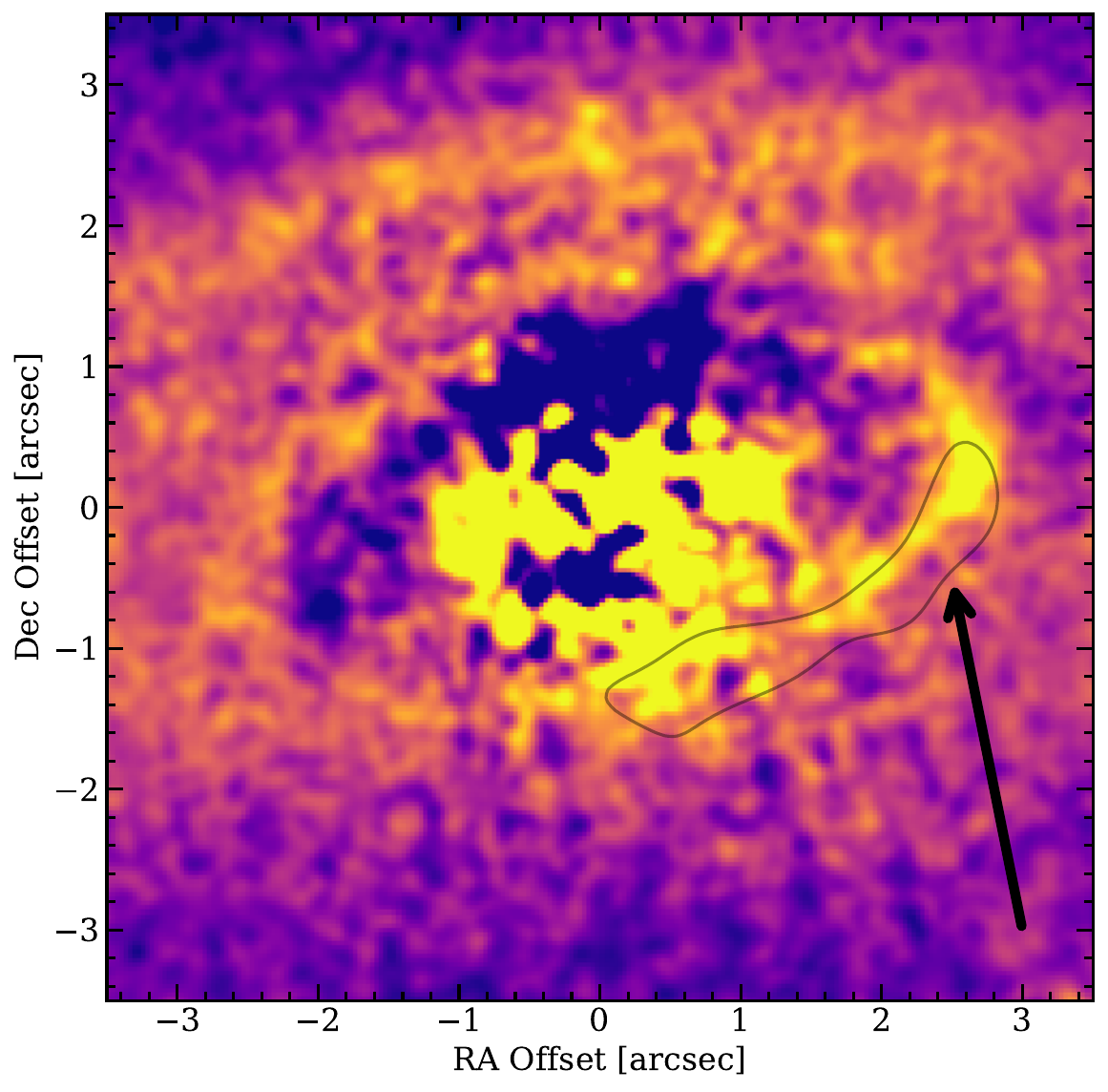} 
        \label{fig:sub5}
    \end{subfigure}

    \caption{Smoothed images of the residuals obtained after modeling and subtracting the disk in the F444W (left panel) and F200W (right panel) filters. The disk luminosity asymmetry detected in both filters in the west portion of the disk is pointed by the arrow. For the both filters, we also added the $2\times10^{-4}$ Jy/beam contour level from ALMA data \citep{Marino2}.}
    \label{lumasym}
\end{figure*}

Finally, we highlight a surface brightness asymmetry in the southwestern portion of the inner disk component that is discussed in details in Section 3.3. 
For the discussion in Section 3.3, we adopt the parameters of the disk retrieved from this last modeling to compare the disk characteristics at different wavelengths while we use the residuals shown in the second row of Figure \ref{F200model} for source vetting and to calculate contrast curves in Section 4. Moreover, the corner plot presented in Figure \ref{corner} refers to the modeling of masked images.

\subsection{F444W filter}

Due to the lower spatial resolution imposed by the NIRCam pixel scale at 4 $\mu$m (0.062''/pixel versus 0.031''/pixel at 2 $\mu$m), and especially due to the larger PSF (2.2 times larger in the F444W filter than in the F200W), the disk's fine structures are more difficult to constrain compared to the 2 $\mu$m data. To overcome this limitation, we generated a disk model at 4 $\mu$m by adopting the morphological parameters derived from the F200W filter, where the resolution is higher and the disk structure better resolved.

We began by varying only the parameters that are strongly wavelength dependent, namely those governing the scattering phase function ($g_1$, $g_2$, $w_{g_1}$) and the peak surface brightness ($F$). The remaining 12 parameters were initially fixed to the average values listed in Table \ref{tabdisk}, either from column two (single reference PSF) or column three (PSF library). The best-fit values for the four wavelength-dependent parameters were then obtained through a non-linear least-squares $\chi^2$ minimization of the residuals.

As done for the F200W filter, we computed two separate models at 4 $\mu$m, one using a single reference PSF and the other based on the full PSF library employed in the RDI process. To mitigate contamination from unrelated sources, the optimization excluded the three bright compact sources identified in the field (C2, C3, and C4, see Section 4), and considered only regions within 8'' of the star to avoid the neutral density squares in the corners.

\begin{figure*}
    \centering
\includegraphics[width=0.85\textwidth]{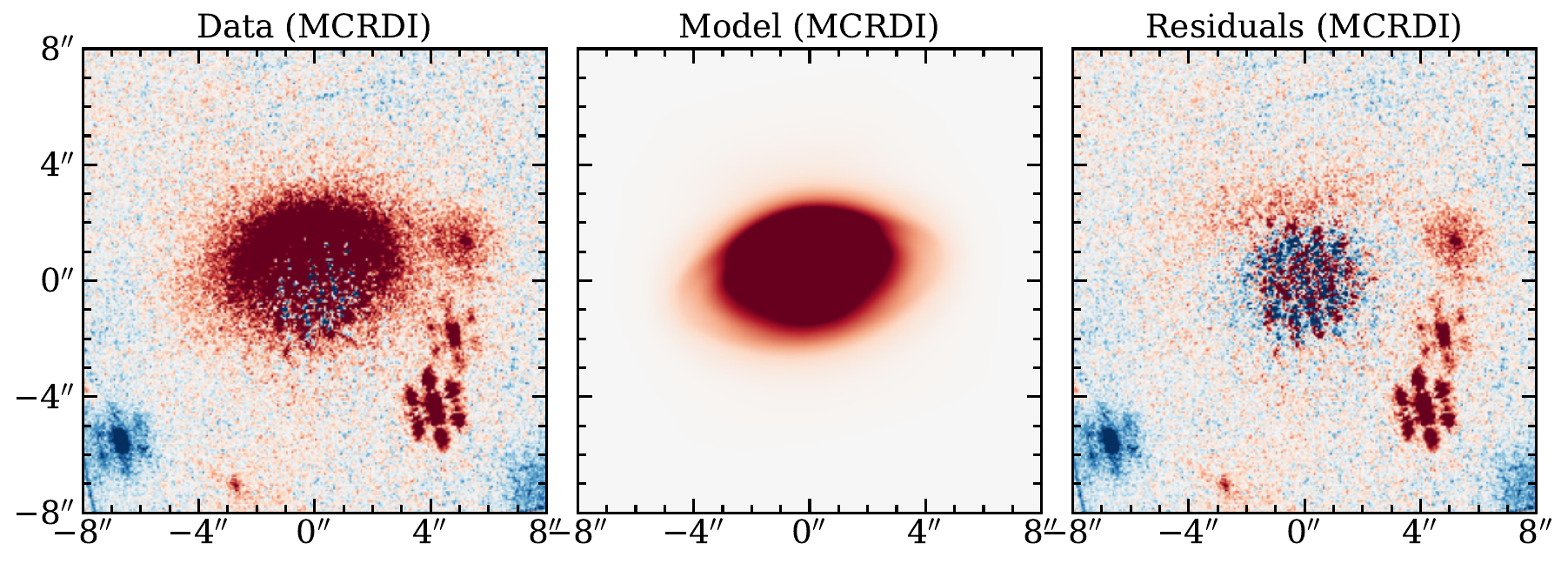} 
\includegraphics[width=0.85\textwidth]{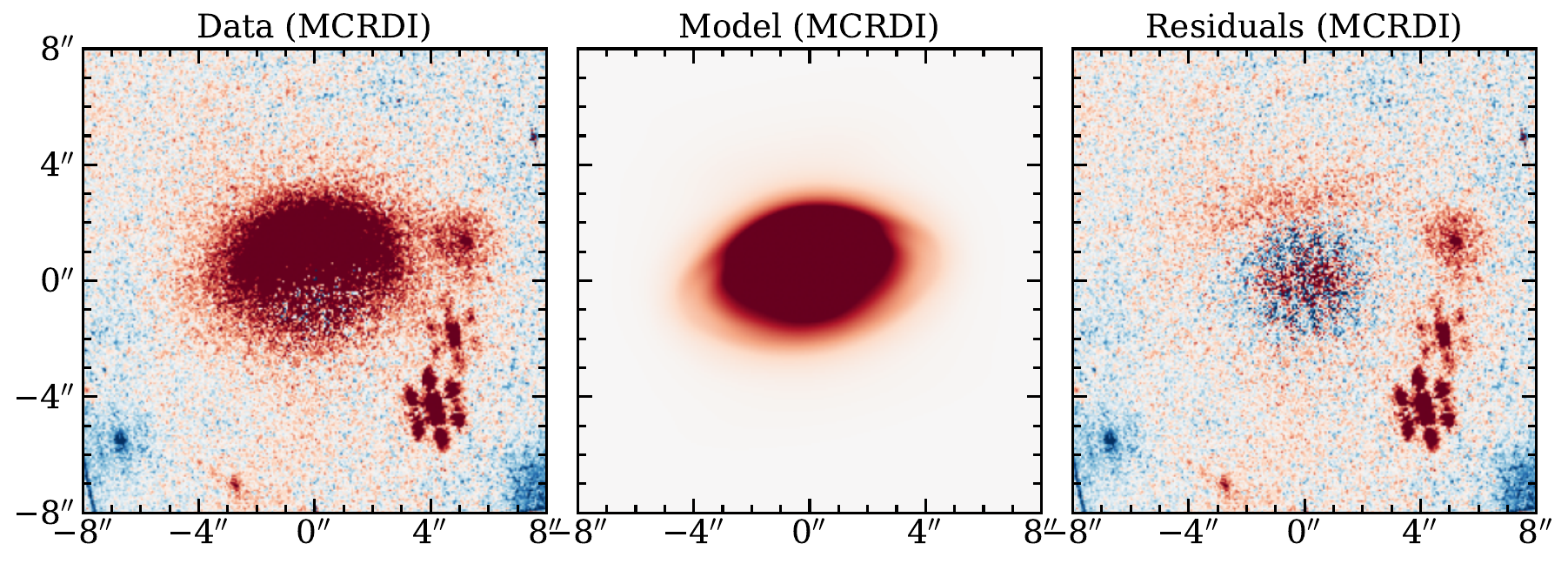}
\includegraphics[width=0.85\textwidth]{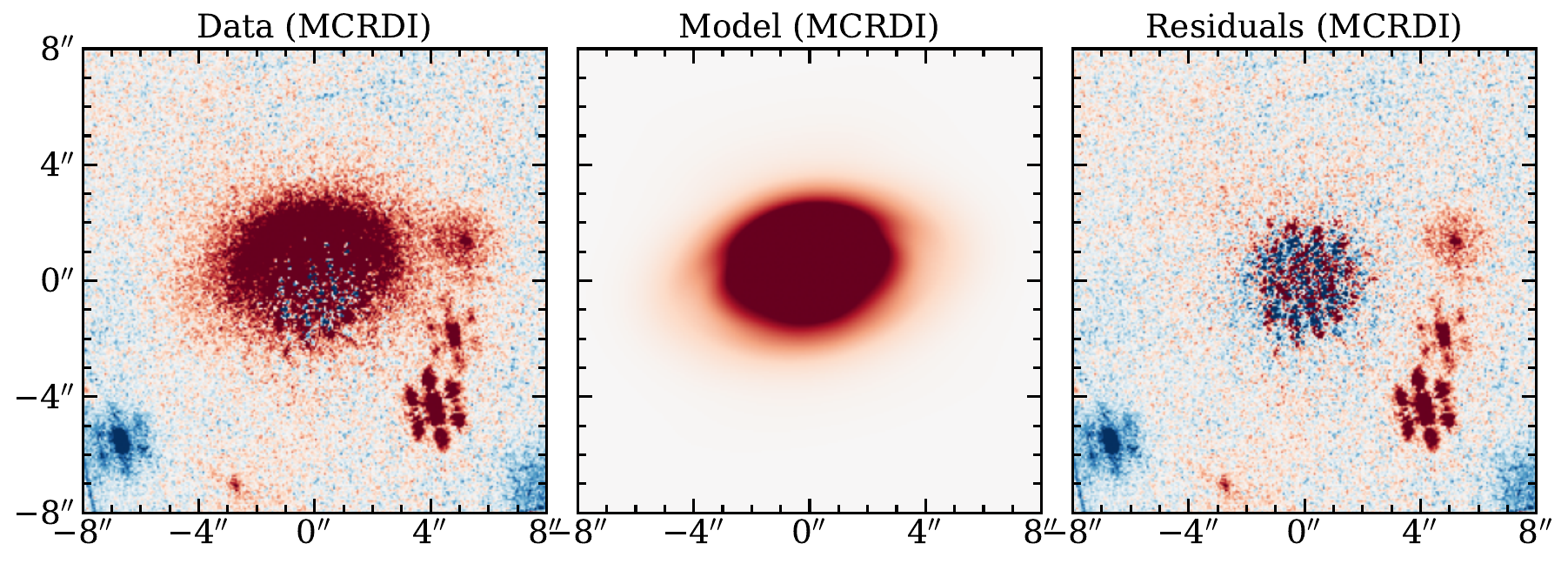}
\includegraphics[width=0.85\textwidth]{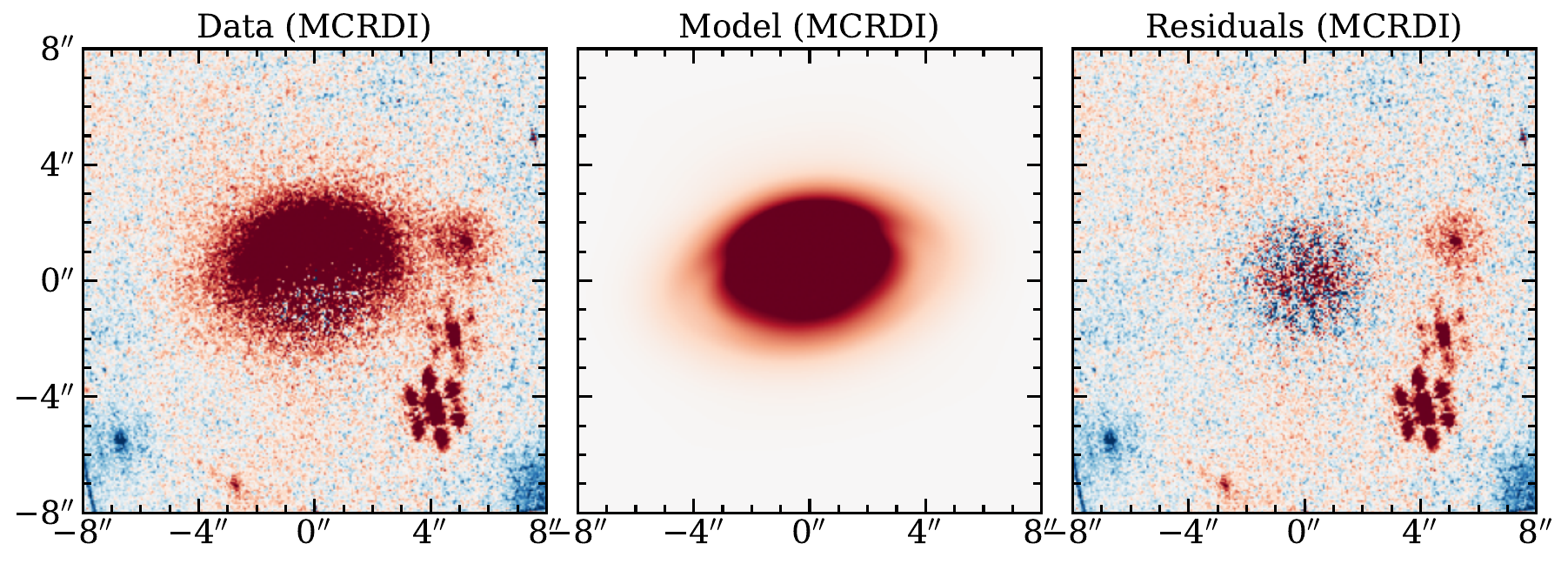}

\caption{From left to right, F444W data, model, and residuals of HD\, 92945's disk using: a) the same morphological parameters adopted for the disk as seen at 2 $\mu$m and one reference, HD\, 92921, (first row) or the PSF library (second row); b) the same morphological parameters adopted at 2 $\mu$m with the exception of $l_{out}$, optimized for these data, using the same single reference, (mid-bottom panels) or the PSF library (bottom panels) }
\label{F444}
\end{figure*}

The resulting models are shown in the first two rows of Figure \ref{F444}. In both cases, the models reproduce the observed disk well, suggesting a radial profile consistent with the one derived at 2 $\mu$m. However, residual emission is visible in the northern part of the image, near the semiminor axis. We explored whether adjusting the remaining 12 parameters could reduce these residuals and found that increasing the parameter that regulates the steepness of the outer edge, $l_{\rm out}$, improved the fit significantly. This was obtained again through a non-linear least-squares $\chi^2$ minimization of the residuals. This behavior is consistent with ALMA observations and likely reflects the improved modeling of the outer disk regions, which are better captured at 4 $\mu$m thanks to the wider field of view. For both the single reference and PSF library cases, values of $l_{\rm out} \sim 28-30$ were required to minimize residuals.  These refined models are presented in the third and fourth rows of Figure \ref{F444}.

As noticed for the F200W filter, the use of the PSF library provides a cleaner residual image, particularly in the central regions. In fact, residual of over and undersubtractions that are visible in the central area in the right panels of the first and third rows, are strongly mitigated when considering multiple PSFs (right panels of second and last rows). These artifacts likely arise from slight misalignments between the science and reference images, potentially caused by the bright northern region of the disk biasing the estimated stellar centroid. When using a PSF library, the subtraction algorithm effectively builds a synthetic reference PSF from a combination of many reference frames. This linear combination can better reproduce the actual centroid shift and speckle morphology of the science frame.

\subsection{Comparison with ALMA and HST}

Our HD\, 92945's disk modeling results at 2 and 4.4 $\mu$m show a morphology that is qualitatively similar to that found with ALMA and HST, that is a wide disk with a gap. In this Section we discuss their similarities and differences.

First, and as expected, the disk orientation that we find at 2 $\mu$m agrees  with that found at 0.59-0.8 $\mu$m \citep[ACS/HST,][]{Golimowski} and at $0.86$ mm \citep[ALMA,]{Marino2}. The disk's scale height $h$ in the near infrared, on the other hand,  is nearly twice that measured for larger grains with ALMA ($0.077^{+0.010}_{-0.011}$ versus $0.043^{+0.014}_{-0.014}$). Although the two estimates are consistent within $2\sigma$, the discrepancy can be attributed to underlying physical processes. The scale height of planetesimals and their general orbital excitation is thought to be set by gravitational interactions with planets \citep[e.g.,][]{Mustill} or disk self-stirring \citep[e.g.,][]{Krivov1}. The micron-size dust grains, detected at near infrared wavelengths, are also subject to radiation pressure that could play a role in increasing their eccentricities and inclinations  \citep{Thebault}, therefore increasing the scale height of the disk as seen at 2 $\mu$m.

\begin{figure}
    \centering
\includegraphics[width=\columnwidth]{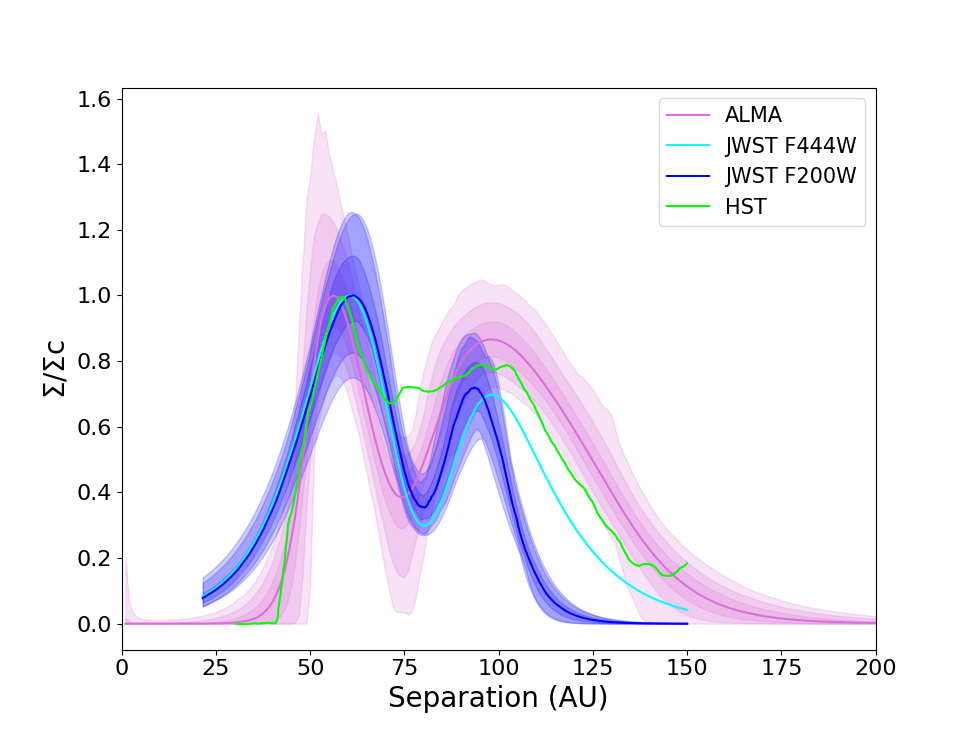} 
\caption{Radial density distribution of the dust obtained from PSF libray+mask modeling of F200W and F444W JWST data (dark blue and light blue curves) and comparison with ALMA (pink curves) and HST (green curve). For F200W JWST filter and ALMA, the solid line represents the median, while the shaded areas represent the 68, 95, and 99.7 per cent confidence regions as obtained by \cite{Marino3}. F444W JWST profile was obtained using the mean parameters derived at 2 $\mu$m (column four of Table \ref{tabdisk}), with the exception of $l_{\rm out}$, set at 30. For HST, the curve is taken from Fig. 7 of \cite{Golimowski}.}
\label{density}
\end{figure}

Figure \ref{density} illustrates in shades of blue the radial density distribution of the dust as obtained analyzing NIRCam F200W (dark blue) and F444W (light blue) filters and their comparison with previous results from ALMA in the submillimeter \citep[shades of pink,][]{Marino3} and HST at 0.59 and 0.83 $\mu$m \citep[green, adapted from][]{Golimowski}. 

For ALMA (pink) and 2 $\mu$m JWST (blue) data, the solid line represents the median, while the shaded areas represent the 68, 95, and 99.7 per cent confidence regions. We note that the 4 $\mu$m profile is only qualitative, as the F444W images do not provide sufficient spatial resolution to perform a reliable optimization of the disk at this wavelength. As discussed in the previous Section, we therefore adopted the mean parameters derived at 2 $\mu$m (column four of Table \ref{tabdisk}), with the exception of $l_{\rm out}$, which was set to 30. The 4 $\mu$m results are not included in the subsequent discussion.
Finally, since both the ALMA and JWST data were modeled using the same functional form, the comparison between the results from this work and those at submillimeter wavelengths presented by \citet{Marino3} is more robust than comparisons between JWST and HST models outcomes.

As given by the best-fit results presented in the fourth column of Table \ref{tabdisk}, the inner and outer edges of the disk are placed at $58.8^{+3.3}_{-2.6}$ au and $97.7^{+2.9}_{-3.6}$ au. These results can be compared with what was found by \cite{Golimowski} in HST data, where HD\,  92945's disk was modeled using two separate components, the inner one located between 43–65 au and the outer one placed between 65–145 au, with a steep decrease in the surface brightness of the latter starting from 110 au. 
Considering the full extent of the disk, defined where the profiles in Figure \ref{density} drop to 90\% below the peak density, the disk appears shifted inward by approximately 20 au at 2 $\mu$m (extending from $\sim$ 25 to $\sim$ 112 au) compared to shorter HST wavelengths (extending from 43 to 145 au). 

The different width of the inner portion of the disk at 2 and 0.59–0.83 $\mu$m might arise from residuals left in the JWST data after PSF subtraction, affecting the retrieval of the density profile close to the star. A combination of the limited field of view ($\sim$ 100 au at 2 $\mu$m) and the faintness of the outer portion of the disk could also explain the apparent inward shift of the outer edge seen in the JWST data with respect to the HST images. Although uncertainties for the HST-derived profile are not provided, it is likely that the two datasets are consistent within the expected uncertainties, especially when considering the limitations mentioned above.

From ALMA images, instead, the inner and outer edges are placed at $50.3^{+3.0}_{-2.4}$ and $128.3^{+7.3}_{-14.4}$ au, respectively. Once more, the disk looks less extended in its outer regions at 2$\mu$m than in the submillimeter images. Although our estimates agree with the ALMA model within 2.1 and 2.7 $\sigma$ for the inner and outer disk radii, respectively, these differences may arise from the distinct dust populations probed at submillimeter and micron wavelengths. Moreover, when comparing the peaks of the radial density distribution for the inner and outer components at 2 $\mu$m ($61.0^{+0.8}_{-0.4}$ au and $93.7^{+3.6}_{-3.6}$ au), we observe outward and inward shifts, respectively, relative to ALMA's profile, which reports values of $56^{+1}_{-4}$ au and $98^{+1}_{-3}$ au. We also note that these shifts cannot be attributed to gas drag, as no gas has been detected in the disk of HD\, 92945 and upper limits on the $^{12}$CO gas mass were placed at $3\times10^{-5}$ M$_\oplus$ \citep{Marino2}. This implies that if gas is present it is not sufficient to influence the dust distribution.

Another mechanism that may explain differences between the spatial distribution of small and large grains is radiation pressure. However, for HD\, 92945's low luminosity \citep[0.37 $L_{\odot}$,][]{Matra}, radiation pressure is not strong enough to remove the smallest grains \citep{Kirchschlager}. Alternatively, stellar wind drag could be more important and explain the tentative grain size segregation \citep{Plavchan}.

As shown in Figure \ref{density}, the inner and outer slopes of the disk differ in 2 $\mu$m JWST data with respect to ALMA and HST models. In fact at 2 $\mu$m, $l_{\rm in}/r_{\rm in}$ and $l_{\rm out}/r_{\rm out}$, which control how smooth or sharp the inner and outer edge are, appear more shallow in the inner regions of the disk and steeper at its outer extents, with values of $0.32^{+0.04}_{-0.03}$ and $0.11^{+0.02}_{-0.02}$, respectively. This is the opposite of what was found with ALMA, for which $l_{\rm in}/r_{\rm in} = 0.12^{+0.05}_{-0.06}$  and $l_{\rm out}/r_{\rm out}= 0.20^{+0.06}_{-0.06}$. However, the values derived at 2 $\mu$m and with ALMA differ marginally by 3.3 and 1.4 $\sigma$, respectively.

In particular, the discrepancy in the outer slope may be attributed to limitations of the NIRCam field of view at 2 $\mu$m, as previously discussed. This interpretation is further supported by the modeling at 4.4 $\mu$m, where the field of view extends well beyond 150 au and the derived outer slope is significantly shallower than at 2 $\mu$m. At this wavelength, the dust density profile shows a much closer agreement with the ALMA and HST results for the outer disk, as illustrated by the light-blue curve in Figure \ref{density}.

The gap parameters are mostly comparable for JWST and ALMA, although the central radius of the gap (i.e., where the surface density minimum is) is slightly shifted outward in the 2 $\mu$m data, at $79.8^{+1.1}_{-0.97}$ au compared to the $73.7^{+1.8}_{-1.7}$ au retrieved in the submillimeter, which corresponds to a 3 $\sigma$ difference between the two estimates. This result might once more be related to the challenges of modeling the broad and faint outer component in a limited field of view. Regarding the gap depth ($\delta_g$), instead, we may expect that Pointing-Robertson (PR) drag could push dust toward the gap, making $\delta_g$ smaller (i.e., a shallower gap) at this wavelength than in submillimeter. Nevertheless, the two $\delta_g$ are comparable, pointing to an efficient dust removal process or gap clearing on timescales shorter than PR drag.

\begin{figure}
    \centering
\includegraphics[width=\columnwidth]{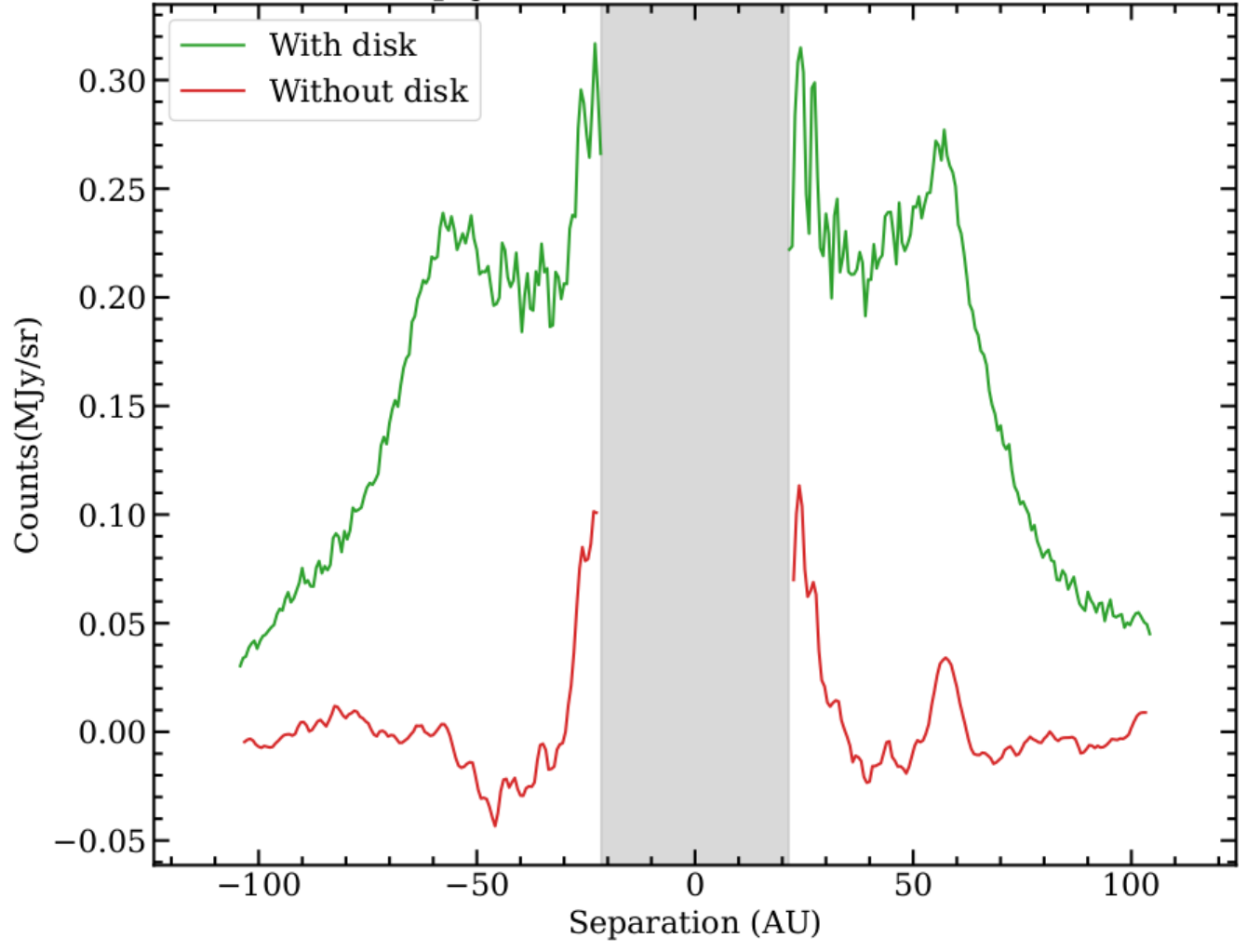} 
\caption{Deprojected radial profile at 2 $\mu$m before (green curves) and after (red curves) disk model subtraction. The grey area represents the 1 arcsec region excluded from the modeling.}
\label{depradprof}
\end{figure}

Finally, we highlight a surface brightness asymmetry in the southwestern portion of the inner disk component. The asymmetry is detected both at 4.4 $\mu$m, although marginally, and at 2 $\mu$m.  In Figure \ref{depradprof} we quantify such asymmetry for the F200W filter. The Figure shows the de-projected radial profile of the disk obtained taking two equivalent wedges with an aperture of $\pm 30^{\circ}$ along the semimajor axis on the left (eastern) and right (western) sides of the disk. The green profile was calculated on the images where the disk was present whereas the red one is estimated on disk-subtracted images. At $\sim$ 57 au, roughly compatible with the position of the inner edge of the disk, we estimate that the western side of the disk is 10\% brighter than the eastern side.

This asymmetry closely resembles an asymmetry previously observed in the ALMA data at 0.86 mm and published in \cite{Marino2}.  
This is shown in Figure \ref{lumasym}, where we superimposed on the JWST images the $2\times10^{-4}$ Jy/beam contour which identifies the brightest portion of the disk as seen by ALMA. The overlap of the asymmetry identified in ALMA and NIRCam observations is particularly evident at 2 $\mu$m (right panel). This feature was probably already detected in HST data acquired with the F606W filter. Infact, looking at Figure 4 of \citet{Golimowski}, we can notice that the western side of the disk is brighter that the eastern side between 50 and 60 au, at separations comparable with what found by ALMA and NIRCam.

In \citet{Marino2}, the authors discuss different origins for the disc asymmetry. One considers the asymmetry and the gap in the disk to be produced by the interaction with one or more planets. In particular, the secular apsidal resonance model, in which two planets interior to the disk carve a gap via secular interactions, could explain both the presence of the gap and the asymmetry observed with ALMA. However, they also raised the possibility that the apparent asymmetry in the ALMA data could be due to a background object, such as a submillimeter galaxy.

The detection of this asymmetry in JWST data at multiple wavelengths (F444W and F200W) and at a different epoch provides strong evidence that this feature is intrinsic to the disk structure. While the possibility of a background contaminant must be considered, several arguments disfavor this scenario. First, the location and morphology of the feature are aligned with the disk’s semimajor axis and appear smoothly connected to the inner edge of the belt, which would be highly coincidental for a background source. Moreover, if this feature was caused by a background source, it would have moved significantly in the 8~yr between the ALMA and JWST observations. Accounting for the proper motion of HD\, 92945, the suggested source in the ALMA image should be observed at present days at $\Delta RA=$0.5'' and $\Delta Dec=$-0.6'', significantly far from the detected asymmetry in the NIRCam data. Moreover, since no source is detected at the aforementioned position, we can confirm that the asymmetry detected in the ALMA image is actually associated with an intrinsic disk feature.

Therefore, the new JWST observations presented here strongly support the interpretation that HD~92945's gap and asymmetry could be caused via secular interactions between one or two planets and the disc \citep{Yelverton, Marino2, Sefilian}. 

In summary, the morphology of the HD\, 92945 disk observed at near infrared wavelengths with JWST and HST, and at submillimeter wavelengths with ALMA, shows remarkable consistency. This agreement is especially noteworthy between JWST and ALMA, as both datasets were modeled using the same underlying radial density profile. Most of the derived disk parameters agree within 3 $\sigma$, with the majority falling within 2 $\sigma$.

The main discrepancy arises in the characterization of the outer regions of the disk. Despite this, the overall agreement between the datasets reinforces the robustness of the derived disk morphology and highlights the importance of multiwavelength modeling in tracing the spatial distribution of dust grain populations.

\section{Constraints on the presence of planets}

\subsection{Candidates vetting}
\begin{figure*}
    \centering
\includegraphics[trim=0.2cm 0.2cm 0.2cm 0.2cm, clip=true, width=\textwidth]{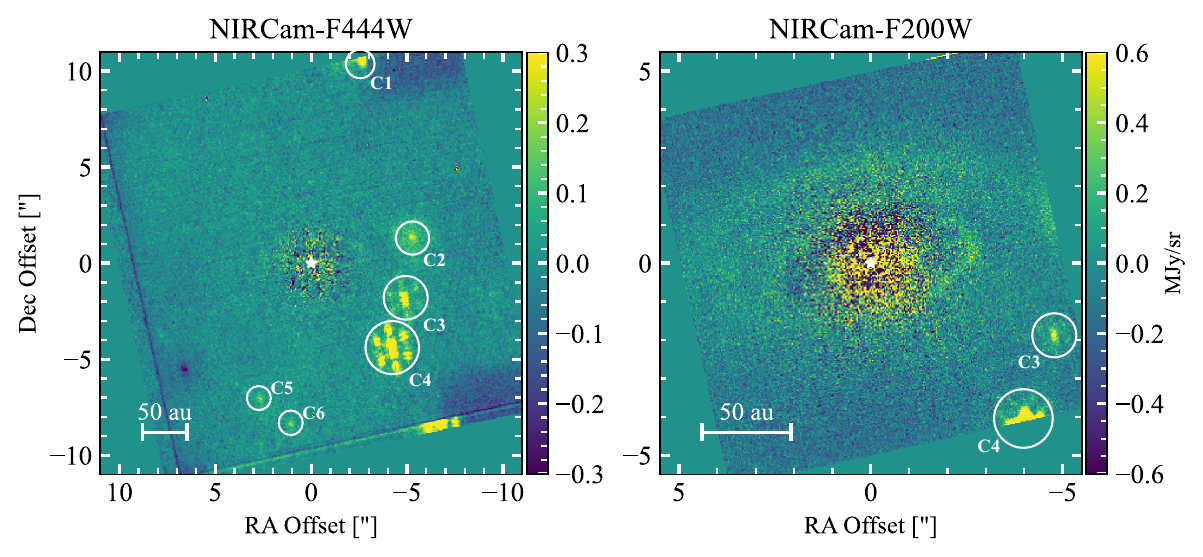}
\caption{Candidates in the field of view of NIRCam F444W (left panel) and F200W (right panel) as seen after the subtraction of the disk via the modeling. Candidates are encircled in white and named Cx with x=1,2,3,4,5,6. The scale bar in the bottom left corner represents a projected distance of 50~au.}
\label{candidates}
\end{figure*}

After modeling and subtracting the disk from the two NIRCam filters, we examined the presence of sources in the field of view. As shown in Figure \ref{candidates}, there are multiple candidates found in each filter (named Cx with x=1,2,3,4,5,6). F444W presents candidates 1 to 6, with C1 only marginally in the field of view, while only the closer ones, C2 and marginally C3, are detected at 2 $\mu$m. No further point sources were detected after the subtraction of the disk. 

By tracing back the proper motion of HD\, 92945 and comparing with data from Gaia DR3 \citep{GAIA2}, as well as ALMA \citep{Marino2} and HST images \citep{Schneider1}, we evaluated whether the observed candidates are consistent with being background objects. Only C1, C2, and C3 have previously detected counterparts in archival observations (ALMA and HST), placing these sources as being consistent with background objects. Searching through the Gaia archive, only the star Gaia DR3 5455707157212258048 is found within a 20'' radius of HD\, 92945, which when accounting for the proper motion of HD\, 92945, is found to be within $\sim$0.6'' of the location of C4. The lack of position overlap could be due to the missing proper motion value for the Gaia DR3 star. C5 and C6 have no archival counterparts.

To investigate the nature of the observed candidates, we performed PSF fitting for the sources fully contained within the NIRCam field of view, specifically, C2, C4, C5, and C6 in the F444W data and C3 in both the F200W and F444W data. For this purpose, we used the \textit{extract\_companions()} function provided by \texttt{spaceKLIP}, which generates a forward model of a point-like PSF based on the host star’s spectral type and photometry to simulate a realistic PSF, aiding in the retrieval of the sources' astrometry and photometry. A recent feature of \texttt{spaceKLIP} also allows for the convolution of the synthetic PSF, which can improve modeling of extended sources. Specifically, we employed the Powell minimization method from the Python package \texttt{scipy}, convolving the PSF with a 2D Gaussian. It is expected that point-like sources are attributed to planetary candidates or background stars, when more extended sources would be more likely to be background galaxies.

Figures \ref{sources444} and \ref{source200} show the best-fit results obtained using a point-like forward model (left column) and a convolved extended PSF (right column). Based on the observed fits, the majority of the detected sources appear to be extended (C2, C3, C5, C6). A visual analysis of the residuals indicates that C2 and C5 are best modeled with an extended PSF, as reflected in the broader, more diffuse structure of their best-fit models. Although less pronounced, C3 also appears more consistent with an extended source, given the oversubtraction in the F444W residuals (second row, left column of Figure \ref{sources444}) and the spread residuals remaining in the F200W images (left column of Figure \ref{source200}). Although C6 does not exhibit significant differences between the forward and convolved models, nor in the residuals, the standard deviations computed from the residual images show a general improvement when convolving the PSF with a 2D Gaussian.

C4 is most likely the only point-like object in the HD\, 92945 field of view, as shown by the lack of blurring in the mid-panel of the right column in the third row of Figure \ref{sources444}. However, the best-fit retrieved when using a convolved model results in a 2D Gaussian extended along the y-axis (see Table \ref{compTab}). This outcome may arise from discrepancies between the synthetic PSF and the actual source being subtracted.

Table \ref{compTab} lists the parameters retrieved for each object. The second and third columns show the coordinates of the sources, given as relative right ascension ($\Delta$RA) and declination ($\Delta$Dec). The fourth column provides the contrast, in units of magnitudes, relative to the star, where
$M^*_{444}=5.68$ and $M^*_{200}=5.70$, as given by the \textit{extract\_companions()} function. We note that these three parameters, $\Delta$RA, $\Delta$Dec, and magnitudes, are the same for both fitting procedures. The fifth and sixth columns report the standard deviations computed from the residual images shown in the left and right columns of Figures \ref{sources444} and \ref{source200}. Finally, the last columns present the best-fit parameters for the 2D Gaussian: $\sigma_x$, $\sigma_y$, and the rotation angle $\theta$. Additionally, we included a final entry, assessing whether each source is extended (C2, C3, and C5) or point-like (C4). C6 is likely a background galaxy, although too faint to be properly characterized and modeled.

\begin{table*}
   \centering
   \caption{Parameters for the detected sources.}
    \begin{tabular}{cccccccccc}
        \hline
        Source & $\Delta RA$ ('') & $\Delta Dec$ ('') & Mag (contrast) &  $res_{\sigma_{bf}}$ & $res_{\sigma_{bfconv}}$ & $\sigma_x$ & $\sigma_y$ & $\Theta$ & Extended?  \\ 
        \hline
       \textbf{F444W} &  & &&&&&&\\
       C2 & $-5.16\pm0.01$ & $1.185\pm0.007$ & $13.84\pm0.05$ & 0.047 & 0.044 & 5 & 5 & 89.6 & Y \\ 
       C3 & $-4.756\pm0.003$ & $-1.859\pm0.003$ & $12.92\pm0.02$ & 0.041 & 0.040 & 1.5 & 1.2 & 48.3 & Y \\ 
       C4 & $-4.1464\pm0.0006$ & $-4.4279\pm0.0005$ & $11.078\pm0.005$ & 0.083 & 0.079 & 0.14 & 2.1 & -46.9 & N \\ 
       C5 & $2.71\pm0.01$ & $-7.03\pm0.01$ & $14.75\pm0.07$ & 0.029 & 0.029 & 1.9 & 1.3 &-87.1 & Y \\ 
       C6 & $1.14\pm0.01$ & $-8.37\pm0.01$ & $15.2\pm0.1$ & 0.023 & 0.023 & 0.35 & 0.36 & 15.5 & Y? \\ 
       \textbf{F200W} &  & &&&&&& \\
       C3 & $-4.730\pm0.003$ & $-1.862\pm0.002$ & $16.34\pm0.05$ & 0.0998 & 0.0998 & 0.26 & 0.11 & 74.2 & Y\\ 
        \hline
    \end{tabular}
    \tablefoot{Parameters for the sources detected in the F444W (C2, C3, C4, C5, and C6) and F200W (C3) field of views. In the second and third column we list the coordinates for each companion and in the fourth column their contrasts with respect to the star. In the fifth and sixth columns are shown the standard deviations calculated on the residual images from left and right column of Figure \ref{sources444} and \ref{source200}. Finally the last columns list the elongation and orientation of the convolved PSF.}
    \label{compTab}
\end{table*}

\subsection{Contrast curves}

\begin{figure*}
    \centering

    \begin{subfigure}[b]{0.49\textwidth}
        \includegraphics[width=\textwidth]{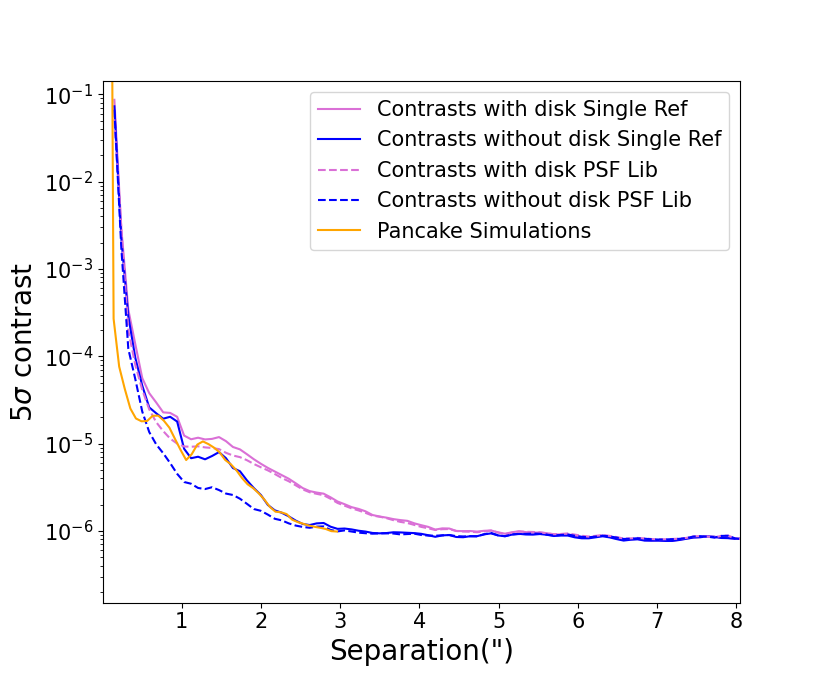}
    \end{subfigure}
    \hfill
    \begin{subfigure}[b]{0.49\textwidth}
        \includegraphics[width=\textwidth]{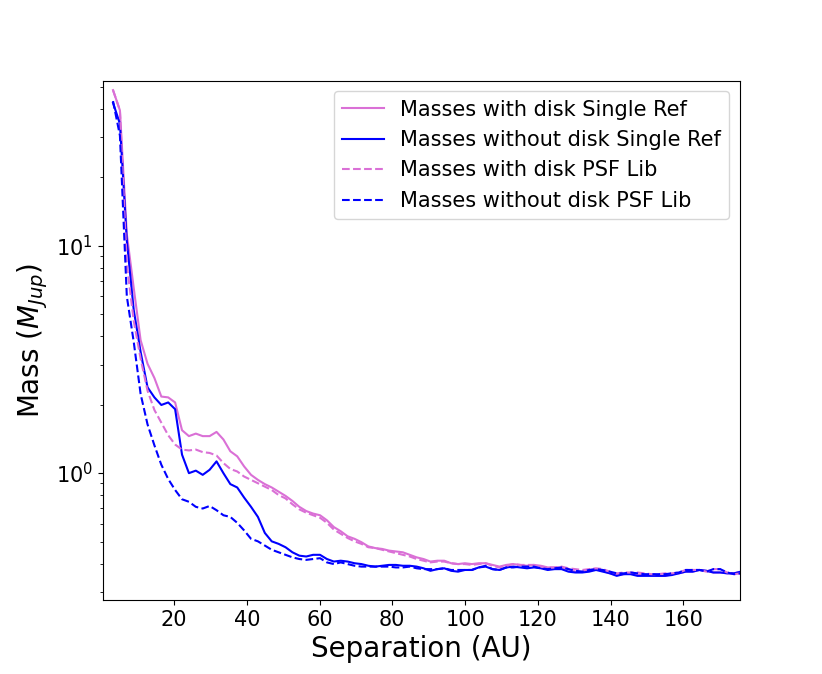}
    \end{subfigure}

    \begin{subfigure}[b]{0.49\textwidth}
        \includegraphics[width=\textwidth]{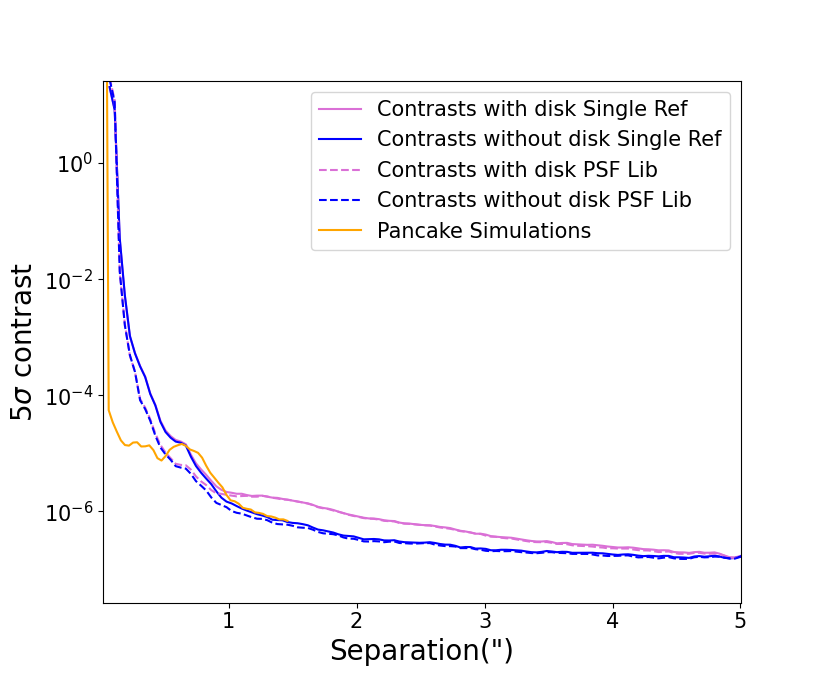}
    \end{subfigure}
    \hfill
    \begin{subfigure}[b]{0.49\textwidth}
        \includegraphics[width=\textwidth]{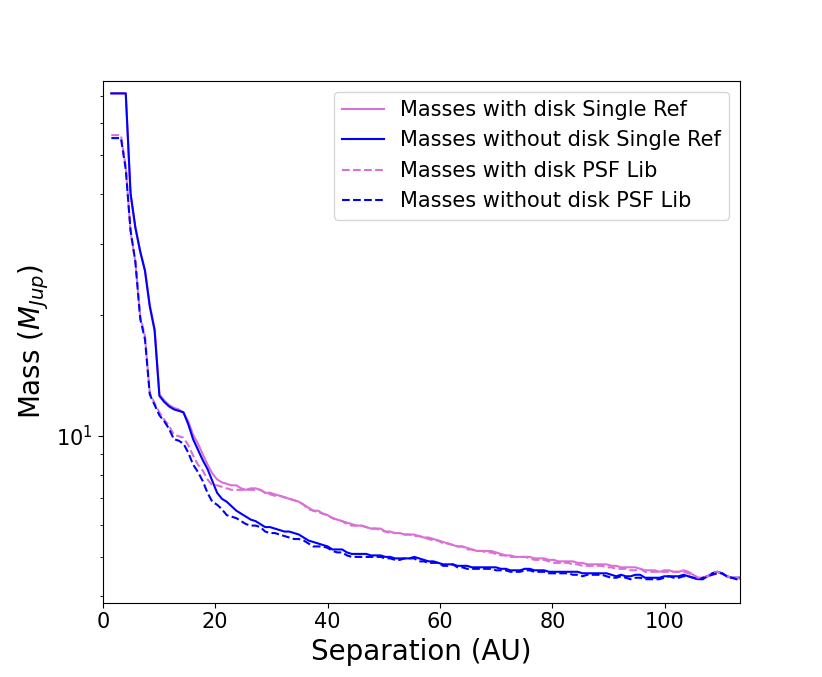}
    \end{subfigure}

    \caption{Contrast curves (left panels) and mass detection limits (right panel) for F444W (top) and F200W (bottom) filters, as a function of projected separation. The yellow curve represents the contrast simulated by PanCAKE. The solid and dashed lines are obtained using one reference and the PSF library, respectively. Finally, The pink curves are retrieved from data where the disk was present and blue curves from disk-subtracted images.}
    \label{contcurves}
\end{figure*}

Figure \ref{contcurves} presents the contrast curves and detection limits obtained for the two NIRCam filters, both in the presence of the disk (pink curves) and after disk subtraction (blue curves). Moreover, we show the contrasts derived using one reference (solid lines) or the PSF library (dashed lines). The contrasts were computed using the \textit{raw\_contrast()} function from the \texttt{spaceKLIP} package, applied to the outputs generated by the \texttt{Winnie} pipeline \citep{Lawson1,Lawson2}. In \texttt{spaceKLIP}, raw contrasts are computed following the standard procedure of measuring the variance in pixel values within narrow annuli centered on the corresponding radial separation and then correcting these measurements to account for small sample statistics \citep{Mawet}. For data where the disk was present, the contrast was calculated on an image processed with the model constrained reference differential imaging (MCRDI) algorithm \citep{Lawson}, while to calculate the contrast in the absence of the disk, we used the corresponding disk-subtracted image. Of these, the latter should be considered to be more representative of the true sensitivity of the data — as the azimuthal brightness variations from the disk serve to artificially inflate the measured noise levels in the former. The contrast performance, in both cases, takes into account the throughput of the coronagraph. Conventionally, these raw contrasts would then be corrected for flux loss resulting from stellar PSF-subtraction by injecting point sources at various positions and then assessing peak throughput following PSF-subtraction as a function of stellocentric separation. However, using MCRDI, throughput loss for any point sources could be avoided simply by including the source in the model of the circumstellar scene (noting that sources outside the PSF optimization region cannot induce oversubtraction). It is therefore common to adopt contrast curves assuming full algorithmic throughput for MCRDI results \citep[e.g.,][]{Lawson1}. As a comparison, we also added to the plot the contrasts estimate obtained with PanCAKE, a simulation tool that extends the official JWST exposure time calculator (Pandeia) to produce more accurate predictions of JWST coronagraphic performance \citep{Carter2}. We note that this simulation tool can only use one reference (HD\, 92921 in our case) to estimate the constrasts. Unsuprisingly, PanCAKE predicts better contrasts within 0.5''. This is likely related to the photon noise introduced by the disk that cannot be accounted for by PanCAKE. At wider separations, Pancake predictions for both filters align with the contrast obtained with a single reference after disk subtraction but they are outperformed when using the PSF library.
As already noted from a visual investigation of the images, we confirm that the use of a PSF library significantly improves the contrasts with respect to the single reference case, in both cases where the disk is present, and especially, after modeling and subtraction.

The contrast limits were subsequently converted into planetary masses using a combination of chemical equilibrium, cloud-free models from \texttt{ATMO} \citep{Phillips} and \texttt{BEX} \citep{Linder}, following the methodology described in \cite{Carter2}. We use the \texttt{MADYS} package\footnote{\url{https://github.com/vsquicciarini/madys}} \citep{Squicciarini2022} to convert from contrast to mass using the \verb|bex-atmo2023-ceq| evolutionary model. For these conversions, the age of the system is fixed to 200 Myrs \citep{Song, Plavchan}. It is important to note, however, that cloud formation and chemical disequilibrium likely play a significant role in shaping the emission spectra of substellar atmospheres, and may affect the inferred mass limits. This was, for example, the case for TWA 7b \citep{Lagrange6, Crotts}, whose characterization, combining JWST/MIRI and JWST/NIRCam data, required the inclusion of water clouds and high atmospheric metallicity to reproduce the observed mid-infrared flux, highlighting the limitations of cloud-free equilibrium models for young, sub-Jupiter mass planets. Detection limits for F444W filter (top) and F200W (bottom) are shown in the right panels of Figure \ref{contcurves}

The gain in contrast and, consequently, in planet's mass that can be detected when subtracting the disk is significant at both wavelengths, taking to a maximum gain of $\sim 0.5$ M\textsubscript{Jup} ($\sim 8.9$ M\textsubscript{Jup}) beyond (within) 1 arcsec at 4 $\mu$m and $\sim 1.4$ M\textsubscript{Jup} at 2 $\mu$m, when using the PSF library. For example, for the F444W filter at $\sim$ 25 au, the minimum detectable masses are $\sim$ 1.2 M\textsubscript{Jup} and $\sim$ 0.7 M\textsubscript{Jup} before and after disk subtraction, with a gain of $\sim$ 0.5 M\textsubscript{Jup}. At the same separation, with the F200W filter instead, the gain is $\sim$ 1.4 M\textsubscript{Jup}, going from $\sim$ 7.1 M\textsubscript{Jup} to $\sim$ 5.7 M\textsubscript{Jup} minimum detectable mass. We also note that the term "gain" might not be accurate in this context. 
This is because the disk emission behaves like an additional noise source when searching for the faint signal of a planet. Therefore, subtracting the disk does not significantly enhance the intrinsic contrast, but rather improves the accuracy of the noise estimation, leading to more reliable detection limits.

\begin{figure*}
    \begin{subfigure}{0.49\textwidth}
        \includegraphics[trim=0.1cm 0.1cm 0.cm 0.97cm, clip=true, width=\textwidth]{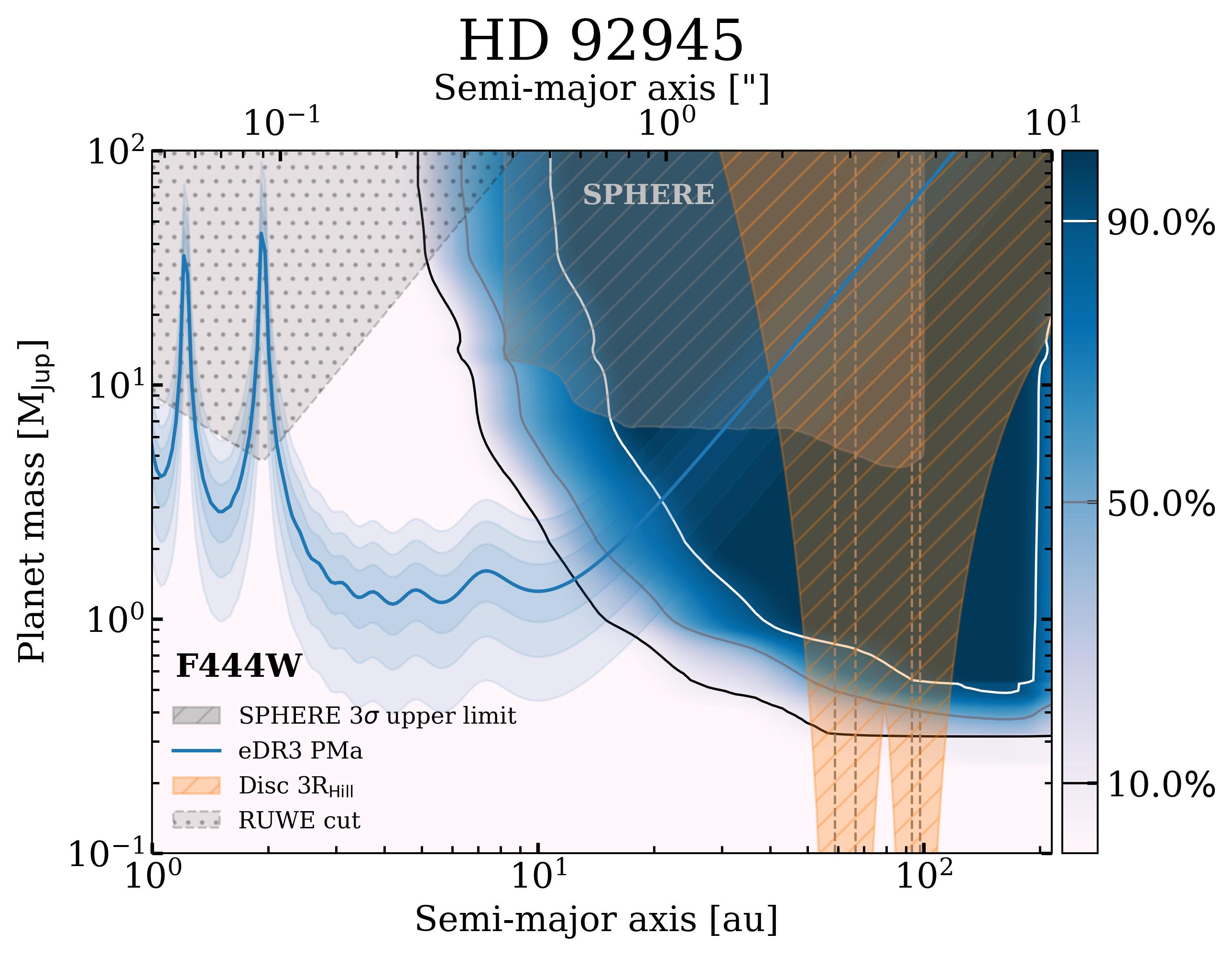} 
    \end{subfigure}
    \hfill
    \begin{subfigure}{0.49\textwidth}
        \includegraphics[trim=0.1cm 0.1cm 0.cm 0.97cm, clip=true, width=\textwidth]{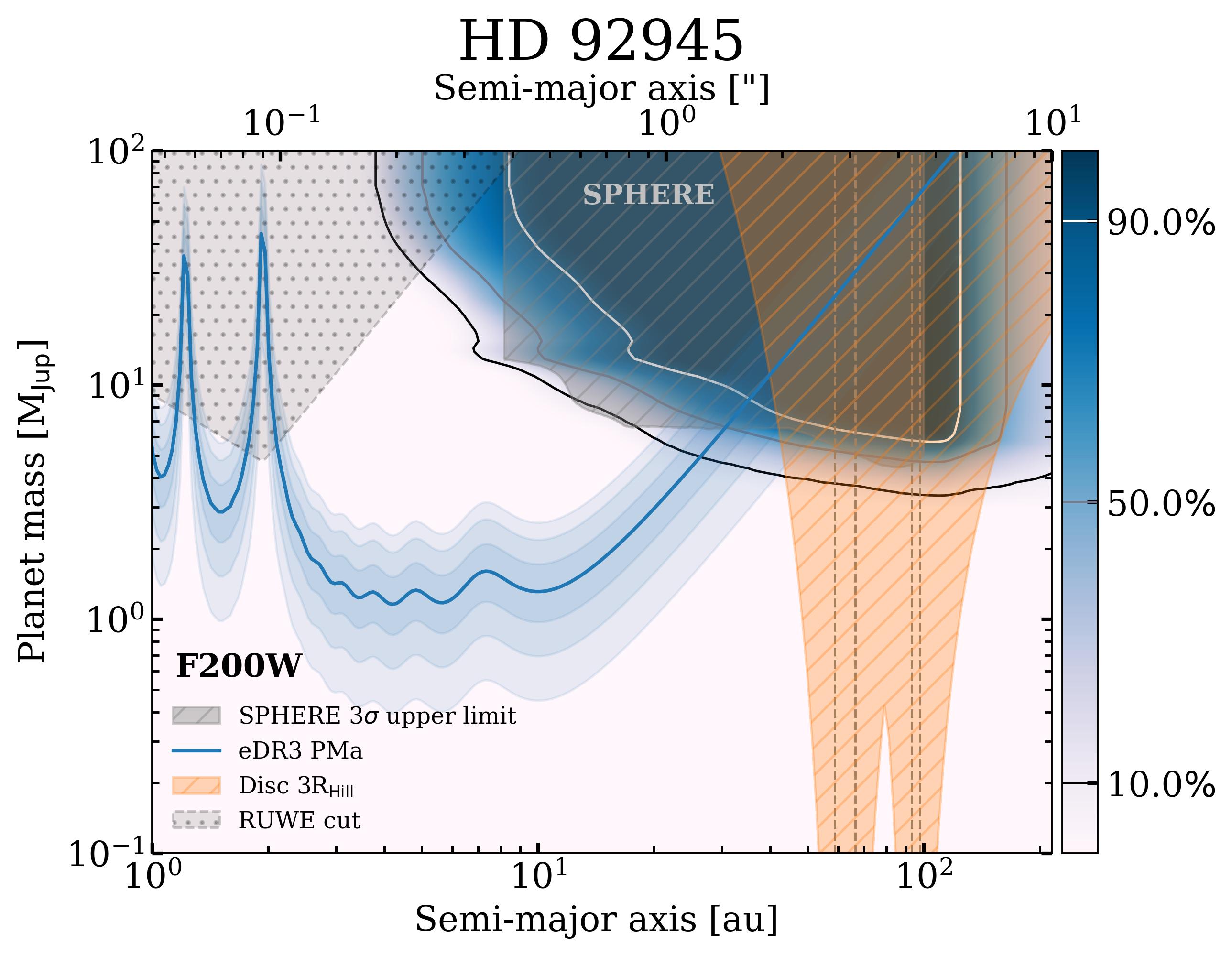} 
    \end{subfigure}

    \caption{Detection probability maps for F444W (left) and F200W (right) filters. JWST results are shown in shades of blue with the white, gray, and black contours representing the 90, 50, and 10 per cent detection probabilities, respectively. The orange dashed vertical lines sign the mean inner and outer edges of the disk, along with the extension of the gap and, shaded in orange, the regions not allowed for planets. The hatched gray area represents planetary parameters excluded by archival SPHERE observations. The gray dotted region illustrates the planet parameter space excluded by Gaia RUWE. The blue curve was obtained modeling the Gaia DR3 proper motion anomaly as generated by a single planet orbiting HD\, 92945.}
    \label{DPM}
\end{figure*}

In Figure \ref{DPM} we show the detection probability maps (DPMs) for both filters, calculated combining both \texttt{MADYS} and \texttt{ExoDMC} \citep{2020ExoDMC}. Detection probabilities for planets from this study are shown in shades of blue and were calculated in the following way: a grid of masses ([0.1,100] M\textsubscript{Jup}) and semimajor axis ([0.1,10] arcsec) was generated and at each point of this grid was attributed a number of orbital phases azimuthally distributed. For each configuration, the separation of the planet was then projected taking into account the inclination of the system. The planetary eccentricity is drawn from a broken Gaussian distribution centered at zero with a standard deviation of 0.1. This choice reflects the assumption that highly eccentric planets are unlikely to be present in the system, as such orbits would likely induce measurable eccentricity in the disk, which is not observed. The mass of the planet was then converted into contrast assuming an age for the system and the \verb|bex-atmo2023-ceq| evolutionary model from \texttt{MADYS}. At each projected separations, the calculated contrast is compared to the 1D contrast curve (PSF library + disk subtracted) shown in Figure \ref{contcurves}. The planet is considered detected if its contrast is above such curve. Moreover, these steps were repeated sampling the different ages taken in the range [100,300] Myrs. The final detection probability for each combination of mass and semimajor axis was computed as the fraction of configurations that resulted in successful detections.

The white, grey, and black contours in Figure \ref{DPM}, associated with JWST/NIRCam detection probability, represent the 90, 50, and 10 per cent chance of retrieving a 5$\sigma$ detection of a planet. In the DPM for the F444W filter, there is a noticeable hard limit at 0.3 M\textsubscript{Jup}. This cutoff arises from gaps in the isochrone grid, which does not extend to certain mass ranges at the adopted system age. As a result, detection probabilities are underestimated at the low-mass end, leading to the appearance of an artificial sharp boundary in the map.

In addition to the DPMs, complementary constraints based on the disk morphology can be overplotted, as shown in orange in Figure \ref{DPM}. The vertical dashed lines indicate the inner and outer edges of the debris disk, as well as the inner and outer boundaries of the central gap. The shaded orange region marks the dynamically unstable zone, where planets located within three Hill radii ($R_{\rm Hill}$) would likely disrupt the surrounding disk material \citep[see, e.g.,][]{Pearce, pearce2022, Pearce1}. 

Additionally, the gray hatched region correspond to areas of the parameter space already ruled out by archival ground-based direct imaging detection limits from SPHERE/IRDIS in the H2 band \citep{Mesa5}, converted into DPM following the same procedure described for JWST data.
The dotted gray area delineates planet parameters excluded by Gaia’s Renormalised Unit Weight Error (RUWE): where RUWE<1.4, we follow the methodology from \citet{Limbach2024} to exclude companions with orbital periods shorter than Gaia DR3’s 1038-day baseline, and that of \citet{Kiefer2024} for longer-period orbits. 
Lastly, the blue curve represents the locus of planet parameters capable of reproducing the observed proper motion anomaly (PMa) in HD\, 92945. This curve is derived using the formalism of \citet{Kervella1}, based on the PMa catalog of \citet{Kervella}, assuming a single planet in a circular orbit responsible for the astrometric signal as in \cite{Marino1}. 

The new detection limits provided by JWST/NIRCam give access to a portion of parameter space that was previously unexplored. Thanks to these observations, we can exclude the presence of Jupiter-mass planets starting from 20–40 au and down to 0.4–0.5 M\textsubscript{Jup} beyond 100 au with probabilities of 50\% and 90\%, respectively. More specifically, we can exclude the presence of planets that could shape the inner edge of the disk that are more massive than 0.5-0.8 M\textsubscript{Jup} around 50 au, with probabilities of 50\% and 90\%, respectively. The same applies for the gap location at 80 au, where no planet is observed more massive than 0.4-0.6 M\textsubscript{Jup} at the 50\% and 90\% confidence level, respectively. On the other hand, comparison between the DPM and the PMa curve suggests that a planet capable of producing the observed proper motion anomaly is expected to lie between 2.5 and 20 au, with a mass in the range of 0.4 to 4.5 M\textsubscript{Jup}. Larger semimajor axes are ruled out, as planets in those regions with sufficient mass to account for the PMa would have been detected by our observations.
We also note that the detection limits derived for putative planets are highly model-dependent, particularly at this wavelength. For example, it is possible that a more massive planet, for instance one with a cloudy atmosphere, could still remain faint enough to fall below our current detection thresholds.

Using the gap width and location derived in this work (see Table \ref{tabdisk}), we can also estimate that a planet with a mass between 0.3 and 0.7 M\textsubscript{Jup}, located at the gap center, could account for the depletion of solids through dynamical clearing. These masses are derived assuming a massless disk and the depletion zone of particles within 3 Hill radii on each side of the planet to be equal to the width of the gap \citep{Gladman,Ida, Kirsh, Friebe}. These values can be compared to the sensitivity reached by JWST/NIRCam in the F444W filter (Figure \ref{DPM}) to see if we can further constrain, or even rule out, this scenario. At $\sim80$ au, the white and grey contours in the DPM allow the detection of planets of 0.65 and 0.45 M\textsubscript{Jup}, with a probability of 90 and 50 per cent, respectively. Thus, ruling out this gap-carving scenario, would require deeper observations to probe lower planetary masses. 
However, as mentioned previously, this scenario is not the only one suitable to explain the disk's features. In particular, multiple planets could explain at the same time the observed gap and luminosity asymmetry when invoking secular apsidal resonances. However, a detailed characterization of possible planetary architectures is beyond the scope of this paper and will be presented in Bendahan-West et al., submitted.

Overall, these results demonstrate that while JWST/NIRCam provide high sensitivity at wide separations, the parameter space for low-mass planets within or near the disk remains only partially explored. Future dynamical modeling, deeper multiwavelength imaging, and precision astrometry will be crucial to fully uncover the planetary architecture of HD\, 92945.

\section{Conclusions}

We presented the first JWST/NIRCam observations of the debris disk surrounding HD\, 92945. The disk was successfully imaged at both 2 $\mu$m and 4.4 $\mu$m, revealing a broad, inclined structure with a prominent gap centered at $\sim$80 au, consistent with previous findings from ALMA and HST. Forward modeling using PSF libraries and MCMC sampling allowed us to constrain the disk’s geometry and dust distribution with fine details. At 2 $\mu$m, we recover inner and outer edges at $\sim$59 au and $\sim$98 au, and a gap with a width of $\sim$27 au and a fractional depth of $\sim$0.79. The disk appears to have a scale height nearly twice that inferred from submillimeter observations, indicating a vertical puffiness likely driven by radiation pressure on small grains. The inner slope of the dust density profile is shallower than that observed at longer wavelengths, while the outer slope is steeper, possibly reflecting differences in the grain size distribution and the influence of dynamical interactions.
At both wavelengths, the disk shows a brightness asymmetry in the southwest portion of the inner ring. This feature is consistent with previous ALMA findings and may be explained by the presence of inner planetary companions perturbing the disk structure \citep{Marino2}. Despite this, no comoving sources were identified in the field of view, and field sources were consistent with background galaxies or stars.

Contrast curves obtained after disk subtraction demonstrate a significant improvement in detection limits, enabling the exclusion of planets down to $\sim$0.7 and 5.7 M\textsubscript{Jup} at 25 au at 4.4 and 2 $\mu$m, respectively. Detection probability maps indicate that Jupiter-mass planets can be excluded with high confidence beyond 20–40 au, and that the presence of a single massive planet beyond 20 au causing the proper motion anomaly is unlikely.

When combining the observed disk features with potential planetary scenarios, a single planet carving the gap at 80 au cannot be entirely ruled out. However, detection limits obtained with NIRCam constrain the upper mass of any such planet to $\sim0.65$ M\textsubscript{Jup}. Alternatively, the detection of the azimuthal asymmetry at both 4 and 2 microns may point toward a more complex scenario involving two planets located interior to the disk, as already suggested by \citet{Marino2}. In this case, the planets would be responsible not only for carving the gap but also for inducing the observed asymmetry through secular apsidal resonances. This latter mechanism appears to provide a more comprehensive explanation for the full set of observational constraints. However, explaining the true origin of the gap is very degenerate and other scenarios could also explain the full set of observational constraints (see Bendahan-West et al. (in prep.) for a more comprehensive analysis).

This work highlights the power of JWST in combining high-contrast imaging with precise disk modeling, pushing the boundaries of planet detection and disk characterization at near infrared wavelengths. Future multiepoch observations and dynamical modeling will help clarify the nature of the inner disk asymmetry and further constrain the architecture of the HD\, 92945 planetary system.

\begin{acknowledgements}
   
Cecilia Lazzoni acknowledges the financial contribution from PRIN MUR 2022 (code 2022YP5ACE) funded by the European Union – NextGenerationEU.
RBW is supported by a Royal Society grant (RF-ERE-221025).
We acknowledge a humble dog named Winnie, whose companionship contributed significantly to the creation of the software called \texttt{Winnie}, which is utilized herein.
We also acknowledge support from ANID -- Millennium Science Initiative Program -- Center Code NCN2024\_001.
We gratefully acknowledge support from the "Programma di Ricerca Fondamentale INAF 2023" of the National Institute of Astrophysics (Large Grant 2023 NextSTEPS). 

This work is based on observations with the NASA/ESA/CSA JWST, obtained at the Space Telescope Science Institute, which is operated by AURA, Inc., under NASA contract NAS 5-03127. These observations are associated with the JWST program 3989 (PI: S. Hinkley). The JWST data presented in this article were obtained from the Mikulski Archive for Space Telescopes (MAST) at the Space Telescope Science Institute. Support for program 3989 was provided by NASA through a grant from the Space Telescope Science Institute, which is operated by the Association of Universities for Research in Astronomy, Inc., under NASA contract NAS 5-03127. 
\end{acknowledgements}

\bibliographystyle{aa} 
\bibliography{bibliography}

@ARTICLE{Adams,
       author = {{Adams}, Jea and {Wang}, Jason},
        title = "{Quantifying the Effect of Coronagraphs on Planet Photometry with the James Webb Space Telescope}",
      journal = {Research Notes of the American Astronomical Society},
         year = 2020,
        month = dec,
       volume = {4},
       number = {12},
          eid = {227},
        pages = {227},
          doi = {10.3847/2515-5172/abd1d4},
       adsurl = {https://ui.adsabs.harvard.edu/abs/2020RNAAS...4..227A}
}

@ARTICLE{Squicciarini2022,
       author = {{Squicciarini}, V. and {Bonavita}, M.},
        title = "{MADYS: the Manifold Age Determination for Young Stars. I. Isochronal age estimates and model comparison}",
      journal = {\aap},
         year = 2022,
        month = oct,
       volume = {666},
          eid = {A15},
        pages = {A15},
          doi = {10.1051/0004-6361/202244193},
archivePrefix = {arXiv},
       eprint = {2206.02446},
 primaryClass = {astro-ph.SR},
       adsurl = {https://ui.adsabs.harvard.edu/abs/2022A&A...666A..15S}
}

@software{2020ExoDMC,
       author = {{Bonavita}, Mariangela},
        title = "{Exo-DMC: Exoplanet Detection Map Calculator}",
 howpublished = {Astrophysics Source Code Library, record ascl:2010.008},
         year = 2020,
        month = oct,
          eid = {ascl:2010.008},
       adsurl = {https://ui.adsabs.harvard.edu/abs/2020ascl.soft10008B}
}

@ARTICLE{Augereau,
       author = {{Augereau}, J.~C. and {Lagrange}, A.~M. and {Mouillet}, D. and {Papaloizou}, J.~C.~B. and {Grorod}, P.~A.},
        title = "{On the HR 4796 A circumstellar disk}",
      journal = {\aap},
         year = 1999,
        month = aug,
       volume = {348},
        pages = {557-569},
          doi = {10.48550/arXiv.astro-ph/9906429},
archivePrefix = {arXiv},
       eprint = {astro-ph/9906429},
 primaryClass = {astro-ph},
       adsurl = {https://ui.adsabs.harvard.edu/abs/1999A&A...348..557A}
}

@ARTICLE{Beuzit,
       author = {{Beuzit}, J. -L. and {Vigan}, A. and {Mouillet}, D. and {Dohlen}, K. and
         {Gratton}, R. and {Boccaletti}, A. and {Sauvage}, J. -F. and
         {Schmid}, H.~M. and {Langlois}, M. and {Petit}, C. and {Baruffolo}, A. and
         {Feldt}, M. and {Milli}, J. and {Wahhaj}, Z. and {Abe}, L. and
         {Anselmi}, U. and {Antichi}, J. and {Barette}, R. and {Baudrand}, J. and
         {Baudoz}, P. and {Bazzon}, A. and {Bernardi}, P. and {Blanchard}, P. and
         {Brast}, R. and {Bruno}, P. and {Buey}, T. and {Carbillet}, M. and
         {Carle}, M. and {Cascone}, E. and {Chapron}, F. and {Charton}, J. and
         {Chauvin}, G. and {Claudi}, R. and {Costille}, A. and {De Caprio}, V. and
         {de Boer}, J. and {Delboulb{\'e}}, A. and {Desidera}, S. and
         {Dominik}, C. and {Downing}, M. and {Dupuis}, O. and {Fabron}, C. and
         {Fantinel}, D. and {Farisato}, G. and {Feautrier}, P. and
         {Fedrigo}, E. and {Fusco}, T. and {Gigan}, P. and {Ginski}, C. and
         {Girard}, J. and {Giro}, E. and {Gisler}, D. and {Gluck}, L. and
         {Gry}, C. and {Henning}, T. and {Hubin}, N. and {Hugot}, E. and
         {Incorvaia}, S. and {Jaquet}, M. and {Kasper}, M. and {Lagadec}, E. and
         {Lagrange}, A. -M. and {Le Coroller}, H. and {Le Mignant}, D. and
         {Le Ruyet}, B. and {Lessio}, G. and {Lizon}, J. -L. and {Llored}, M. and
         {Lundin}, L. and {Madec}, F. and {Magnard}, Y. and {Marteaud}, M. and
         {Martinez}, P. and {Maurel}, D. and {M{\'e}nard}, F. and {Mesa}, D. and
         {M{\"o}ller-Nilsson}, O. and {Moulin}, T. and {Moutou}, C. and
         {Orign{\'e}}, A. and {Parisot}, J. and {Pavlov}, A. and {Perret}, D. and
         {Pragt}, J. and {Puget}, P. and {Rabou}, P. and {Ramos}, J. and
         {Reess}, J. -M. and {Rigal}, F. and {Rochat}, S. and {Roelfsema}, R. and
         {Rousset}, G. and {Roux}, A. and {Saisse}, M. and {Salasnich}, B. and
         {Santambrogio}, E. and {Scuderi}, S. and {Segransan}, D. and
         {Sevin}, A. and {Siebenmorgen}, R. and {Soenke}, C. and {Stadler}, E. and
         {Suarez}, M. and {Tiph{\`e}ne}, D. and {Turatto}, M. and {Udry}, S. and
         {Vakili}, F. and {Waters}, L.~B.~F.~M. and {Weber}, L. and {Wildi}, F. and
         {Zins}, G. and {Zurlo}, A.},
        title = "{SPHERE: the exoplanet imager for the Very Large Telescope}",
      journal = {A\&A},
         year = "2019",
        month = "Nov",
       volume = {631},
          eid = {A155},
        pages = {A155},
          doi = {10.1051/0004-6361/201935251},
       adsurl = {https://ui.adsabs.harvard.edu/abs/2019A&A...631A.155B}
}

@ARTICLE{Carter,
       author = {{Carter}, Aarynn L. and {Hinkley}, Sasha and {Kammerer}, Jens and {Skemer}, Andrew and {Biller}, Beth A. and {Leisenring}, Jarron M. and {Millar-Blanchaer}, Maxwell A. and {Petrus}, Simon and {Stone}, Jordan M. and {Ward-Duong}, Kimberly and {Wang}, Jason J. and {Girard}, Julien H. and {Hines}, Dean C. and {Perrin}, Marshall D. and {Pueyo}, Laurent and {Balmer}, William O. and {Bonavita}, Mariangela and {Bonnefoy}, Mickael and {Chauvin}, Gael and {Choquet}, Elodie and {Christiaens}, Valentin and {Danielski}, Camilla and {Kennedy}, Grant M. and {Matthews}, Elisabeth C. and {Miles}, Brittany E. and {Patapis}, Polychronis and {Ray}, Shrishmoy and {Rickman}, Emily and {Sallum}, Steph and {Stapelfeldt}, Karl R. and {Whiteford}, Niall and {Zhou}, Yifan and {Absil}, Olivier and {Boccaletti}, Anthony and {Booth}, Mark and {Bowler}, Brendan P. and {Chen}, Christine H. and {Currie}, Thayne and {Fortney}, Jonathan J. and {Grady}, Carol A. and {Greebaum}, Alexandra Z. and {Henning}, Thomas and {Hoch}, Kielan K.~W. and {Janson}, Markus and {Kalas}, Paul and {Kenworthy}, Matthew A. and {Kervella}, Pierre and {Kraus}, Adam L. and {Lagage}, Pierre-Olivier and {Liu}, Michael C. and {Macintosh}, Bruce and {Marino}, Sebastian and {Marley}, Mark S. and {Marois}, Christian and {Matthews}, Brenda C. and {Mawet}, Dimitri and {McElwain}, Michael W. and {Metchev}, Stanimir and {Meyer}, Michael R. and {Molliere}, Paul and {Moran}, Sarah E. and {Morley}, Caroline V. and {Mukherjee}, Sagnick and {Pantin}, Eric and {Quirrenbach}, Andreas and {Rebollido}, Isabel and {Ren}, Bin B. and {Schneider}, Glenn and {Vasist}, Malavika and {Worthen}, Kadin and {Wyatt}, Mark C. and {Briesemeister}, Zackery W. and {Bryan}, Marta L. and {Calissendorff}, Per and {Cantalloube}, Faustine and {Cugno}, Gabriele and {De Furio}, Matthew and {Dupuy}, Trent J. and {Factor}, Samuel M. and {Faherty}, Jacqueline K. and {Fitzgerald}, Michael P. and {Franson}, Kyle and {Gonzales}, Eileen C. and {Hood}, Callie E. and {Howe}, Alex R. and {Kuzuhara}, Masayuki and {Lagrange}, Anne-Marie and {Lawson}, Kellen and {Lazzoni}, Cecilia and {Lew}, Ben W.~P. and {Liu}, Pengyu and {Llop-Sayson}, Jorge and {Lloyd}, James P. and {Martinez}, Raquel A. and {Mazoyer}, Johan and {Palma-Bifani}, Paulina and {Quanz}, Sascha P. and {Redai}, Jea Adams and {Samland}, Matthias and {Schlieder}, Joshua E. and {Tamura}, Motohide and {Tan}, Xianyu and {Uyama}, Taichi and {Vigan}, Arthur and {Vos}, Johanna M. and {Wagner}, Kevin and {Wolff}, Schuyler G. and {Ygouf}, Marie and {Zhang}, Xi and {Zhang}, Keming and {Zhang}, Zhoujian},
        title = "{The JWST Early Release Science Program for Direct Observations of Exoplanetary Systems I: High-contrast Imaging of the Exoplanet HIP 65426 b from 2 to 16 {\ensuremath{\mu}}m}",
      journal = {\apjl},
         year = 2023,
        month = jul,
       volume = {951},
       number = {1},
          eid = {L20},
        pages = {L20},
          doi = {10.3847/2041-8213/acd93e},
archivePrefix = {arXiv},
       eprint = {2208.14990},
 primaryClass = {astro-ph.EP},
       adsurl = {https://ui.adsabs.harvard.edu/abs/2023ApJ...951L..20C}
}

@INPROCEEDINGS{Carter2,
       author = {{Carter}, Aarynn L. and {Skemer}, Andrew J.~I. and {Danielski}, Camilla and {Leisenring}, Jarron and {Wang}, Jason J. and {Van Gorkom}, Kyle and {York}, Brian and {Adams}, Jea and {Biller}, Beth and {Girard}, Julien H. and {Hinkley}, Sasha and {Nickson}, Bryony and {Perrin}, Marshall and {Pueyo}, Laurent},
        title = "{Simulating JWST high contrast observations with PanCAKE}",
    booktitle = {Techniques and Instrumentation for Detection of Exoplanets X},
         year = 2021,
       editor = {{Shaklan}, Stuart B. and {Ruane}, Garreth J.},
       series = {Society of Photo-Optical Instrumentation Engineers (SPIE) Conference Series},
       volume = {11823},
        month = sep,
          eid = {118230H},
        pages = {118230H},
          doi = {10.1117/12.2594501},
       adsurl = {https://ui.adsabs.harvard.edu/abs/2021SPIE11823E..0HC}
}

@ARTICLE{Chen1,
       author = {{Chen}, C.~H. and {Patten}, B.~M. and {Werner}, M.~W. and {Dowell}, C.~D. and {Stapelfeldt}, K.~R. and {Song}, I. and {Stauffer}, J.~R. and {Blaylock}, M. and {Gordon}, K.~D. and {Krause}, V.},
        title = "{A Spitzer Study of Dusty Disks around Nearby, Young Stars}",
      journal = {\apj},
         year = 2005,
        month = dec,
       volume = {634},
       number = {2},
        pages = {1372-1384},
          doi = {10.1086/497124},
       adsurl = {https://ui.adsabs.harvard.edu/abs/2005ApJ...634.1372C}
}

@ARTICLE{Crotts,
       author = {{Crotts}, Katie A. and {Carter}, Aarynn L. and {Lawson}, Kellen and {Mang}, James and {Biller}, Beth and {Booth}, Mark and {Ferrer-Chavez}, Rodrigo and {Girard}, Julien H. and {Lagrange}, Anne-Marie and {Liu}, Michael C. and {Marino}, Sebastian and {Millar-Blanchaer}, Maxwell A. and {Skemer}, Andy and {Strampelli}, Giovanni M. and {Wang}, Jason and {Absil}, Olivier and {Balmer}, William O. and {Bendahan-West}, Rapha{\"e}l and {Bogat}, Ellis and {Bowens-Rubin}, Rachel and {Chauvin}, Ga{\"e}l and {Fontanive}, Cl{\'e}mence and {Franson}, Kyle and {Kammerer}, Jens and {Leisenring}, Jarron and {Morley}, Caroline V. and {Rebollido}, Isabel and {Skaf}, Nour and {Sutlieff}, Ben J. and {Bruinsma}, Evelyn L. and {Hinkley}, Sasha and {Hoch}, Kielan and {James}, Andrew D. and {Kane}, Rohan and {Mawet}, Dimitri and {Meyer}, Michael R. and {Palatnick}, Skyler and {Perrin}, Marshall D. and {Ray}, Shrishmoy and {Rickman}, Emily and {Sanghi}, Aniket and {Stephenson}, Klaus Subbotina},
        title = "{Follow-up Exploration of the TWA 7 Planet{\textendash}Disk System with JWST NIRCam}",
      journal = {\apjl},
         year = 2025,
        month = jul,
       volume = {987},
       number = {2},
          eid = {L41},
        pages = {L41},
          doi = {10.3847/2041-8213/ade798},
archivePrefix = {arXiv},
       eprint = {2506.19932},
 primaryClass = {astro-ph.EP},
       adsurl = {https://ui.adsabs.harvard.edu/abs/2025ApJ...987L..41C}
}

@ARTICLE{Dahlqvist,
       author = {{Dahlqvist}, C. -H. and {Milli}, J. and {Absil}, O. and {Cantalloube}, F. and {Matra}, L. and {Choquet}, E. and {del Burgo}, C. and {Marshall}, J.~P. and {Wyatt}, M. and {Ertel}, S.},
        title = "{The SHARDDS survey: limits on planet occurrence rates based on point sources analysis via the Auto-RSM framework}",
      journal = {\aap},
         year = 2022,
        month = oct,
       volume = {666},
          eid = {A33},
        pages = {A33},
          doi = {10.1051/0004-6361/202244145},
archivePrefix = {arXiv},
       eprint = {2208.09204},
 primaryClass = {astro-ph.EP},
       adsurl = {https://ui.adsabs.harvard.edu/abs/2022A&A...666A..33D}
}

@ARTICLE{Foreman-Mackey,
       author = {{Foreman-Mackey}, Daniel and {Hogg}, David W. and {Lang}, Dustin and
         {Goodman}, Jonathan},
        title = "{emcee: The MCMC Hammer}",
      journal = {PASP},
         year = 2013,
        month = mar,
       volume = {125},
       number = {925},
        pages = {306},
          doi = {10.1086/670067},
archivePrefix = {arXiv},
       eprint = {1202.3665},
 primaryClass = {astro-ph.IM},
       adsurl = {https://ui.adsabs.harvard.edu/abs/2013PASP..125..306F}
}

@ARTICLE{Friebe,
       author = {{Friebe}, Marc F. and {Pearce}, Tim D. and {L{\"o}hne}, Torsten},
        title = "{Gap carving by a migrating planet embedded in a massive debris disc}",
      journal = {\mnras},
         year = 2022,
        month = may,
       volume = {512},
       number = {3},
        pages = {4441-4454},
          doi = {10.1093/mnras/stac664},
archivePrefix = {arXiv},
       eprint = {2203.03611},
 primaryClass = {astro-ph.EP},
       adsurl = {https://ui.adsabs.harvard.edu/abs/2022MNRAS.512.4441F}
}

@ARTICLE{GAIA2,
       author = {{Gaia Collaboration} and {Brown}, A.~G.~A. and {Vallenari}, A. and {Prusti}, T. and {de Bruijne}, J.~H.~J. and {Babusiaux}, C. and {Biermann}, M. and {Creevey}, O.~L. and {Evans}, D.~W. and {Eyer}, L. and {Hutton}, A. and {Jansen}, F. and {Jordi}, C. and {Klioner}, S.~A. and {Lammers}, U. and {Lindegren}, L. and {Luri}, X. and {Mignard}, F. and {Panem}, C. and {Pourbaix}, D. and {Randich}, S. and {Sartoretti}, P. and {Soubiran}, C. and {Walton}, N.~A. and {Arenou}, F. and {Bailer-Jones}, C.~A.~L. and {Bastian}, U. and {Cropper}, M. and {Drimmel}, R. and {Katz}, D. and {Lattanzi}, M.~G. and {van Leeuwen}, F. and {Bakker}, J. and {Cacciari}, C. and {Casta{\~n}eda}, J. and {De Angeli}, F. and {Ducourant}, C. and {Fabricius}, C. and {Fouesneau}, M. and {Fr{\'e}mat}, Y. and {Guerra}, R. and {Guerrier}, A. and {Guiraud}, J. and {Jean-Antoine Piccolo}, A. and {Masana}, E. and {Messineo}, R. and {Mowlavi}, N. and {Nicolas}, C. and {Nienartowicz}, K. and {Pailler}, F. and {Panuzzo}, P. and {Riclet}, F. and {Roux}, W. and {Seabroke}, G.~M. and {Sordo}, R. and {Tanga}, P. and {Th{\'e}venin}, F. and {Gracia-Abril}, G. and {Portell}, J. and {Teyssier}, D. and {Altmann}, M. and {Andrae}, R. and {Bellas-Velidis}, I. and {Benson}, K. and {Berthier}, J. and {Blomme}, R. and {Brugaletta}, E. and {Burgess}, P.~W. and {Busso}, G. and {Carry}, B. and {Cellino}, A. and {Cheek}, N. and {Clementini}, G. and {Damerdji}, Y. and {Davidson}, M. and {Delchambre}, L. and {Dell'Oro}, A. and {Fern{\'a}ndez-Hern{\'a}ndez}, J. and {Galluccio}, L. and {Garc{\'\i}a-Lario}, P. and {Garcia-Reinaldos}, M. and {Gonz{\'a}lez-N{\'u}{\~n}ez}, J. and {Gosset}, E. and {Haigron}, R. and {Halbwachs}, J. -L. and {Hambly}, N.~C. and {Harrison}, D.~L. and {Hatzidimitriou}, D. and {Heiter}, U. and {Hern{\'a}ndez}, J. and {Hestroffer}, D. and {Hodgkin}, S.~T. and {Holl}, B. and {Jan{\ss}en}, K. and {Jevardat de Fombelle}, G. and {Jordan}, S. and {Krone-Martins}, A. and {Lanzafame}, A.~C. and {L{\"o}ffler}, W. and {Lorca}, A. and {Manteiga}, M. and {Marchal}, O. and {Marrese}, P.~M. and {Moitinho}, A. and {Mora}, A. and {Muinonen}, K. and {Osborne}, P. and {Pancino}, E. and {Pauwels}, T. and {Petit}, J. -M. and {Recio-Blanco}, A. and {Richards}, P.~J. and {Riello}, M. and {Rimoldini}, L. and {Robin}, A.~C. and {Roegiers}, T. and {Rybizki}, J. and {Sarro}, L.~M. and {Siopis}, C. and {Smith}, M. and {Sozzetti}, A. and {Ulla}, A. and {Utrilla}, E. and {van Leeuwen}, M. and {van Reeven}, W. and {Abbas}, U. and {Abreu Aramburu}, A. and {Accart}, S. and {Aerts}, C. and {Aguado}, J.~J. and {Ajaj}, M. and {Altavilla}, G. and {{\'A}lvarez}, M.~A. and {{\'A}lvarez Cid-Fuentes}, J. and {Alves}, J. and {Anderson}, R.~I. and {Anglada Varela}, E. and {Antoja}, T. and {Audard}, M. and {Baines}, D. and {Baker}, S.~G. and {Balaguer-N{\'u}{\~n}ez}, L. and {Balbinot}, E. and {Balog}, Z. and {Barache}, C. and {Barbato}, D. and {Barros}, M. and {Barstow}, M.~A. and {Bartolom{\'e}}, S. and {Bassilana}, J. -L. and {Bauchet}, N. and {Baudesson-Stella}, A. and {Becciani}, U. and {Bellazzini}, M. and {Bernet}, M. and {Bertone}, S. and {Bianchi}, L. and {Blanco-Cuaresma}, S. and {Boch}, T. and {Bombrun}, A. and {Bossini}, D. and {Bouquillon}, S. and {Bragaglia}, A. and {Bramante}, L. and {Breedt}, E. and {Bressan}, A. and {Brouillet}, N. and {Bucciarelli}, B. and {Burlacu}, A. and {Busonero}, D. and {Butkevich}, A.~G. and {Buzzi}, R. and {Caffau}, E. and {Cancelliere}, R. and {C{\'a}novas}, H. and {Cantat-Gaudin}, T. and {Carballo}, R. and {Carlucci}, T. and {Carnerero}, M.~I. and {Carrasco}, J.~M. and {Casamiquela}, L. and {Castellani}, M. and {Castro-Ginard}, A. and {Castro Sampol}, P. and {Chaoul}, L. and {Charlot}, P. and {Chemin}, L. and {Chiavassa}, A. and {Cioni}, M. -R.~L. and {Comoretto}, G. and {Cooper}, W.~J. and {Cornez}, T. and {Cowell}, S. and {Crifo}, F. and {Crosta}, M. and {Crowley}, C. and {Dafonte}, C. and {Dapergolas}, A. and {David}, M. and {David}, P.},
        title = "{Gaia Early Data Release 3. Summary of the contents and survey properties}",
      journal = {\aap},
         year = 2021,
        month = may,
       volume = {649},
          eid = {A1},
        pages = {A1},
          doi = {10.1051/0004-6361/202039657},
archivePrefix = {arXiv},
       eprint = {2012.01533},
 primaryClass = {astro-ph.GA},
       adsurl = {https://ui.adsabs.harvard.edu/abs/2021A&A...649A...1G}
}

@ARTICLE{Gaspar1,
       author = {{G{\'a}sp{\'a}r}, Andr{\'a}s and {Wolff}, Schuyler Grace and {Rieke}, George H. and {Leisenring}, Jarron M. and {Morrison}, Jane and {Su}, Kate Y.~L. and {Ward-Duong}, Kimberly and {Aguilar}, Jonathan and {Ygouf}, Marie and {Beichman}, Charles and {Llop-Sayson}, Jorge and {Bryden}, Geoffrey},
        title = "{Spatially resolved imaging of the inner Fomalhaut disk using JWST/MIRI}",
      journal = {Nature Astronomy},
         year = 2023,
        month = jul,
       volume = {7},
        pages = {790-798},
          doi = {10.1038/s41550-023-01962-6},
archivePrefix = {arXiv},
       eprint = {2305.03789},
 primaryClass = {astro-ph.EP},
       adsurl = {https://ui.adsabs.harvard.edu/abs/2023NatAs...7..790G}
}

@article{Gladman,
	Author = {{Gladman}, B.},
	Journal = {ICARUS},
	Month = may,
	Pages = {247-263},
	Title = {{Dynamics of systems of two close planets}},
	Volume = 106,
	Year = 1993}

@ARTICLE{Golimowski,
       author = {{Golimowski}, D.~A. and {Krist}, J.~E. and {Stapelfeldt}, K.~R. and {Chen}, C.~H. and {Ardila}, D.~R. and {Bryden}, G. and {Clampin}, M. and {Ford}, H.~C. and {Illingworth}, G.~D. and {Plavchan}, P. and {Rieke}, G.~H. and {Su}, K.~Y.~L.},
        title = "{Hubble and Spitzer Space Telescope Observations of the Debris Disk around the nearby K Dwarf HD 92945}",
      journal = {\aj},
         year = 2011,
        month = jul,
       volume = {142},
       number = {1},
          eid = {30},
        pages = {30},
          doi = {10.1088/0004-6256/142/1/30},
archivePrefix = {arXiv},
       eprint = {1105.0888},
 primaryClass = {astro-ph.SR},
       adsurl = {https://ui.adsabs.harvard.edu/abs/2011AJ....142...30G}
}

@ARTICLE{Gonzalez,
       author = {{Gomez Gonzalez}, Carlos Alberto and {Wertz}, Olivier and
         {Absil}, Olivier and {Christiaens}, Valentin and {Defr{\`e}re}, Denis and
         {Mawet}, Dimitri and {Milli}, Julien and {Absil}, Pierre-Antoine and
         {Van Droogenbroeck}, Marc and {Cantalloube}, Faustine and
         {Hinz}, Philip M. and {Skemer}, Andrew J. and {Karlsson}, Mikael and
         {Surdej}, Jean},
        title = "{VIP: Vortex Image Processing Package for High-contrast Direct Imaging}",
      journal = {AJ},
         year = "2017",
        month = "Jul",
       volume = {154},
       number = {1},
          eid = {7},
        pages = {7},
          doi = {10.3847/1538-3881/aa73d7},
archivePrefix = {arXiv},
       eprint = {1705.06184},
 primaryClass = {astro-ph.IM},
       adsurl = {https://ui.adsabs.harvard.edu/abs/2017AJ....154....7G}
}

@article{Gray,
	Adsurl = {http://adsabs.harvard.edu/abs/1999AJ....118.2993G},
	Author = {{Gray}, R.~O. and {Kaye}, A.~B.},
	Doi = {10.1086/301134},
	Journal = {AJ},
	Month = dec,
	Pages = {2993-2996},
	Title = {{HR 8799: A Link between {$\gamma$} Doradus Variables and {$\lambda$} Bootis Stars}},
	Volume = 118,
	Year = 1999}

@ARTICLE{Hinkley1,
       author = {{Hinkley}, Sasha and {Carter}, Aarynn L. and {Ray}, Shrishmoy and {Skemer}, Andrew and {Biller}, Beth and {Choquet}, Elodie and {Millar-Blanchaer}, Maxwell A. and {Sallum}, Stephanie and {Miles}, Brittany and {Whiteford}, Niall and {Patapis}, Polychronis and {Perrin}, Marshall and {Pueyo}, Laurent and {Schneider}, Glenn and {Stapelfeldt}, Karl and {Wang}, Jason and {Ward-Duong}, Kimberly and {Bowler}, Brendan P. and {Boccaletti}, Anthony and {Girard}, Julien H. and {Hines}, Dean and {Kalas}, Paul and {Kammerer}, Jens and {Kervella}, Pierre and {Leisenring}, Jarron and {Pantin}, Eric and {Zhou}, Yifan and {Meyer}, Michael and {Liu}, Michael C. and {Bonnefoy}, Mickael and {Currie}, Thayne and {McElwain}, Michael and {Metchev}, Stanimir and {Wyatt}, Mark and {Absil}, Olivier and {Adams}, Jea and {Barman}, Travis and {Baraffe}, Isabelle and {Bonavita}, Mariangela and {Booth}, Mark and {Bryan}, Marta and {Chauvin}, Gael and {Chen}, Christine and {Danielski}, Camilla and {De Furio}, Matthew and {Factor}, Samuel M. and {Fitzgerald}, Michael P. and {Fortney}, Jonathan J. and {Grady}, Carol and {Greenbaum}, Alexandra and {Henning}, Thomas and {Hoch}, Kielan K.~W. and {Janson}, Markus and {Kennedy}, Grant and {Kenworthy}, Matthew and {Kraus}, Adam and {Kuzuhara}, Masayuki and {Lagage}, Pierre-Olivier and {Lagrange}, Anne-Marie and {Launhardt}, Ralf and {Lazzoni}, Cecilia and {Lloyd}, James and {Marino}, Sebastian and {Marley}, Mark and {Martinez}, Raquel and {Marois}, Christian and {Matthews}, Brenda and {Matthews}, Elisabeth C. and {Mawet}, Dimitri and {Mazoyer}, Johan and {Phillips}, Mark and {Petrus}, Simon and {Quanz}, Sascha P. and {Quirrenbach}, Andreas and {Rameau}, Julien and {Rebollido}, Isabel and {Rickman}, Emily and {Samland}, Matthias and {Sargent}, B. and {Schlieder}, Joshua E. and {Sivaramakrishnan}, Anand and {Stone}, Jordan M. and {Tamura}, Motohide and {Tremblin}, Pascal and {Uyama}, Taichi and {Vasist}, Malavika and {Vigan}, Arthur and {Wagner}, Kevin and {Ygouf}, Marie},
        title = "{The JWST Early Release Science Program for the Direct Imaging and Spectroscopy of Exoplanetary Systems}",
      journal = {\pasp},
         year = 2022,
        month = sep,
       volume = {134},
       number = {1039},
          eid = {095003},
        pages = {095003},
          doi = {10.1088/1538-3873/ac77bd},
archivePrefix = {arXiv},
       eprint = {2205.12972},
 primaryClass = {astro-ph.EP},
       adsurl = {https://ui.adsabs.harvard.edu/abs/2022PASP..134i5003H}
}

@article{Ida,
	title = {Orbital {Migration} of {Neptune} and {Orbital} {Distribution} of {Trans}-{Neptunian} {Objects}},
	volume = {534},
	issn = {0004-637X},
	url = {https://ui.adsabs.harvard.edu/abs/2000ApJ...534..428I},
	doi = {10.1086/308720},
	urldate = {2024-03-30},
	journal = {The Astrophysical Journal},
	author = {Ida, Shigeru and Bryden, Geoffrey and Lin, D. N. C. and Tanaka, Hidekazu},
	month = may,
	year = {2000},
	note = {ADS Bibcode: 2000ApJ...534..428I},
	pages = {428--445},
}

@INPROCEEDINGS{Kammerer,
       author = {{Kammerer}, Jens and {Girard}, Julien and {Carter}, Aarynn L. and {Perrin}, Marshall D. and {Cooper}, Rachel and {Thatte}, Deepashri and {Vandal}, Thomas and {Leisenring}, Jarron and {Wang}, Jason and {Balmer}, William O. and {Sivaramakrishnan}, Anand and {Pueyo}, Laurent and {Ward-Duong}, Kimberly and {Sunnquist}, Ben and {Adams Redai}, J{\'e}a.},
        title = "{Performance of near-infrared high-contrast imaging methods with JWST from commissioning}",
    booktitle = {Space Telescopes and Instrumentation 2022: Optical, Infrared, and Millimeter Wave},
         year = 2022,
       editor = {{Coyle}, Laura E. and {Matsuura}, Shuji and {Perrin}, Marshall D.},
       series = {Society of Photo-Optical Instrumentation Engineers (SPIE) Conference Series},
       volume = {12180},
        month = aug,
          eid = {121803N},
        pages = {121803N},
          doi = {10.1117/12.2628865},
archivePrefix = {arXiv},
       eprint = {2208.00996},
 primaryClass = {astro-ph.EP},
       adsurl = {https://ui.adsabs.harvard.edu/abs/2022SPIE12180E..3NK}
}

@ARTICLE{Kervella,
       author = {{Kervella}, Pierre and {Arenou}, Fr{\'e}d{\'e}ric and {Th{\'e}venin}, Fr{\'e}d{\'e}ric},
        title = "{Stellar and substellar companions from Gaia EDR3. Proper-motion anomaly and resolved common proper-motion pairs}",
      journal = {\aap},
         year = 2022,
        month = jan,
       volume = {657},
          eid = {A7},
        pages = {A7},
          doi = {10.1051/0004-6361/202142146},
archivePrefix = {arXiv},
       eprint = {2109.10912},
 primaryClass = {astro-ph.SR},
       adsurl = {https://ui.adsabs.harvard.edu/abs/2022A&A...657A...7K}
}

@ARTICLE{Kervella1,
       author = {{Kervella}, Pierre and {Arenou}, Fr{\'e}d{\'e}ric and {Mignard}, Fran{\c{c}}ois and {Th{\'e}venin}, Fr{\'e}d{\'e}ric},
        title = "{Stellar and substellar companions of nearby stars from Gaia DR2. Binarity from proper motion anomaly}",
      journal = {\aap},
         year = 2019,
        month = mar,
       volume = {623},
          eid = {A72},
        pages = {A72},
          doi = {10.1051/0004-6361/201834371},
archivePrefix = {arXiv},
       eprint = {1811.08902},
 primaryClass = {astro-ph.SR},
       adsurl = {https://ui.adsabs.harvard.edu/abs/2019A&A...623A..72K}
}

@ARTICLE{Kirchschlager,
       author = {{Kirchschlager}, F. and {Wolf}, S.},
        title = "{Porous dust grains in debris disks}",
      journal = {\aap},
         year = 2013,
        month = apr,
       volume = {552},
          eid = {A54},
        pages = {A54},
          doi = {10.1051/0004-6361/201220486},
archivePrefix = {arXiv},
       eprint = {1302.5275},
 primaryClass = {astro-ph.SR},
       adsurl = {https://ui.adsabs.harvard.edu/abs/2013A&A...552A..54K}
}

@article{Kirsh,
	title = {Simulations of planet migration driven by planetesimal scattering},
	volume = {199},
	issn = {0019-1035},
	url = {https://ui.adsabs.harvard.edu/abs/2009Icar..199..197K},
	doi = {10.1016/j.icarus.2008.05.028},
	urldate = {2024-03-30},
	journal = {Icarus},
	author = {Kirsh, David R. and Duncan, Martin and Brasser, Ramon and Levison, Harold F.},
	month = jan,
	year = {2009},
	note = {ADS Bibcode: 2009Icar..199..197K},
	pages = {197--209},
}

@ARTICLE{Krivov1,
       author = {{Krivov}, Alexander V. and {Booth}, Mark},
        title = "{Self-stirring of debris discs by planetesimals formed by pebble concentration}",
      journal = {\mnras},
         year = 2018,
        month = sep,
       volume = {479},
       number = {3},
        pages = {3300-3307},
          doi = {10.1093/mnras/sty1607},
archivePrefix = {arXiv},
       eprint = {1806.05431},
 primaryClass = {astro-ph.EP},
       adsurl = {https://ui.adsabs.harvard.edu/abs/2018MNRAS.479.3300K}
}

@ARTICLE{Lafreniere,
       author = {{Lafreni{\`e}re}, David and {Marois}, Christian and {Doyon}, Ren{\'e} and {Barman}, Travis},
        title = "{HST/NICMOS Detection of HR 8799 b in 1998}",
      journal = {\apjl},
         year = 2009,
        month = apr,
       volume = {694},
       number = {2},
        pages = {L148-L152},
          doi = {10.1088/0004-637X/694/2/L148},
archivePrefix = {arXiv},
       eprint = {0902.3247},
 primaryClass = {astro-ph.EP},
       adsurl = {https://ui.adsabs.harvard.edu/abs/2009ApJ...694L.148L}
}

@ARTICLE{Lagrange6,
       author = {{Lagrange}, A.-M. and {Wilkinson}, C. and {M{\^a}lin}, M. and {Boccaletti}, A. and {Perrot}, C. and {Matr{\`a}}, L. and {Combes}, F. and {Beust}, H. and {Rouan}, D. and {Chomez}, A. and {Milli}, J. and {Charnay}, B. and {Mazevet}, S. and {Flasseur}, O. and {Olofsson}, J. and {Bayo}, A. and {Kral}, Q. and {Carter}, A. and {Crotts}, K.~A. and {Delorme}, P. and {Chauvin}, G. and {Thebault}, P. and {Rubini}, P. and {Kiefer}, F. and {Radcliffe}, A. and {Mazoyer}, J. and {Bodrito}, T. and {Stasevic}, S. and {Langlois}, M.},
        title = "{Evidence for a sub-Jovian planet in the young TWA 7 disk}",
      journal = {\nat},
         year = 2025,
        month = jun,
       volume = {642},
       number = {8069},
        pages = {905-908},
          doi = {10.1038/s41586-025-09150-4},
archivePrefix = {arXiv},
       eprint = {2502.15081},
 primaryClass = {astro-ph.EP},
       adsurl = {https://ui.adsabs.harvard.edu/abs/2025Natur.642..905L}
}

@ARTICLE{Lawson,
       author = {{Lawson}, Kellen and {Currie}, Thayne and {Wisniewski}, John P. and {Groff}, Tyler D. and {McElwain}, Michael W. and {Schlieder}, Joshua E.},
        title = "{Constrained Reference Star Differential Imaging: Enabling High-fidelity Imagery of Highly Structured Circumstellar Disks}",
      journal = {\apjl},
         year = 2022,
        month = aug,
       volume = {935},
       number = {2},
          eid = {L25},
        pages = {L25},
          doi = {10.3847/2041-8213/ac853b},
archivePrefix = {arXiv},
       eprint = {2208.01606},
 primaryClass = {astro-ph.EP},
       adsurl = {https://ui.adsabs.harvard.edu/abs/2022ApJ...935L..25L}
}

@ARTICLE{Lawson1,
       author = {{Lawson}, Kellen and {Schlieder}, Joshua E. and {Leisenring}, Jarron M. and {Bogat}, Ell and {Beichman}, Charles A. and {Bryden}, Geoffrey and {G{\'a}sp{\'a}r}, Andr{\'a}s and {Groff}, Tyler D. and {McElwain}, Michael W. and {Meyer}, Michael R. and {Barclay}, Thomas and {Calissendorff}, Per and {De Furio}, Matthew and {Ygouf}, Marie and {Boccaletti}, Anthony and {Greene}, Thomas P. and {Krist}, John and {Plavchan}, Peter and {Rieke}, Marcia J. and {Roellig}, Thomas L. and {Stansberry}, John and {Wisniewski}, John P. and {Young}, Erick T.},
        title = "{JWST/NIRCam Coronagraphy of the Young Planet-hosting Debris Disk AU Microscopii}",
      journal = {\aj},
         year = 2023,
        month = oct,
       volume = {166},
       number = {4},
          eid = {150},
        pages = {150},
          doi = {10.3847/1538-3881/aced08},
archivePrefix = {arXiv},
       eprint = {2308.02486},
 primaryClass = {astro-ph.EP},
       adsurl = {https://ui.adsabs.harvard.edu/abs/2023AJ....166..150L}
}

@ARTICLE{Lawson2,
       author = {{Lawson}, Kellen and {Schlieder}, Joshua E. and {Leisenring}, Jarron M. and {Bogat}, Ell and {Beichman}, Charles A. and {Bryden}, Geoffrey and {G{\'a}sp{\'a}r}, Andr{\'a}s and {Groff}, Tyler D. and {McElwain}, Michael W. and {Meyer}, Michael R. and {Barclay}, Thomas and {Calissendorff}, Per and {De Furio}, Matthew and {Li}, Yiting and {Rieke}, Marcia J. and {Ygouf}, Marie and {Greene}, Thomas P. and {Girard}, Julien H. and {Gennaro}, Mario and {Kammerer}, Jens and {Rest}, Armin and {Roellig}, Thomas L. and {Sunnquist}, Ben},
        title = "{JWST/NIRCam Detection of the Fomalhaut C Debris Disk in Scattered Light}",
      journal = {\apjl},
         year = 2024,
        month = may,
       volume = {967},
       number = {1},
          eid = {L8},
        pages = {L8},
          doi = {10.3847/2041-8213/ad4496},
archivePrefix = {arXiv},
       eprint = {2405.00573},
 primaryClass = {astro-ph.EP},
       adsurl = {https://ui.adsabs.harvard.edu/abs/2024ApJ...967L...8L}
}

@ARTICLE{Linder,
       author = {{Linder}, Esther F. and {Mordasini}, Christoph and {Molli{\`e}re}, Paul and {Marleau}, Gabriel-Dominique and {Malik}, Matej and {Quanz}, Sascha P. and {Meyer}, Michael R.},
        title = "{Evolutionary models of cold and low-mass planets: cooling curves, magnitudes, and detectability}",
      journal = {\aap},
         year = 2019,
        month = mar,
       volume = {623},
          eid = {A85},
        pages = {A85},
          doi = {10.1051/0004-6361/201833873},
archivePrefix = {arXiv},
       eprint = {1812.02027},
 primaryClass = {astro-ph.EP},
       adsurl = {https://ui.adsabs.harvard.edu/abs/2019A&A...623A..85L}
}

@ARTICLE{Marino1,
       author = {{Marino}, S. and {Zurlo}, A. and {Faramaz}, V. and {Milli}, J. and
         {Henning}, Th and {Kennedy}, G.~M. and {Matr{\`a}}, L. and
         {P{\'e}rez}, S. and {Delorme}, P. and {Cieza}, L.~A. and
         {Hughes}, A.~M.},
        title = "{Insights into the planetary dynamics of HD 206893 with ALMA}",
      journal = {\mnras},
         year = 2020,
        month = aug,
       volume = {498},
       number = {1},
        pages = {1319-1334},
          doi = {10.1093/mnras/staa2386},
       adsurl = {https://ui.adsabs.harvard.edu/abs/2020MNRAS.498.1319M}
}

@ARTICLE{Marino2,
       author = {{Marino}, S. and {Yelverton}, B. and {Booth}, M. and {Faramaz}, V. and {Kennedy}, G.~M. and {Matr{\`a}}, L. and {Wyatt}, M.~C.},
        title = "{A gap in HD 92945's broad planetesimal disc revealed by ALMA}",
      journal = {\mnras},
         year = 2019,
        month = mar,
       volume = {484},
       number = {1},
        pages = {1257-1269},
          doi = {10.1093/mnras/stz049},
archivePrefix = {arXiv},
       eprint = {1901.01406},
 primaryClass = {astro-ph.EP},
       adsurl = {https://ui.adsabs.harvard.edu/abs/2019MNRAS.484.1257M}
}

@ARTICLE{Marino3,
       author = {{Marino}, Sebastian},
        title = "{Constraining planetesimal stirring: how sharp are debris disc edges?}",
      journal = {\mnras},
         year = 2021,
        month = may,
       volume = {503},
       number = {4},
        pages = {5100-5114},
          doi = {10.1093/mnras/stab771},
archivePrefix = {arXiv},
       eprint = {2104.02072},
 primaryClass = {astro-ph.EP},
       adsurl = {https://ui.adsabs.harvard.edu/abs/2021MNRAS.503.5100M}
}

@ARTICLE{Marino4,
       author = {{Marino}, S. and {Carpenter}, J. and {Wyatt}, M.~C. and {Booth}, M. and {Casassus}, S. and {Faramaz}, V. and {Guzman}, V. and {Hughes}, A.~M. and {Isella}, A. and {Kennedy}, G.~M. and {Matr{\`a}}, L. and {Ricci}, L. and {Corder}, S.},
        title = "{A gap in the planetesimal disc around HD 107146 and asymmetric warm dust emission revealed by ALMA}",
      journal = {\mnras},
         year = 2018,
        month = oct,
       volume = {479},
       number = {4},
        pages = {5423-5439},
          doi = {10.1093/mnras/sty1790},
archivePrefix = {arXiv},
       eprint = {1805.01915},
 primaryClass = {astro-ph.EP},
       adsurl = {https://ui.adsabs.harvard.edu/abs/2018MNRAS.479.5423M}
}

@ARTICLE{Marois5,
       author = {{Marois}, Christian and {Lafreni{\`e}re}, David and {Doyon}, Ren{\'e} and
         {Macintosh}, Bruce and {Nadeau}, Daniel},
        title = "{Angular Differential Imaging: A Powerful High-Contrast Imaging Technique}",
      journal = {ApJ},
         year = "2006",
        month = "Apr",
       volume = {641},
       number = {1},
        pages = {556-564},
          doi = {10.1086/500401},
archivePrefix = {arXiv},
       eprint = {astro-ph/0512335},
 primaryClass = {astro-ph},
       adsurl = {https://ui.adsabs.harvard.edu/abs/2006ApJ...641..556M}
}

@ARTICLE{Matra,
       author = {{Matr{\`a}}, L. and {Marino}, S. and {Wilner}, D.~J. and {Kennedy}, G.~M. and {Booth}, M. and {Krivov}, A.~V. and {Williams}, J.~P. and {Hughes}, A.~M. and {del Burgo}, C. and {Carpenter}, J. and {Davies}, C.~L. and {Ertel}, S. and {Kral}, Q. and {Lestrade}, J. -F. and {Marshall}, J.~P. and {Milli}, J. and {{\"O}berg}, K.~I. and {Pawellek}, N. and {Sepulveda}, A.~G. and {Wyatt}, M.~C. and {Matthews}, B.~C. and {MacGregor}, M.},
        title = "{REsolved ALMA and SMA Observations of Nearby Stars (REASONS): A population of 74 resolved planetesimal belts at millimetre wavelengths}",
      journal = {\aap},
         year = 2025,
        month = jan,
       volume = {693},
          eid = {A151},
        pages = {A151},
          doi = {10.1051/0004-6361/202451397},
archivePrefix = {arXiv},
       eprint = {2501.09058},
 primaryClass = {astro-ph.EP},
       adsurl = {https://ui.adsabs.harvard.edu/abs/2025A&A...693A.151M}
}

@ARTICLE{Matthews2,
       author = {{Matthews}, E.~C. and {Carter}, A.~L. and {Pathak}, P. and {Morley}, C.~V. and {Phillips}, M.~W. and {P.~M.}, S. Krishanth and {Feng}, F. and {Bonse}, M.~J. and {Boogaard}, L.~A. and {Burt}, J.~A. and {Crossfield}, I.~J.~M. and {Douglas}, E.~S. and {Henning}, Th. and {Hom}, J. and {Ko}, C. -L. and {Kasper}, M. and {Lagrange}, A. -M. and {Petit dit de la Roche}, D. and {Philipot}, F.},
        title = "{A temperate super-Jupiter imaged with JWST in the mid-infrared}",
      journal = {\nat},
         year = 2024,
        month = sep,
       volume = {633},
       number = {8031},
        pages = {789-792},
          doi = {10.1038/s41586-024-07837-8},
       adsurl = {https://ui.adsabs.harvard.edu/abs/2024Natur.633..789M}
}

@ARTICLE{Mawet,
       author = {{Mawet}, D. and {Milli}, J. and {Wahhaj}, Z. and {Pelat}, D. and
         {Absil}, O. and {Delacroix}, C. and {Boccaletti}, A. and {Kasper}, M. and
         {Kenworthy}, M. and {Marois}, C. and {Mennesson}, B. and {Pueyo}, L.},
        title = "{Fundamental Limitations of High Contrast Imaging Set by Small Sample Statistics}",
      journal = {\apj},
         year = 2014,
        month = sep,
       volume = {792},
       number = {2},
          eid = {97},
        pages = {97},
          doi = {10.1088/0004-637X/792/2/97},
archivePrefix = {arXiv},
       eprint = {1407.2247},
 primaryClass = {astro-ph.IM},
       adsurl = {https://ui.adsabs.harvard.edu/abs/2014ApJ...792...97M}
}

@ARTICLE{Mesa5,
       author = {{Mesa}, D. and {Marino}, S. and {Bonavita}, M. and {Lazzoni}, C. and {Fontanive}, C. and {P{\'e}rez}, S. and {D'Orazi}, V. and {Desidera}, S. and {Gratton}, R. and {Engler}, N. and {Henning}, T. and {Janson}, M. and {Kral}, Q. and {Langlois}, M. and {Messina}, S. and {Milli}, J. and {Pawellek}, N. and {Perrot}, C. and {Rigliaco}, E. and {Rickman}, E. and {Squicciarini}, V. and {Vigan}, A. and {Wahhaj}, Z. and {Zurlo}, A. and {Boccaletti}, A. and {Bonnefoy}, M. and {Chauvin}, G. and {De Caprio}, V. and {Feldt}, M. and {Gluck}, L. and {Hagelberg}, J. and {Keppler}, M. and {Lagrange}, A. -M. and {Launhardt}, R. and {Maire}, A. -L. and {Meyer}, M. and {Moeller-Nilsson}, O. and {Pavlov}, A. and {Samland}, M. and {Schmidt}, T. and {Weber}, L.},
        title = "{Limits on the presence of planets in systems with debris discs: HD 92945 and HD 107146}",
      journal = {\mnras},
         year = 2021,
        month = may,
       volume = {503},
       number = {1},
        pages = {1276-1289},
          doi = {10.1093/mnras/stab438},
archivePrefix = {arXiv},
       eprint = {2102.05353},
 primaryClass = {astro-ph.EP},
       adsurl = {https://ui.adsabs.harvard.edu/abs/2021MNRAS.503.1276M}
}

@ARTICLE{Meshkat,
   author = {{Meshkat}, T. and {Bailey}, V.~P. and {Su}, K.~Y.~L. and {Kenworthy}, M.~A. and 
	{Mamajek}, E.~E. and {Hinz}, P.~M. and {Smith}, P.~S.},
    title = "{Searching for Planets in Holey Debris Disks with the Apodizing Phase Plate}",
  journal = {ApJ},
archivePrefix = "arXiv",
   eprint = {1412.5179},
 primaryClass = "astro-ph.EP",
     year = 2015,
    month = feb,
   volume = 800,
      eid = {5},
    pages = {5},
      doi = {10.1088/0004-637X/800/1/5},
   adsurl = {http://adsabs.harvard.edu/abs/2015ApJ...800....5M}
}

@ARTICLE{Meshkat1,
       author = {{Meshkat}, Tiffany and {Mawet}, Dimitri and {Bryan}, Marta L. and {Hinkley}, Sasha and {Bowler}, Brendan P. and {Stapelfeldt}, Karl R. and {Batygin}, Konstantin and {Padgett}, Deborah and {Morales}, Farisa Y. and {Serabyn}, Eugene and {Christiaens}, Valentin and {Brandt}, Timothy D. and {Wahhaj}, Zahed},
        title = "{A Direct Imaging Survey of Spitzer-detected Debris Disks: Occurrence of Giant Planets in Dusty Systems}",
      journal = {\aj},
         year = 2017,
        month = dec,
       volume = {154},
       number = {6},
          eid = {245},
        pages = {245},
          doi = {10.3847/1538-3881/aa8e9a},
archivePrefix = {arXiv},
       eprint = {1710.04185},
 primaryClass = {astro-ph.EP},
       adsurl = {https://ui.adsabs.harvard.edu/abs/2017AJ....154..245M}
}

@article{Mustill,
	Author = {Mustill, Alexander J. and Wyatt, Mark C.},
	Doi = {10.1111/j.1365-2966.2009.15360.x},
	Eprint = {/oup/backfile/Content_public/Journal/mnras/399/3/10.1111/j.1365-2966.2009.15360.x/3/mnras0399-1403.pdf},
	Journal = {MNRAS},
	Number = {3},
	Pages = {1403},
	Title = {Debris disc stirring by secular perturbations from giant planets},
	Url = {+ http://dx.doi.org/10.1111/j.1365-2966.2009.15360.x},
	Volume = {399},
	Year = {2009}}

@INPROCEEDINGS{Nielsen1,
       author = {{Nielsen}, E.~L. and {De Rosa}, R. and {Macintosh}, B. and {Wang}, J. and
         {Ruffio}, J. and {Chiang}, E. and {Marley}, M. and {Saumon}, D. and
         {Savransky}, D. and {Gemini Planet Imager Exoplanet Survey Team}},
        title = "{The Gemini Planet Imager Exoplanet Survey: Giant Planet and Brown Dwarf Demographics from 10-100 AU}",
    booktitle = {American Astronomical Society Meeting Abstracts \#235},
         year = 2020,
       series = {American Astronomical Society Meeting Abstracts},
       volume = {235},
        month = jan,
          eid = {280.02},
        pages = {280.02},
       adsurl = {https://ui.adsabs.harvard.edu/abs/2020AAS...23528002N}
}

@ARTICLE{Nielsen2,
       author = {{Nielsen}, Eric L. and {De Rosa}, Robert J. and {Macintosh}, Bruce and {Wang}, Jason J. and {Ruffio}, Jean-Baptiste and {Chiang}, Eugene and {Marley}, Mark S. and {Saumon}, Didier and {Savransky}, Dmitry and {Ammons}, S. Mark and {Bailey}, Vanessa P. and {Barman}, Travis and {Blain}, C{\'e}lia and {Bulger}, Joanna and {Burrows}, Adam and {Chilcote}, Jeffrey and {Cotten}, Tara and {Czekala}, Ian and {Doyon}, Rene and {Duch{\^e}ne}, Gaspard and {Esposito}, Thomas M. and {Fabrycky}, Daniel and {Fitzgerald}, Michael P. and {Follette}, Katherine B. and {Fortney}, Jonathan J. and {Gerard}, Benjamin L. and {Goodsell}, Stephen J. and {Graham}, James R. and {Greenbaum}, Alexandra Z. and {Hibon}, Pascale and {Hinkley}, Sasha and {Hirsch}, Lea A. and {Hom}, Justin and {Hung}, Li-Wei and {Dawson}, Rebekah Ilene and {Ingraham}, Patrick and {Kalas}, Paul and {Konopacky}, Quinn and {Larkin}, James E. and {Lee}, Eve J. and {Lin}, Jonathan W. and {Maire}, J{\'e}r{\^o}me and {Marchis}, Franck and {Marois}, Christian and {Metchev}, Stanimir and {Millar-Blanchaer}, Maxwell A. and {Morzinski}, Katie M. and {Oppenheimer}, Rebecca and {Palmer}, David and {Patience}, Jennifer and {Perrin}, Marshall and {Poyneer}, Lisa and {Pueyo}, Laurent and {Rafikov}, Roman R. and {Rajan}, Abhijith and {Rameau}, Julien and {Rantakyr{\"o}}, Fredrik T. and {Ren}, Bin and {Schneider}, Adam C. and {Sivaramakrishnan}, Anand and {Song}, Inseok and {Soummer}, Remi and {Tallis}, Melisa and {Thomas}, Sandrine and {Ward-Duong}, Kimberly and {Wolff}, Schuyler},
        title = "{The Gemini Planet Imager Exoplanet Survey: Giant Planet and Brown Dwarf Demographics from 10 to 100 au}",
      journal = {\aj},
         year = 2019,
        month = jul,
       volume = {158},
       number = {1},
          eid = {13},
        pages = {13},
          doi = {10.3847/1538-3881/ab16e9},
archivePrefix = {arXiv},
       eprint = {1904.05358},
 primaryClass = {astro-ph.EP},
       adsurl = {https://ui.adsabs.harvard.edu/abs/2019AJ....158...13N}
}

@article{Pearce,
	Author = {Pearce, Tim D. and Wyatt, Mark C.},
	Doi = {10.1093/mnras/stu1302},
	Eprint = {/oup/backfile/Content_public/Journal/mnras/443/3/10.1093/mnras/stu1302/2/stu1302.pdf},
	Journal = {Monthly Notices of the Royal Astronomical Society},
	Number = {3},
	Pages = {2541},
	Title = {Dynamical evolution of an eccentric planet and a less massive debris disc},
	Url = {+ http://dx.doi.org/10.1093/mnras/stu1302},
	Volume = {443},
	Year = {2014}}

@ARTICLE{Pearce1,
       author = {{Pearce}, Tim D. and {Krivov}, Alexander V. and {Sefilian}, Antranik A. and {Jankovic}, Marija R. and {L{\"o}hne}, Torsten and {Morgner}, Tobias and {Wyatt}, Mark C. and {Booth}, Mark and {Marino}, Sebastian},
        title = "{The effect of sculpting planets on the steepness of debris-disc inner edges}",
      journal = {\mnras},
         year = 2024,
        month = jan,
       volume = {527},
       number = {2},
        pages = {3876-3899},
          doi = {10.1093/mnras/stad3462},
archivePrefix = {arXiv},
       eprint = {2311.04265},
 primaryClass = {astro-ph.EP},
       adsurl = {https://ui.adsabs.harvard.edu/abs/2024MNRAS.527.3876P}
}

@article{pearce2022,
	title = {Planet populations inferred from debris discs. {Insights} from 178 debris systems in the {ISPY}, {LEECH}, and {LIStEN} planet-hunting surveys},
	volume = {659},
	issn = {0004-6361},
	url = {https://ui.adsabs.harvard.edu/abs/2022A&A...659A.135P},
	doi = {10.1051/0004-6361/202142720},
	urldate = {2023-11-28},
	journal = {Astronomy and Astrophysics},
	author = {Pearce, Tim D. and Launhardt, Ralf and Ostermann, Robert and Kennedy, Grant M. and Gennaro, Mario and Booth, Mark and Krivov, Alexander V. and Cugno, Gabriele and Henning, Thomas K. and Quirrenbach, Andreas and Barcucci, Arianna Musso and Matthews, Elisabeth C. and Ruh, Henrik L. and Stone, Jordan M.},
	month = mar,
	year = {2022},
	note = {ADS Bibcode: 2022A\&A...659A.135P},
	pages = {A135},
}

@article{Limbach2024,
  title = {The {{MIRI Exoplanets Orbiting White}} Dwarfs ({{MEOW}}) {{Survey}}: {{Mid-infrared Excess Reveals}} a {{Giant Planet Candidate}} around a {{Nearby White Dwarf}}},
  shorttitle = {The {{MIRI Exoplanets Orbiting White}} Dwarfs ({{MEOW}}) {{Survey}}},
  author = {Limbach, Mary Anne and Vanderburg, Andrew and Venner, Alexander and Blouin, Simon and Stevenson, Kevin B. and MacDonald, Ryan J. and Jenkins, Sydney and {Bowens-Rubin}, Rachel and {Soares-Furtado}, Melinda and Morley, Caroline and Janson, Markus and Debes, John and Xu, Siyi and Kleisioti, Evangelia and Kenworthy, Matthew and Butler, Paul and Crane, Jeffrey D. and Osip, Dave and Shectman, Stephen and Teske, Johanna},
  year = {2024},
  month = sep,
  journal = {The Astrophysical Journal},
  volume = {973},
  pages = {L11},
  publisher = {IOP},
  issn = {0004-637X},
  doi = {10.3847/2041-8213/ad74ed},
  urldate = {2025-03-17}
}

@ARTICLE{Kiefer2024,
       author = {{Kiefer}, F. and {Lagrange}, A.-M. and {Rubini}, P. and {Philipot}, F.},
        title = "{Searching for substellar companion candidates with Gaia: II. A catalog of 9698 planet candidate solar-type hosts}",
      journal = {\aap},
         year = 2025,
        month = oct,
       volume = {702},
          eid = {A77},
        pages = {A77},
          doi = {10.1051/0004-6361/202451745},
archivePrefix = {arXiv},
       eprint = {2409.16993},
 primaryClass = {astro-ph.EP},
       adsurl = {https://ui.adsabs.harvard.edu/abs/2025A&A...702A..77K}
}

@ARTICLE{Phillips,
       author = {{Phillips}, M.~W. and {Tremblin}, P. and {Baraffe}, I. and {Chabrier}, G. and {Allard}, N.~F. and {Spiegelman}, F. and {Goyal}, J.~M. and {Drummond}, B. and {H{\'e}brard}, E.},
        title = "{A new set of atmosphere and evolution models for cool T-Y brown dwarfs and giant exoplanets}",
      journal = {\aap},
         year = 2020,
        month = may,
       volume = {637},
          eid = {A38},
        pages = {A38},
          doi = {10.1051/0004-6361/201937381},
archivePrefix = {arXiv},
       eprint = {2003.13717},
 primaryClass = {astro-ph.SR},
       adsurl = {https://ui.adsabs.harvard.edu/abs/2020A&A...637A..38P}
}

@ARTICLE{Plavchan,
       author = {{Plavchan}, Peter and {Werner}, M.~W. and {Chen}, C.~H. and {Stapelfeldt}, K.~R. and {Su}, K.~Y.~L. and {Stauffer}, J.~R. and {Song}, I.},
        title = "{New Debris Disks Around Young, Low-Mass Stars Discovered with the Spitzer Space Telescope}",
      journal = {\apj},
         year = 2009,
        month = jun,
       volume = {698},
       number = {2},
        pages = {1068-1094},
          doi = {10.1088/0004-637X/698/2/1068},
archivePrefix = {arXiv},
       eprint = {0904.0819},
 primaryClass = {astro-ph.SR},
       adsurl = {https://ui.adsabs.harvard.edu/abs/2009ApJ...698.1068P}
}

@ARTICLE{Rebollido,
       author = {{Rebollido}, Isabel and {Stark}, Christopher C. and {Kammerer}, Jens and {Perrin}, Marshall D. and {Lawson}, Kellen and {Pueyo}, Laurent and {Chen}, Christine and {Hines}, Dean and {Girard}, Julien H. and {Worthen}, Kadin and {Ingerbretsen}, Carl and {Betti}, Sarah and {Clampin}, Mark and {Golimowski}, David and {Hoch}, Kielan and {Lewis}, Nikole K. and {Lu}, Cicero X. and {van der Marel}, Roeland P. and {Rickman}, Emily and {Seager}, Sara and {Soummer}, R{\'e}mi and {Valenti}, Jeff A. and {Ward-Duong}, Kimberly and {Mountain}, C. Matt},
        title = "{JWST-TST High Contrast: Asymmetries, Dust Populations, and Hints of a Collision in the {\ensuremath{\beta}} Pictoris Disk with NIRCam and MIRI}",
      journal = {\aj},
         year = 2024,
        month = feb,
       volume = {167},
       number = {2},
          eid = {69},
        pages = {69},
          doi = {10.3847/1538-3881/ad1759},
archivePrefix = {arXiv},
       eprint = {2401.05271},
 primaryClass = {astro-ph.EP},
       adsurl = {https://ui.adsabs.harvard.edu/abs/2024AJ....167...69R}
}

@article{Schneider1,
	Author = {Glenn Schneider and Carol A. Grady and Dean C. Hines and Christopher C. Stark and John H. Debes and Joe Carson and Marc J. Kuchner and Marshall D. Perrin and Alycia J. Weinberger and John P. Wisniewski and Murray D. Silverstone and Hannah Jang-Condell and Thomas Henning and Bruce E. Woodgate and Eugene Serabyn and Amaya Moro-Martin and Motohide Tamura and Phillip M. Hinz and Timothy J. Rodigas},
	Journal = {ApJ},
	Number = {4},
	Pages = {59},
	Title = {Probing for Exoplanets Hiding in Dusty Debris Disks: Disk Imaging, Characterization, and Exploration with HST/STIS Multi-roll Coronagraphy},
	Url = {http://stacks.iop.org/1538-3881/148/i=4/a=59},
	Volume = {148},
	Year = {2014}}

@ARTICLE{Sefilian,
       author = {{Sefilian}, Antranik A. and {Rafikov}, Roman R. and {Wyatt}, Mark C.},
        title = "{Formation of Gaps in Self-gravitating Debris Disks by Secular Resonance in a Single-planet System. II. Toward a Self-consistent Model}",
      journal = {\apj},
         year = 2023,
        month = sep,
       volume = {954},
       number = {1},
          eid = {100},
        pages = {100},
          doi = {10.3847/1538-4357/ace68e},
archivePrefix = {arXiv},
       eprint = {2305.00951},
 primaryClass = {astro-ph.EP},
       adsurl = {https://ui.adsabs.harvard.edu/abs/2023ApJ...954..100S}
}

@PHDTHESIS{Silverstone,
       author = {{Silverstone}, Murray Daniel},
        title = "{The Vega Phenomenon: Evolution and multiplicity}",
       school = {University of California, Los Angeles},
         year = 2000,
        month = aug,
       adsurl = {https://ui.adsabs.harvard.edu/abs/2000PhDT........17S}
}

@ARTICLE{Song,
       author = {{Song}, Inseok and {Zuckerman}, B. and {Bessell}, M.~S.},
        title = "{On Ca II Emission as an Indicator of the Age of Young Stars}",
      journal = {\apjl},
         year = 2004,
        month = oct,
       volume = {614},
       number = {2},
        pages = {L125-L127},
          doi = {10.1086/425683},
       adsurl = {https://ui.adsabs.harvard.edu/abs/2004ApJ...614L.125S}
}

@ARTICLE{Soummer2,
       author = {{Soummer}, R{\'e}mi and {Pueyo}, Laurent and {Larkin}, James},
        title = "{Detection and Characterization of Exoplanets and Disks Using Projections on Karhunen-Lo{\`e}ve Eigenimages}",
      journal = {ApJ},
         year = "2012",
        month = "Aug",
       volume = {755},
       number = {2},
          eid = {L28},
        pages = {L28},
          doi = {10.1088/2041-8205/755/2/L28},
archivePrefix = {arXiv},
       eprint = {1207.4197},
 primaryClass = {astro-ph.IM},
       adsurl = {https://ui.adsabs.harvard.edu/abs/2012ApJ...755L..28S}
}

@ARTICLE{Tabeshian,
       author = {{Tabeshian}, Maryam and {Wiegert}, Paul A.},
        title = "{Detection and Characterization of Extrasolar Planets through Mean-motion Resonances. I. Simulations of Hypothetical Debris Disks}",
      journal = {\apj},
         year = 2016,
        month = feb,
       volume = {818},
       number = {2},
          eid = {159},
        pages = {159},
          doi = {10.3847/0004-637X/818/2/159},
archivePrefix = {arXiv},
       eprint = {1507.02661},
 primaryClass = {astro-ph.EP},
       adsurl = {https://ui.adsabs.harvard.edu/abs/2016ApJ...818..159T}
}

@INPROCEEDINGS{Tamura,
       author = {{Tamura}, Motohide},
        title = "{SEEDS: Strategic Explorations of Exoplanets and Disks with Subaru}",
    booktitle = {Exploring the Formation and Evolution of Planetary Systems},
         year = 2014,
       editor = {{Booth}, Mark and {Matthews}, Brenda C. and {Graham}, James R.},
       series = {IAU Symposium},
       volume = {299},
        month = jan,
        pages = {12-16},
          doi = {10.1017/S1743921313007679},
       adsurl = {https://ui.adsabs.harvard.edu/abs/2014IAUS..299...12T}
}

@ARTICLE{Thebault,
       author = {{Th{\'e}bault}, P.},
        title = "{Vertical structure of debris discs}",
      journal = {\aap},
         year = 2009,
        month = oct,
       volume = {505},
       number = {3},
        pages = {1269-1276},
          doi = {10.1051/0004-6361/200912396},
archivePrefix = {arXiv},
       eprint = {0906.5524},
 primaryClass = {astro-ph.EP},
       adsurl = {https://ui.adsabs.harvard.edu/abs/2009A&A...505.1269T}
}

@ARTICLE{Yelverton,
       author = {{Yelverton}, Ben and {Kennedy}, Grant M.},
        title = "{Empty gaps? Depleting annular regions in debris discs by secular resonance with a two-planet system}",
      journal = {\mnras},
         year = 2018,
        month = sep,
       volume = {479},
       number = {2},
        pages = {2673-2691},
          doi = {10.1093/mnras/sty1678},
archivePrefix = {arXiv},
       eprint = {1806.08802},
 primaryClass = {astro-ph.EP},
       adsurl = {https://ui.adsabs.harvard.edu/abs/2018MNRAS.479.2673Y}
}

@ARTICLE{Vigan2021,
       author = {{Vigan}, A. and {Fontanive}, C. and {Meyer}, M. and {Biller}, B. and {Bonavita}, M. and {Feldt}, M. and {Desidera}, S. and {Marleau}, G. -D. and {Emsenhuber}, A. and {Galicher}, R. and {Rice}, K. and {Forgan}, D. and {Mordasini}, C. and {Gratton}, R. and {Le Coroller}, H. and {Maire}, A. -L. and {Cantalloube}, F. and {Chauvin}, G. and {Cheetham}, A. and {Hagelberg}, J. and {Lagrange}, A. -M. and {Langlois}, M. and {Bonnefoy}, M. and {Beuzit}, J. -L. and {Boccaletti}, A. and {D'Orazi}, V. and {Delorme}, P. and {Dominik}, C. and {Henning}, Th. and {Janson}, M. and {Lagadec}, E. and {Lazzoni}, C. and {Ligi}, R. and {Menard}, F. and {Mesa}, D. and {Messina}, S. and {Moutou}, C. and {M{\"u}ller}, A. and {Perrot}, C. and {Samland}, M. and {Schmid}, H.~M. and {Schmidt}, T. and {Sissa}, E. and {Turatto}, M. and {Udry}, S. and {Zurlo}, A. and {Abe}, L. and {Antichi}, J. and {Asensio-Torres}, R. and {Baruffolo}, A. and {Baudoz}, P. and {Baudrand}, J. and {Bazzon}, A. and {Blanchard}, P. and {Bohn}, A.~J. and {Brown Sevilla}, S. and {Carbillet}, M. and {Carle}, M. and {Cascone}, E. and {Charton}, J. and {Claudi}, R. and {Costille}, A. and {De Caprio}, V. and {Delboulb{\'e}}, A. and {Dohlen}, K. and {Engler}, N. and {Fantinel}, D. and {Feautrier}, P. and {Fusco}, T. and {Gigan}, P. and {Girard}, J.~H. and {Giro}, E. and {Gisler}, D. and {Gluck}, L. and {Gry}, C. and {Hubin}, N. and {Hugot}, E. and {Jaquet}, M. and {Kasper}, M. and {Le Mignant}, D. and {Llored}, M. and {Madec}, F. and {Magnard}, Y. and {Martinez}, P. and {Maurel}, D. and {M{\"o}ller-Nilsson}, O. and {Mouillet}, D. and {Moulin}, T. and {Orign{\'e}}, A. and {Pavlov}, A. and {Perret}, D. and {Petit}, C. and {Pragt}, J. and {Puget}, P. and {Rabou}, P. and {Ramos}, J. and {Rickman}, E.~L. and {Rigal}, F. and {Rochat}, S. and {Roelfsema}, R. and {Rousset}, G. and {Roux}, A. and {Salasnich}, B. and {Sauvage}, J. -F. and {Sevin}, A. and {Soenke}, C. and {Stadler}, E. and {Suarez}, M. and {Wahhaj}, Z. and {Weber}, L. and {Wildi}, F.},
        title = "{The SPHERE infrared survey for exoplanets (SHINE). III. The demographics of young giant exoplanets below 300 au with SPHERE}",
      journal = {\aap},
         year = 2021,
        month = jul,
       volume = {651},
          eid = {A72},
        pages = {A72},
          doi = {10.1051/0004-6361/202038107},
archivePrefix = {arXiv},
       eprint = {2007.06573},
 primaryClass = {astro-ph.EP},
       adsurl = {https://ui.adsabs.harvard.edu/abs/2021A&A...651A..72V}
}

\begin{appendix}
\onecolumn

\section{Corner plots}
\begin{figure}[h!]
    \centering
\includegraphics[width=\textwidth]{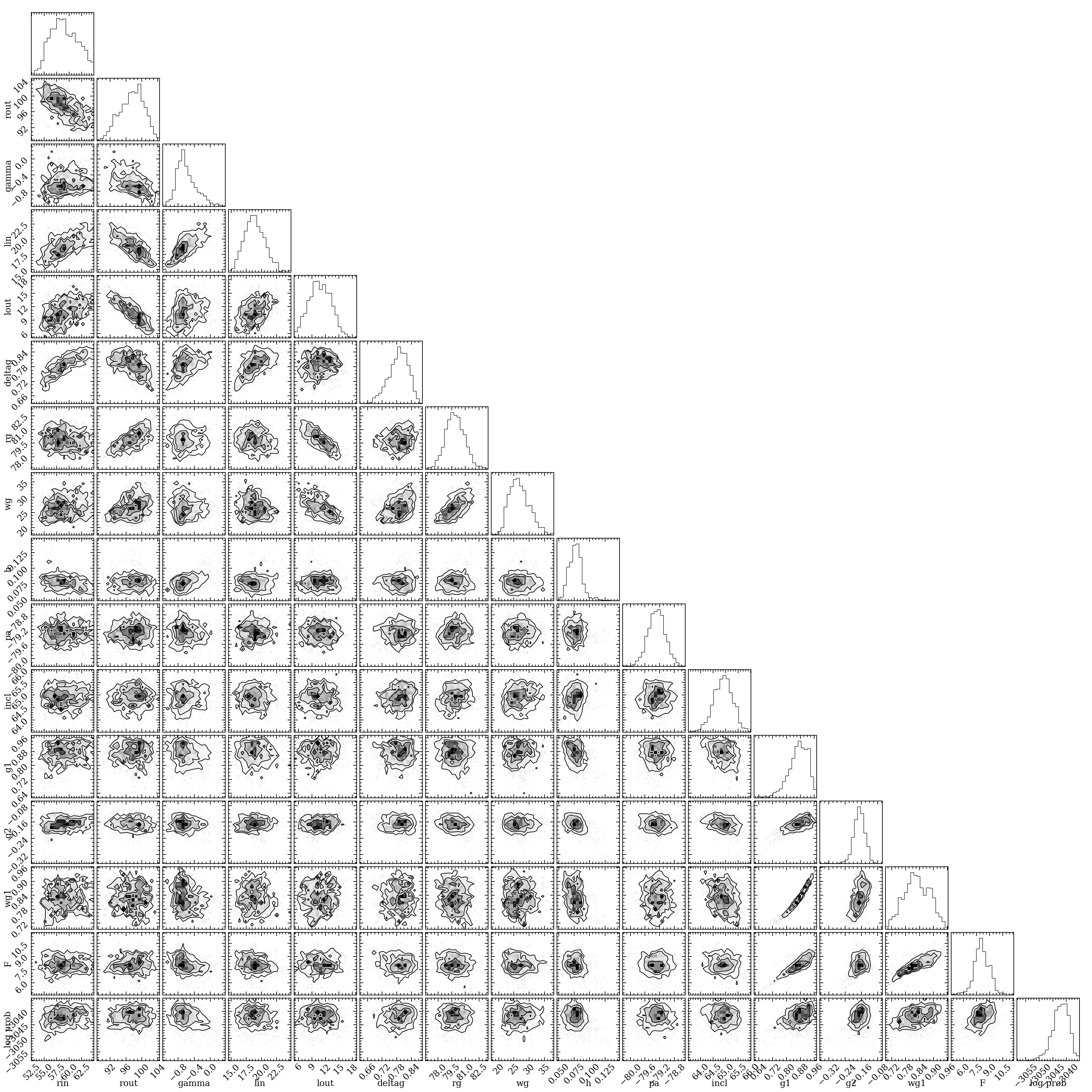} 
\caption{Corner plot of the fourteen free parameters varied in the MCMC simulation run for the F200W disk's images of HD\, 92945 where the bright area close to the minor axis is masked.}
\label{corner}
\end{figure}

In Figure \ref{corner} we show the corner plot for the fourteen free parameters of the MCMC simulation. Specifically, we show the results obtained for the disk as seen at 2 $\mu$m when using the entire PSF library and a butterfly-shaped mask to cover the bright portion arising close to the minor axis of the disk.

\FloatBarrier
\clearpage

\section{Candidates PSFs fitting}
\FloatBarrier
\begin{figure}[h!]
    \centering

    \begin{subfigure}[b]{0.49\textwidth}
        \includegraphics[width=\textwidth]{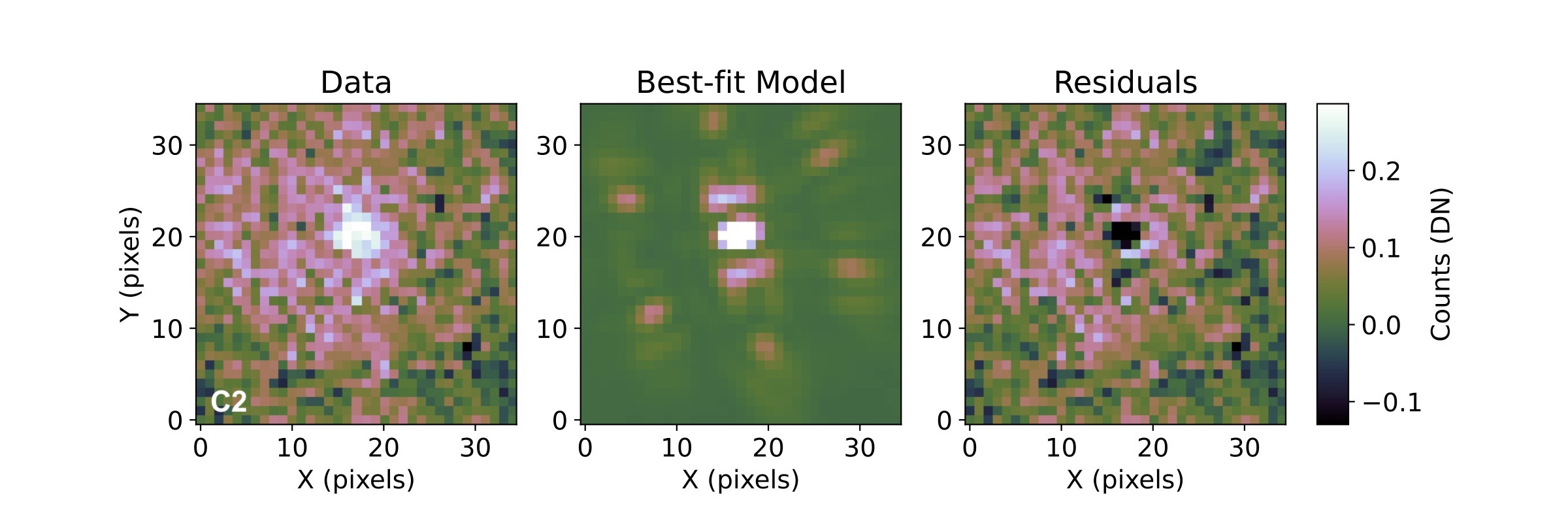}
    \end{subfigure}
    \hfill
    \begin{subfigure}[b]{0.49\textwidth}
        \includegraphics[width=\textwidth]{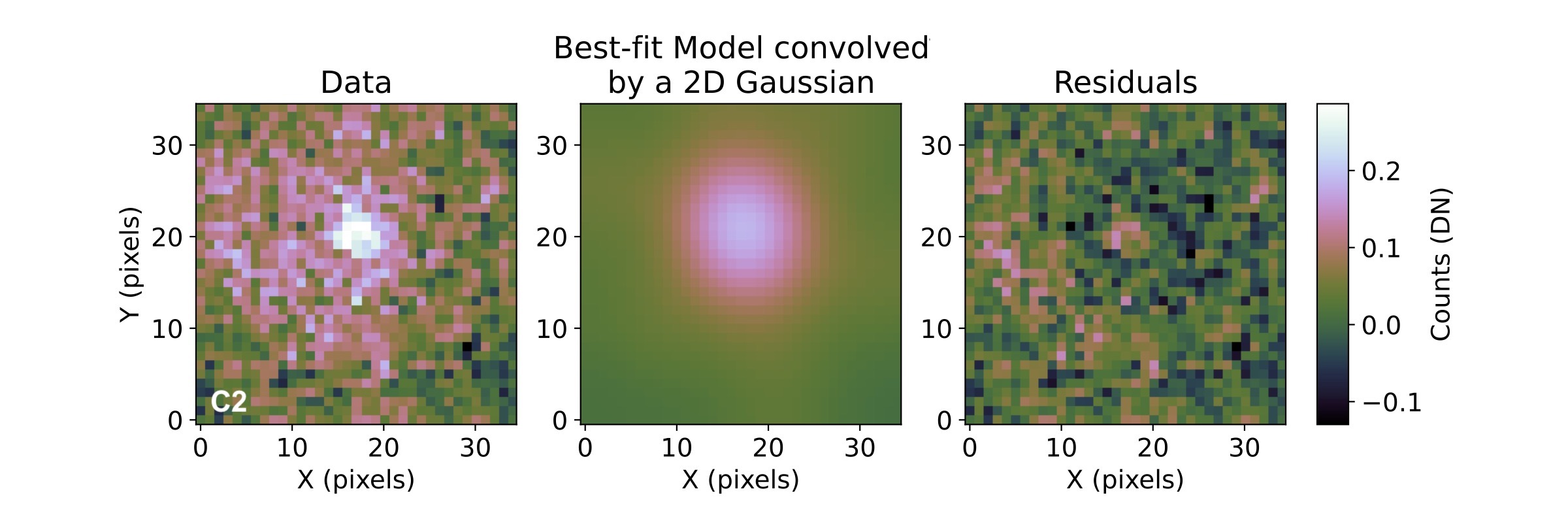}
    \end{subfigure}

    \begin{subfigure}[b]{0.49\textwidth}
        \includegraphics[width=\textwidth]{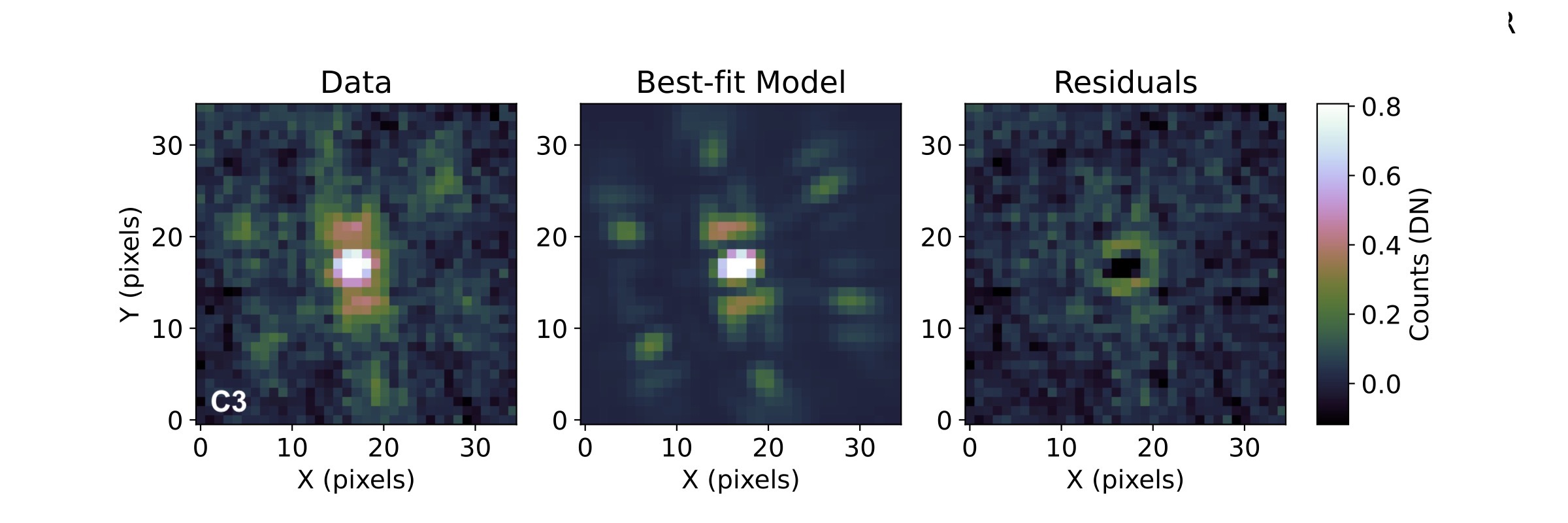}
    \end{subfigure}
    \hfill
    \begin{subfigure}[b]{0.49\textwidth}
        \includegraphics[width=\textwidth]{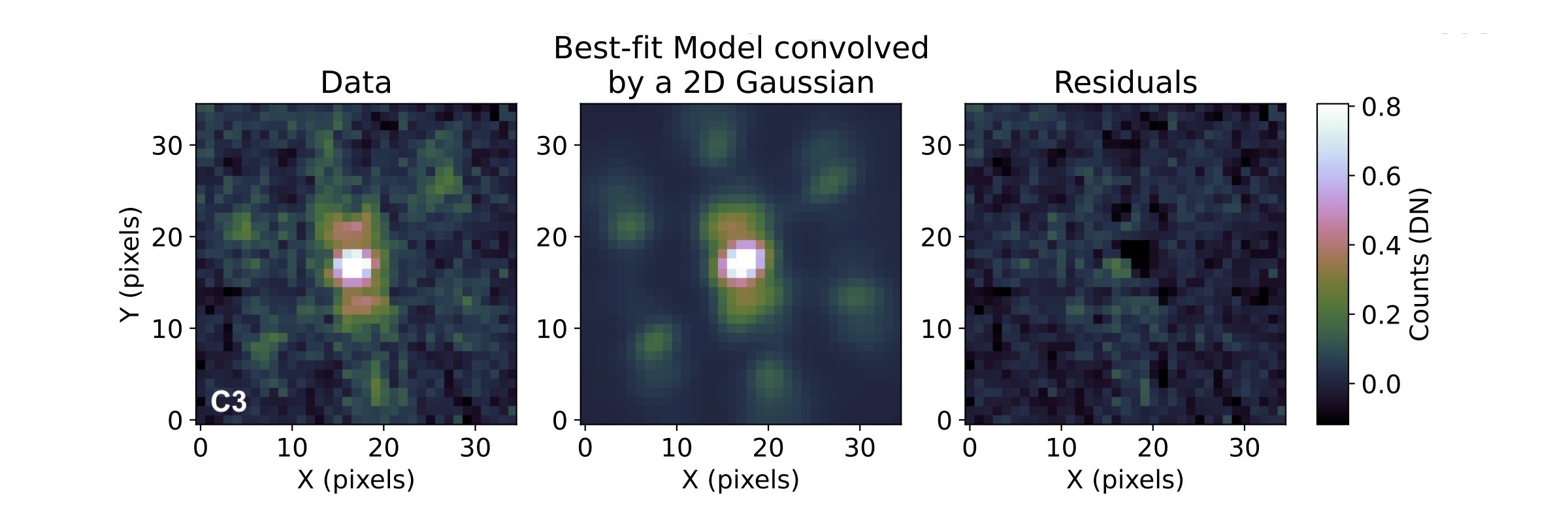}
    \end{subfigure}

    \begin{subfigure}[b]{0.49\textwidth}
        \includegraphics[width=\textwidth]{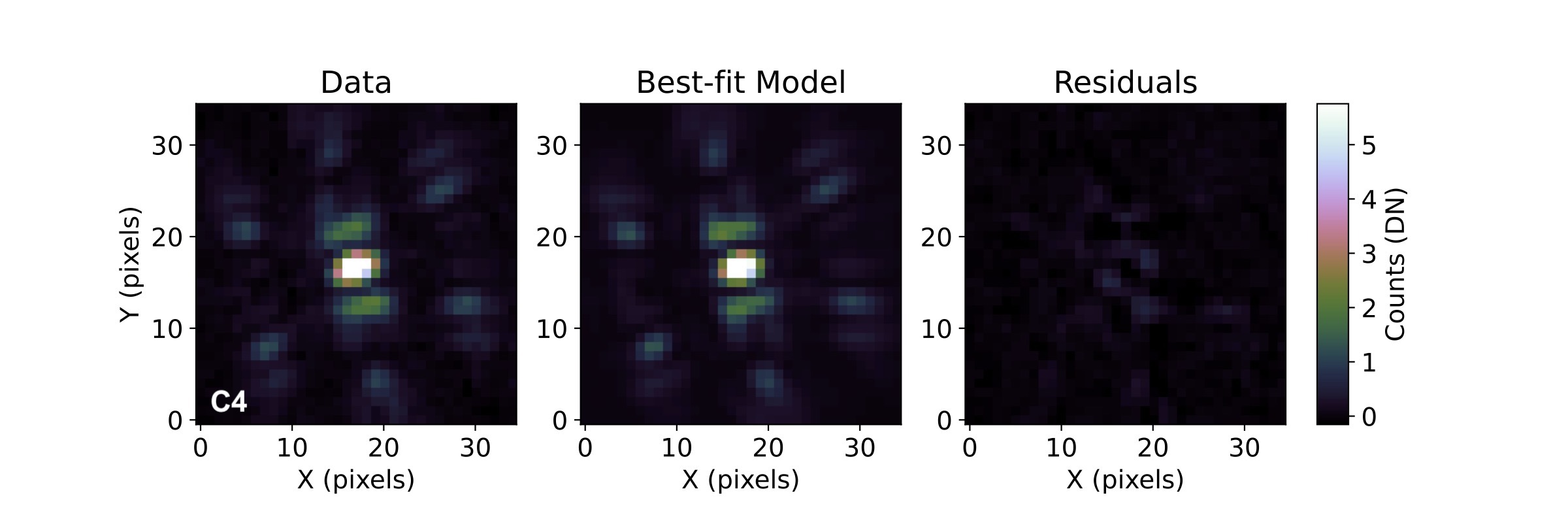}
    \end{subfigure}
    \hfill
    \begin{subfigure}[b]{0.49\textwidth}
        \includegraphics[width=\textwidth]{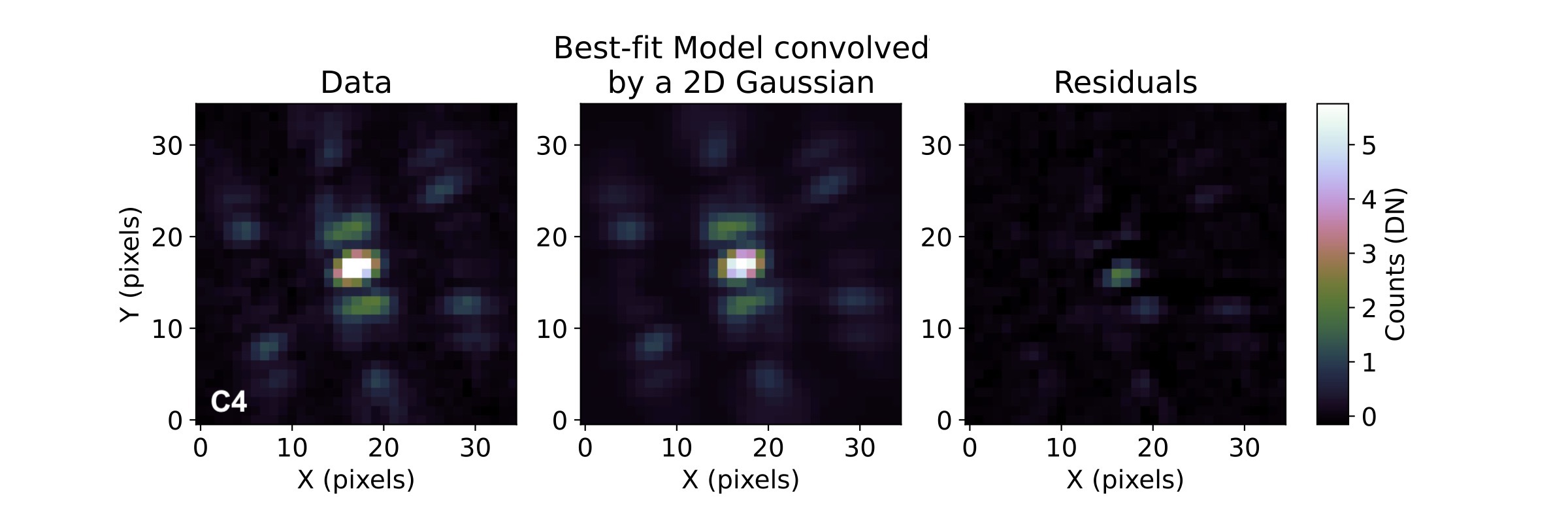}
    \end{subfigure}

    \begin{subfigure}[b]{0.49\textwidth}
        \includegraphics[width=\textwidth]{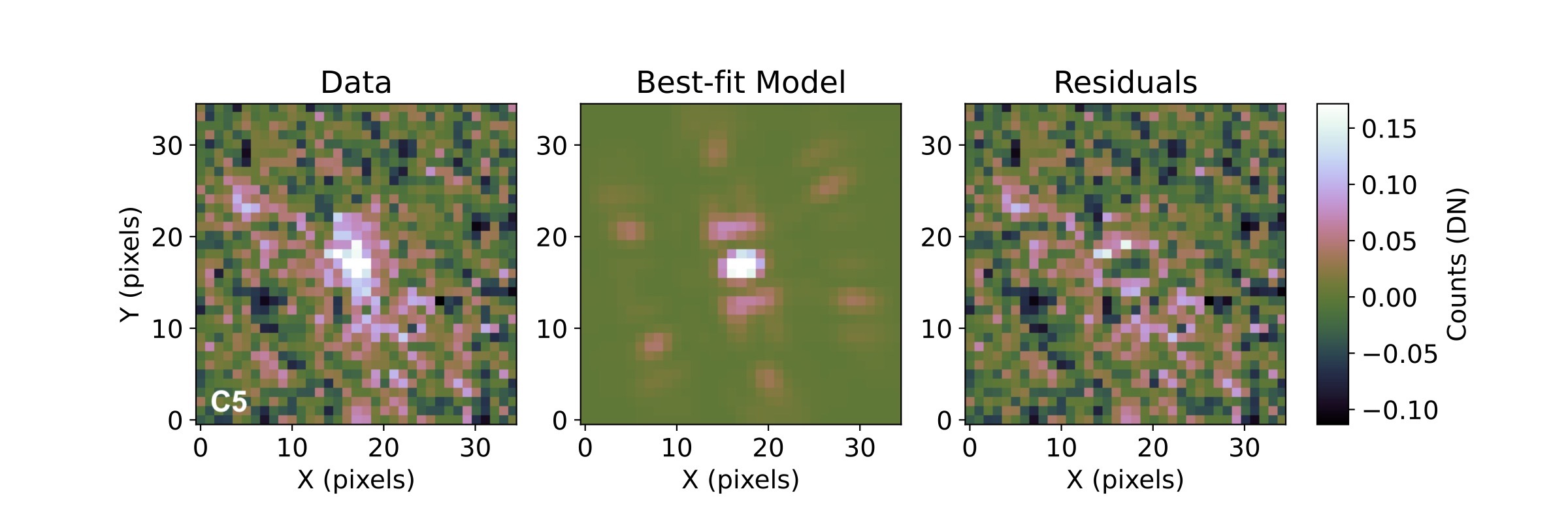}
    \end{subfigure}
    \hfill
    \begin{subfigure}[b]{0.49\textwidth}
        \includegraphics[width=\textwidth]{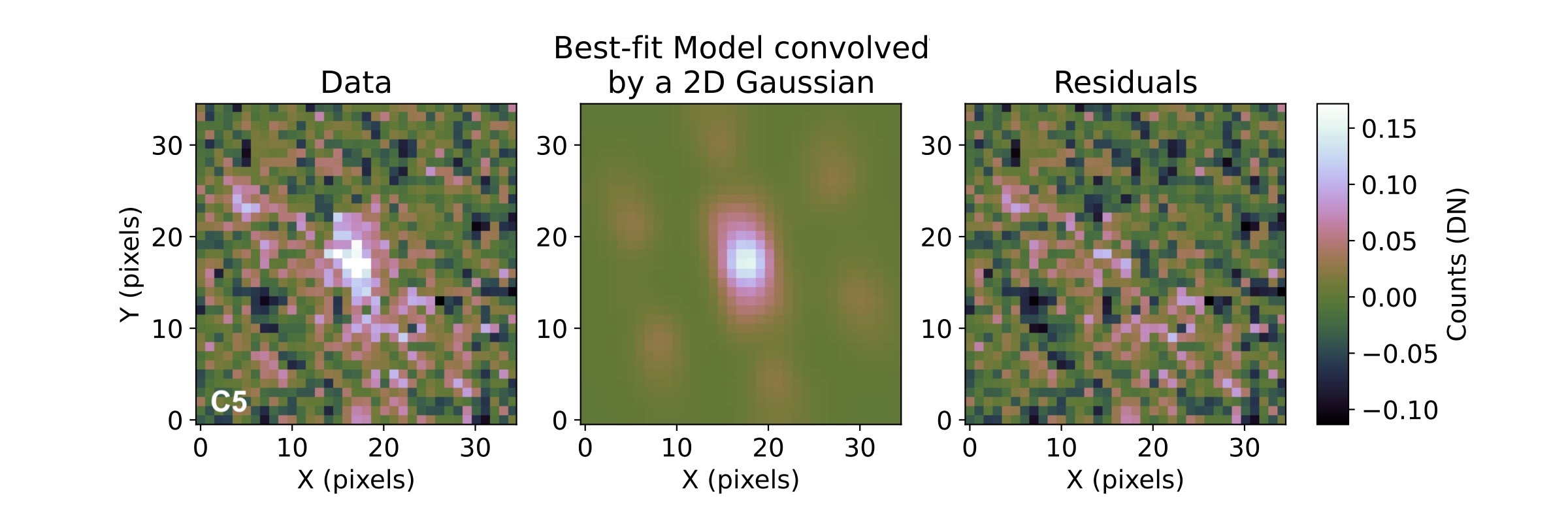}
    \end{subfigure}

    \begin{subfigure}[b]{0.49\textwidth}
        \includegraphics[width=\textwidth]{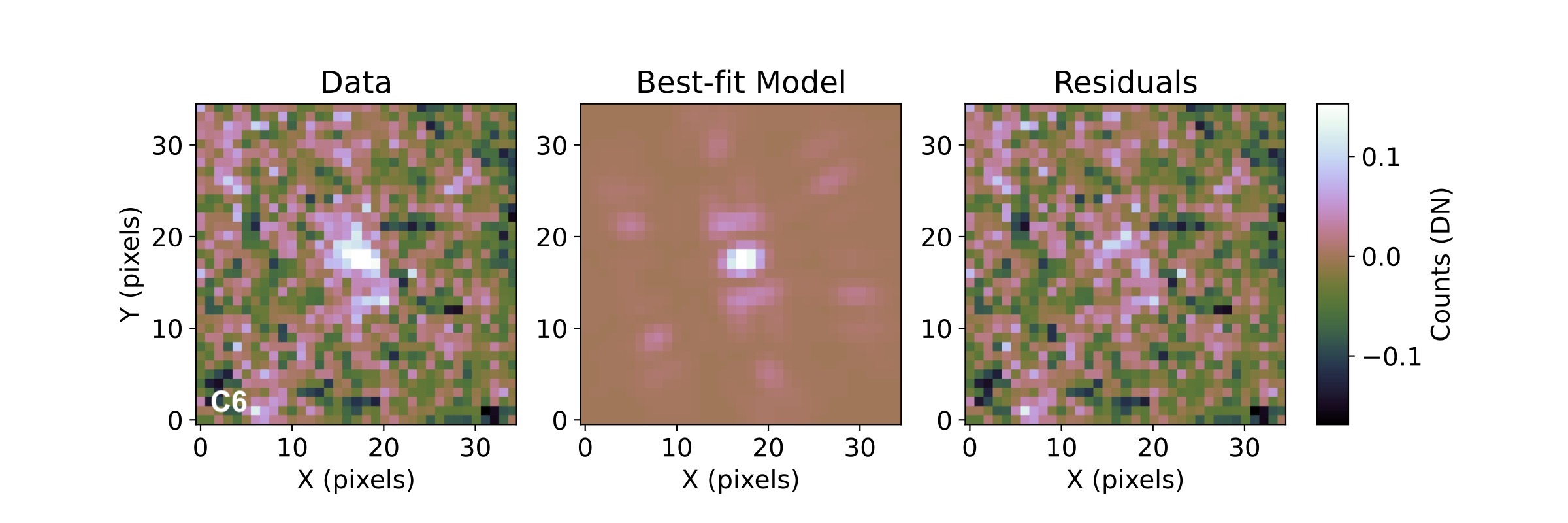}
    \end{subfigure}
    \hfill
    \begin{subfigure}[b]{0.49\textwidth}
        \includegraphics[width=\textwidth]{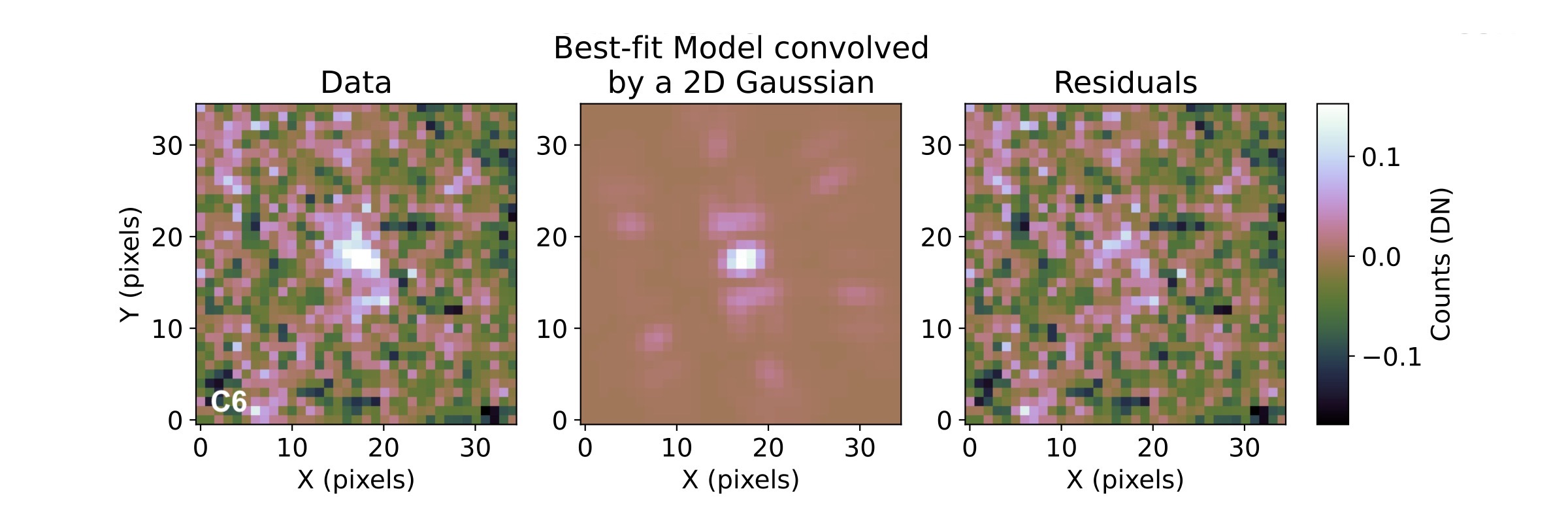}
    \end{subfigure}

    \caption{Characterization, from top to bottom, of the sources C2, C3, C4, C5, and C6 in the F444W field of view. On the left side, modeling coming from forward modeling of a point-like PSF. On the right, convolution of the synthetic PSF, with a 2D Gaussian.}
    \label{sources444}
\end{figure}

\begin{figure*}[h!]
    \centering

    \begin{subfigure}[b]{0.49\textwidth}
        \includegraphics[width=\textwidth]{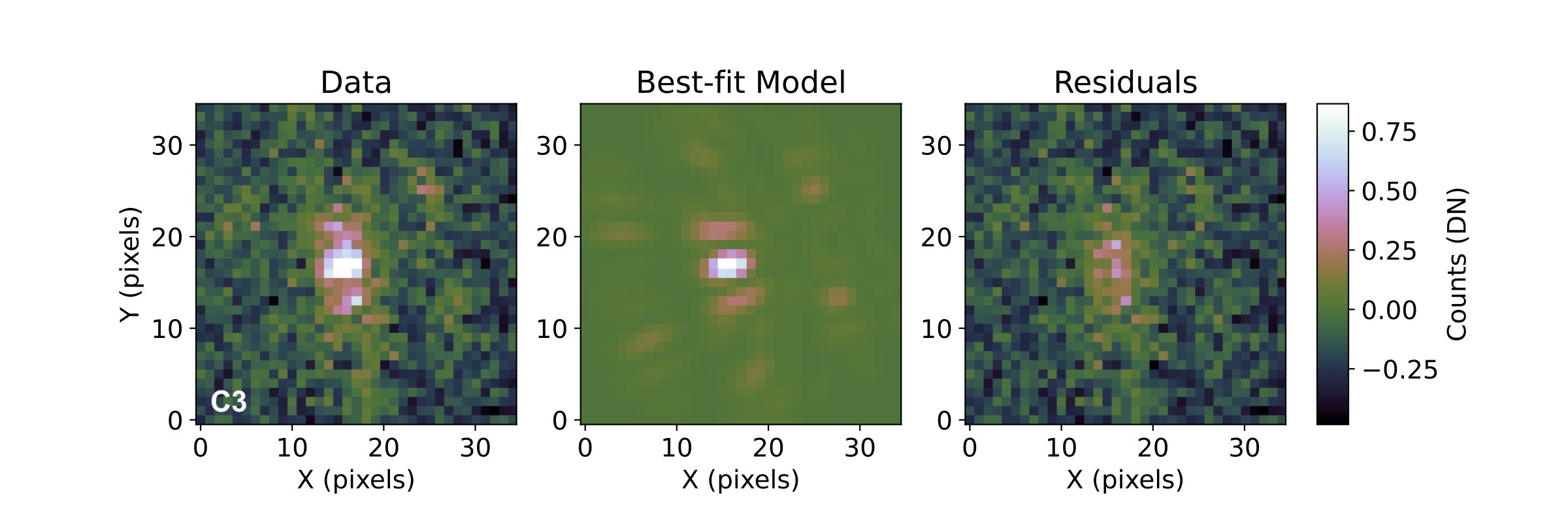}
    \end{subfigure}
    \hfill
    \begin{subfigure}[b]{0.49\textwidth}
        \includegraphics[width=\textwidth]{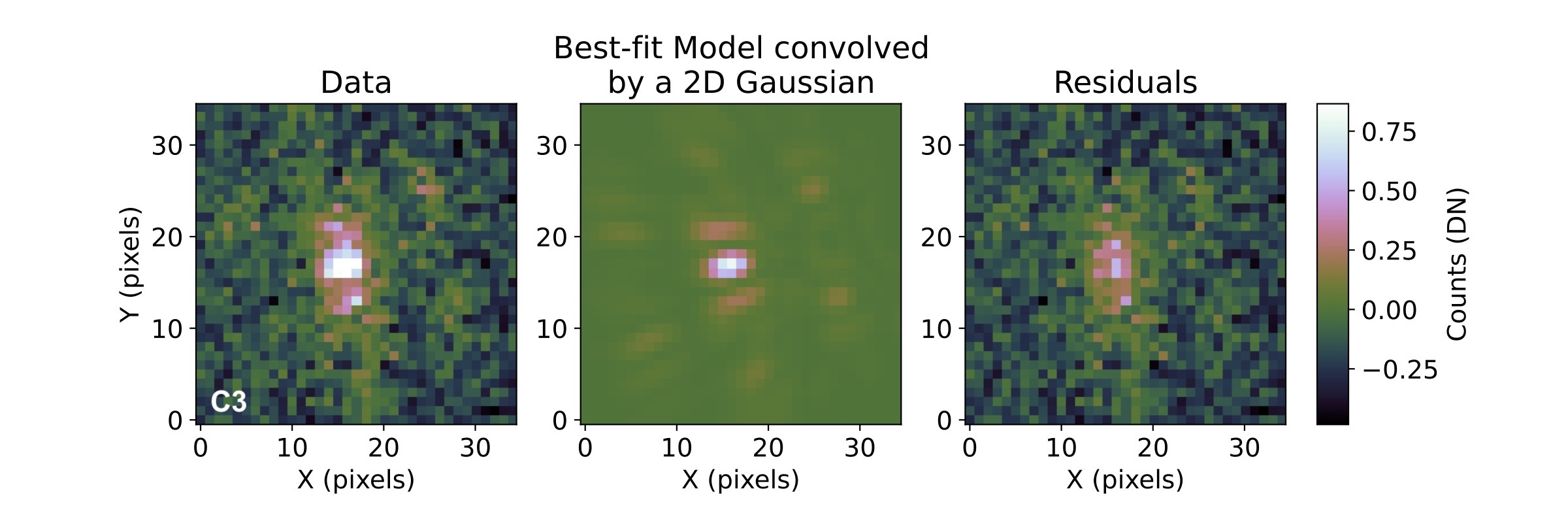}
    \end{subfigure}
    \caption{Characterization of the sources C3 in the F200W field of view. On the left side, modeling coming from forward modeling of a point-like PSF. On the right, convolution of the synthetic PSF, with a 2D Gaussian.}
    \label{source200}
\end{figure*}

\end{appendix}
\end{document}